\documentclass[12pt,onecolumn]{article}
\usepackage{color}
\usepackage{amsmath}
\usepackage{amssymb}
\usepackage{amsfonts}
\usepackage{geometry}
\usepackage{setspace}
\usepackage{amsmath,bm}
\usepackage{graphicx}
\usepackage[section]{placeins}% avoid figures cross the section
\usepackage{subfigure}
\usepackage[sort&compress]{natbib}
\usepackage{booktabs}  
\usepackage{makecell}
\usepackage{caption}% caption
\usepackage[colorlinks, citecolor=blue]{hyperref}% link to the superlink

\author{X. Zhuang$^{1,2}$, S. Zhou$^{1,2,5*}$, G.D. Huynh$^{2}$, P. Aerias$^{4}$, T. Rabczuk$^{3}$}
\title {Phase field modelling and computer implementation: a review}
\UseRawInputEncoding
\begin{document}

%\captionsetup[figure]{labelfont={bf},name={Fig.},labelsep=space,justification=raggedright, singlelinecheck = false}
\captionsetup[figure]{labelfont={bf},name={Fig.},labelsep=space}
% referecne style
%\bibliographystyle{unsrtnat}
%\bibliographystyle{plainnat}
\bibliographystyle{apa}
\setcitestyle{authoryear,round,aysep={},yysep={,}}
% referecne style
%\bibliographystyle{unsrtnat}
%\setcitestyle{numbers,square,aysep={},yysep={,}}
% delte the date
% Increase the section level (declaration}
%\setcounter{tocdepth}{4} 
%\setcounter{secnumdepth}{4}

\date{}
\maketitle

\spacing {1.2}
\noindent
1 Department of Geotechnical Engineering, College of Civil Engineering, Tongji University, Shanghai 200092, P.R. China\\
2 Institute of Continuum Mechanics, Leibniz University Hannover, Hannover 30167, Germany\\
3 Institute of Structural Mechanics, Bauhaus University Weimar, Weimar 99423, Germany\\
4 Department of Mechanical Engineering, Instituto Superior T\'ecnico, University of Lisbon\\
5 Institute for Advanced Study, Tongji University, Shanghai 200092, P.R. China\\
* Corresponding author: Shuwei Zhou (shuwei.zhou@hot.uni-hannover.de)

%\begin{spacing}{2.0}
\begin{abstract}
\noindent This paper presents an overview of the theories and computer implementation aspects of phase field models (PFM) of fracture. The advantage of PFM over discontinuous approaches to fracture is that PFM can elegantly simulate complicated fracture processes including fracture initiation, propagation, coalescence, and branching by using only a scalar field, the phase field. In addition, fracture is a natural outcome of the simulation and obtained through the solution of an additional differential equation related to the phase field. No extra fracture criteria are needed and an explicit representation of a crack surface as well as complex track crack procedures are avoided in PFM for fracture, which in turn dramatically facilitates the implementation. The PFM is thermodynamically consistent and can be easily extended to multi-physics problem by 'changing' the energy functional accordingly. Besides an overview of different PFMs, we also present  comparative numerical benchmark examples to show the capability of PFMs.
\end{abstract}
%\end{spacing}

\noindent Keywords: Phase field, Brittle fracture, Computer implementation, Finite element method, Hydraulic fracture
% Add a content
\tableofcontents

%\twocolumn
\section {Introduction}\label{Introduction}

The prediction of material failure is of major importance in engineering and material science \citep{rabczuk2013computational}. Consequently, many numerical approaches have been developed to handle fracture problems in recent years. They can be classified into two categories namely discontinuous and continuous approaches. Discontinuous approaches introduce a strong discontinuity in the displacement field. Typical discontinuous methods are discrete crack models \citep{ingraffea1985numerical}, the extended finite element method (XFEM) \citep{moes2002extended, moes1999finite}, generalized finite element method (GFEM) \citep{fries2010extended}, and the phantom-node method \citep{chau2012phantom, rabczuk2008new}. Most discontinuous approaches to fracture require a method to represent the crack's topology -- such as triangular facets \citep{zhuang2012fracture} or level sets \citep{zhuang2011accurate} -- and associated crack tracking algorithms. Obtaining complex fracture patterns such as crack branching and crack interactions require additional criteria (the TLS approach \citep{moes2011level} is a discontinuous approach that does not require special criteria for branching). Some of the above methods also employ special schemes to `treat' crack tip singularity, which improves accuracy but imposes other difficulties such as numerical integration.

In contrary to the discontinuous approaches, the continuous approaches to fracture do not introduce discontinuities in the displacement field. Popular continuous approaches include gradient damage models \citep{peerlings1996some}, screened-Poisson models \citep{areias2016damage, areias2016novel} and phase field models (PFMs) \citep{miehe2010thermodynamically, miehe2010phase, borden2012phase}. All these models introduce an intrinsic length scale and smear the fracture over a localization band of finite width. This paper focuses on the phase field model of brittle fracture. Phase-field models of brittle fracture can be traced back to the late 1990s and received extensive development in theory and computer implementation. The crack is diffusively represented by a scalar field (phase field) and the evolution equation of the phase field is used to model crack propagation.

The phase field models recently received extensive attention because they can elegantly simulate complicated fracture processes including crack initiation, propagation, coalescence, and branching quite naturally. In addition, the fracture evolution can be simulated on a fixed mesh. The phase field model avoids the laborious task to track crack surface, which is especially tedious in 3D. 

In the past decade, researchers made a huge effort to develop novel, efficient, and accurate phase field models and achieved enormous progress. This paper aims to review the advances in this direction. It is organized as follows. Section \ref{Theories of phase field models for fracture} gives an overview of the theories used in the phase field modelling. Section \ref{PFMs coupled with different discretization methods} is devoted to the different discretization methods in fracture modelling. Section \ref{Finite element implementation of phase field methods} gives detailed FE discretization and some details about the computer implementation in phase field modeling. An overview of the state-of-the-art applications and extensions of the phase field models is given in Section \ref{Extensions and applications of the phase field methods} followed by Section \ref{Representative numerical examples} which presents some representative numerical examples to show the capability and practicability of PFM in 2D and 3D. Finally, we end with concluding remarks and future directions in Section \ref{Conclusions}.

\section{Theories of phase field models for fracture}\label{Theories of phase field models for fracture}

In this section, we provide the basic theories of the phase field method for fractures. The phase field models are different in the `physics' and `mechanics' communities. The difference between these two models lies in that in the physics community, the models commonly come from the Landau-Ginzburg phase transition \citep{aranson2000continuum} without introducing the idea of length scale to diffuse the crack. In addition, the free energy in the models don't contain the fracture energy. The PFMs in the physics community are often used to model dynamic fractures \citep{aranson2000continuum, karma2001phase, henry2004dynamic}. However, the phase field models in the mechanics community \citep{miehe2010phase, miehe2010thermodynamically}, can be regarded as the extension of Griffith's fracture theory although the crack evolution is similar to the Landau-Ginzburg phase transition while a clear length scale parameter is used. The free energy used for variational analysis contains the fracture energy, which is regularized by using the length scale parameter.

\subsection{Physical models based on Landau-Ginzburg phase transition}
\subsubsection{Aranson, Kalastky, Vinokur model, 2000}

The fracture model presented by \citet{aranson2000continuum} is among the first phase-field like descriptions of crack propagation in brittle materials. Their model  focuses on Mode-I fracture and they implemented their approach in 2D. The displacement $\bm u$ satisfies the standard elastodynamic equation with a damping term:
	\begin{equation}
	\rho \ddot{\bm u}=\eta\Delta \dot{\bm u}+\nabla\cdot\bm\sigma
	\end{equation}
\noindent where $\rho$ is the density of material, $\nabla$ and $\Delta$ are the divergence and Laplace operators, $\bm \sigma$ is the stress tensor, $\eta>0$ is a viscous damping parameter, $\dot {\bm u}=\partial{\bm u}/\partial {t}$, and $\ddot {\bm u}=\partial^2{\bm u}/\partial {t}^2$ with $t$ being the time. 

\citet{aranson2000continuum} defined a local-order parameter $s$ (a field parameter): $s=1$ outside the crack
(no defects) and $s= 0$ inside the crack (all the atomic bonds are broken). The order parameter $s$ is assumed much larger than the inter-atomic distance, thereby obeying the continuum description of fracture. Materials cannot bear tensile stresses and fail at an order parameter below the critical value $s_c$ \citep{aranson2000continuum}. The stress-strain relation subsequently differs from that used in the isotropic linear elasticity $\bm \sigma= \mathbb C : \bm \varepsilon$ by introducing the dependency on the order parameter $s$:
	\begin{equation}
	\bm \sigma=s \mathbb {C} : \bm \varepsilon + \chi\dot s \bm I
	\end{equation}
\noindent where $\mathbb {C}$ is fourth-order elasticity tensor of material, $\bm \varepsilon$ is the strain tensor, $\bm I$ is the second-order identity tensor, and $\chi >0$ is an additional material parameter describing the hydrostatic pressure due to creation of new defects. 

The order parameter $s$ is assumed to be governed by pure dissipative dynamics and $\dot s = -\delta E/ \delta s$ with $E$ being a "free-energy" type functional. Based on the Landau's phase transitions \citep{landau1980statistical}, the simplest form of $E$ reads
	\begin{equation}
	E=\int_\Omega \left[P(s)+D_s|\nabla s|^2 \right] \mathrm{d} \Omega
	\label{free energy 1}
	\end{equation}
\noindent where $P$ is a polynomial functional and $D_s>0$ is an adaptive constant.

In the fracture model of \cite{aranson2000continuum}, the order parameter evolution equation is naturally expressed as
	\begin{equation}
	\dot s = D_s\Delta s- s(1-s)\left[a\left(1+(\mathrm{tr}(\bm\varepsilon)-b)s\right)-c\dot{\bm u}\cdot\nabla s\right]
	\end{equation}
\noindent where $a$, $b$, and $c$ are model parameters.

In general, the Aranson, Kalastky, Vinokur model can simulate multiple fracture behaviors such as crack initiation, propagation, branching, dynamic fracture instability, sound emission and fragmentation. However, some discrepancies still exist between the numerical predications and experimental observations of Mode I crack propagation in a rectangular strip of finite width \citep{aranson2000continuum}.

\subsubsection{Karma, Kessler, Levine model, 2001}

\citet{karma2001phase} proposed a phase-field model for Mode III (antiplane shear) fracture. In their model, only the out-of-plane displacement component exists in a 2D setting. Their approach for fracture emanates from the original phase field models for solidification and the basic free energy function is expressed as \citep{karma2001phase, hakim2009laws}
	\begin{equation}
	E(\bm u, s)=\int_\Omega\left[g(s)\left(\Psi_0(\varepsilon)-\Psi_c\right)+V(s)+\frac 1 2D_s|\nabla s|^2\right]\mathrm{d}\Omega
	\end{equation}
\noindent where $g(s)$ is a function of the order parameter $s$ and satisfies $g(s)>0$ for $0<s\le 1$. $V(s)=s^2(1-s^2)/4$ is the so-called Ginzburg-Landau double-well potential \citep{ambati2015review}. In addition, $\Psi_c$ denotes the strain energy threshold for crack initiation, and $D_s>0$ is the positive constant identical to that in Eq. \eqref{free energy 1}. Note that the function $g(s)=s^{2+a_1}$ is used in \citet{karma2001phase} with $a_1$ a dimensionless coefficient. Subsequently, the energy functional is used to governing the main equations of systems, including the momentum balance equation, the stress-strain relation and the phase-field evolution equation. 

That is, the variation of the functional $E$ with respect to $\bm u$ results in the equilibrium equation: $\rho \ddot{\bm u}=\nabla\cdot \bm\sigma$. The stress-strain form is subsequently represented as follows:
	\begin{equation}
	\bm\sigma(\bm u,s)=g(s)\frac{\partial \Psi_0(\bm\varepsilon)}{\partial{\bm\varepsilon}}=g(s)\mathbb C:\bm\varepsilon
	\end{equation}

The proposed fracture evolution law reads $\tau\dot s=-\frac{\partial E(\bm u, s)}{\partial s}$ with $\tau\ge0$ being a kinetic modulus. The nonlinear evolution equation is thus expressed as follows:
	\begin{equation}
	\tau\dot s=D_s\Delta s-V'(s)-g'(s)\left(\Psi_0(\varepsilon)-\Psi_c\right)
	\end{equation}

\subsubsection{Henry and Levine model, 2004}

\citet{henry2004dynamic} subsequently extended the original model of \citet{karma2001phase}. A modification of the elastic energy is seen in \citet{henry2004dynamic} for better simulating the crack growth of Mode I and II under 2D plane strain condition. The total energy functional retains the basic structure of the Karma, Kessler, Levine model (KKL model) and still prevents the compressed region of the elastic body from cracking. The total energy is expressed as follows,
	\begin{equation}
	E(\bm u, s)=\int_\Omega\left[g(s)\left(\tilde\Psi(\bm\varepsilon)-\Psi_c\right)+V(s)+\frac 1 2D_s|\nabla s|^2\right]\mathrm{d}\Omega
	\end{equation}

\noindent where $\tilde\Psi(\bm\varepsilon)$ is the modified elastic strain energy and it is identical to the standard elastic energy density for a positive volume strain. However, for a negative volume strain, a breaking symmetry term is introduced:
	\begin{equation}
	\tilde\Psi(\bm\varepsilon)=\left\{
		\begin{aligned}
		&\Psi_0(\bm\varepsilon)\hspace{2cm}&\mathrm{if}\hspace{0.2cm}\mathrm{tr}(\bm\varepsilon)>0\\
		&\Psi_0(\bm\varepsilon)-\frac 1 2 a_2 K \mathrm{tr}^2(\bm\varepsilon)\hspace{2cm}&\mathrm{if}\hspace{0.2cm}\mathrm{tr}(\bm\varepsilon)<0
		\end{aligned}\right.
	\end{equation}

\noindent where $\Psi_0(\bm\varepsilon) = \frac 1 2 \lambda \varepsilon_{ii}^2+\mu\varepsilon_{ij}^2$, $\lambda$ and $\mu$ are Lam\'e constants, $K=(\lambda+\mu)/2$, and $a_2>1$ is an arbitrary coefficient. It should be noted that the coupling function $g(s)$ used in the model of \citet{henry2004dynamic} is different from that in \citet{karma2001phase} and here $g(s)=(4-3s)s^3$.

The stress-strain relation used in Henry and Levine model is
	\begin{equation}
	\bm\sigma(\bm u,s)=g(s)\frac{\partial \tilde\Psi(\bm\varepsilon)}{\partial{\bm\varepsilon}}
	\end{equation}

\noindent while the evolution equation of $s$ is modified as
	\begin{equation}
	\tau\dot s=D_s\Delta s-V'(s)-g'(s)\left(\tilde\Psi(\varepsilon)-\Psi_c\right)
	\end{equation}

In summary, the Henry and Levine model can accurately reproduce different behaviors of cracks such as branching, experimentally observed oscillating cracks, and well observed supercritical Hopf bifurcation \citep{henry2004dynamic}.

\subsection{Mechanical models based on Griffith's fracture theory}
\subsubsection{Variational approach to fracture}
The original phase field models \citep{miehe2010phase} in the mechanics community is developed independently for quasi-static fractures. These approaches originate from the variational approach of brittle fracture proposed by \citet{francfort1998revisiting} and refer to the regularization formulation used by \citet{bourdin2000numerical}.

Similar to the models in the physics community, the variational approach to fracture also requires constructing an energy functional that governs the entire fracture process. That is, crack initiation, propagation, and branching is a minimization of the energy functional \citep{francfort1998revisiting}:
	\begin{equation}
	E(\bm u, \Gamma)=\int_\Omega\Psi_0(\bm\varepsilon)\mathrm{d}\Omega+G_c\int_\Gamma\mathrm{d}S
	\end{equation}

\noindent where $G_c$ is the fracture toughness and also referred to as the critical energy release rate. $\Gamma\in\Omega$ is an admissible crack set and the displacement field $\bm u$ is discontinuous across $\Gamma$.

To make the variational formulation amenable to a numerical implementation, \citet{bourdin2000numerical} proposed a regularized version of the variational formulation. Without treating the free discontinuity sets of the displacement field $\bm u$, an auxiliary scalar $s(\bm x,t)$ ($\bm x$ the position vector) is employed to diffusely represent the sharp fracture geometry. Identical to the physical phase field models, the scalar field $s$ continuously transits between $s=1$ (intact material) and $s=0$ (fully damaged material). The main advantage of introduction of the phase-field is that the representation of cracks is no longer mesh or geometry based \citep {kuhn2013numerical}. The regularized approach can be easily implemented in a finite element framework and retain the main advantage of the variational formulation to model cracks. In the regularization formulation of \citet{bourdin2000numerical}, the modified energy functional reads
	\begin{equation}
	E(\bm u, s)=\int_\Omega(s^2+\eta)\Psi_0(\bm\varepsilon(\bm u))\mathrm{d}\Omega+G_c\int_\Omega\left[\frac 1 {4\epsilon}(1-s)^2+\epsilon|\nabla s|^2\right]\mathrm{d}\Omega
	\label{bourdin energy functional}
	\end{equation}

\noindent where the parameter $0<\eta\ll1$ is used to avoid numerical singularity when the material is broken. Another parameter $\epsilon$ has the dimension of a length and controls the transition zone between the fully damaged and intact body. To obtain the displacement $\bm u$ and phase field $s$, minimization of the modified energy functional is also required. It should be noted that the regularized approach can be recovered to the original variational approach to fracture when $\epsilon\rightarrow0$ in the sense of $\Gamma$-convergence \citep{braides2006handbook}.

\subsubsection{Kuhn and M{\"u}ller model, 2008}

\citet{kuhn2008phase} reinterpreted the crack variable as a phase field order parameter and they also regarded cracking as a phase transition problem. By applying the thermodynamics framework of  order parameter based models, \citet{kuhn2008phase} reformulated the minimization problem in Eq. \eqref{bourdin energy functional} and used the stress equilibrium equation $\nabla\cdot\bm\sigma=\bm 0$ with
	\begin{equation}
	\bm\sigma=(s^2+\eta)\frac{\partial\Psi_0(\bm\varepsilon)}{\partial\bm\varepsilon}=(s^2+\eta)\mathbb C:\bm\varepsilon
	\end{equation}

The crack propagation is simulated by using  the evolution equation of the order parameter $s$. The Ginzburg-Landau type evolution equation \citep{kuhn2008phase} then reads
	\begin{equation}
	\dot s =-\dot M\left[2s\Psi_0(\bm\varepsilon)-G_c\left(2\epsilon\Delta s +\frac{1-s}{2\epsilon}\right)\right]
	\label{evolution equation of kuhn}
	\end{equation}

Equation \eqref{evolution equation of kuhn} is used under the irreversibility constraint of crack evolution and with a mobility parameter $M$. $M\ge0$ governs the dissipation of stable crack propagation. At a finite value of $M$, the crack model can be also regarded as a viscous quasi-static model \citep{miehe2010phase} and in a limit case $M\rightarrow\infty$ quasi-static crack propagation thereby obeys
	\begin{equation}
	2s\Psi_0(\bm\varepsilon)-G_c\left(2\epsilon\Delta s +\frac{1-s}{2\epsilon}\right)=0
	\end{equation}

In \citet{kuhn2008phase}, the numerical model is implemented within the finite element framework and an implicit Euler scheme is used for time integration. Note that the Kuhn and M{\"u}ller model \citep{kuhn2008phase} has the similar forms and derivations of the Karma, Kessler, Levine model \citep{karma2001phase} and Henry and Levine model \citep{henry2004dynamic}.

\subsubsection{Amor, Marigo, Maurini model, 2009}

The originally developed phase field model (e.g. Eq. \eqref{bourdin energy functional}) does not distinguish
between fractures due to compression and tension. Therefore, some unrealistic crack patterns are reported in \citet{bourdin2000numerical}. To avoid unrealistic simulations, elastic energy must be decomposed into different parts. \citet{amor2009regularized} made the preliminary contribution to prevent phase field representation of cracks due to compression. \citet{amor2009regularized} modified the regularized formulation of Eq. \eqref{bourdin energy functional} by introducing an additional decomposition of the elastic energy $\Psi_0(\bm\varepsilon)$ into volumetric and deviatoric parts. That is, $\Psi_0=\Psi_0^++\Psi_0^-$ with
	\begin{equation}
	\left\{\begin{aligned}
	&\Psi_0^+(\bm\varepsilon)=\frac 1 2 K_n\langle\mathrm{tr}(\bm\varepsilon)\rangle_+^2+\mu(\bm\varepsilon^D:\bm\varepsilon^D)\\
	&\Psi_0^+(\bm\varepsilon)=\frac 1 2 K_n\langle\mathrm{tr}(\bm\varepsilon)\rangle_-^2
	\label{armor decomposition}
	\end{aligned}\right.
	\end{equation}

\noindent where $K_n=\lambda+\mu$, $\bm\varepsilon^D=\bm\varepsilon-\frac 1 3 \mathrm{tr}(\bm\varepsilon)\bm I$, and operators $\langle \ast \rangle_\pm=\frac 1 2(\ast \pm|\ast|)$. \citet{amor2009regularized} subsequently proposed the modified energy functional as 
	\begin{equation}
	E(\bm u, s)=\int_\Omega\left((s^2+\eta)\Psi_0^+(\bm\varepsilon)+\Psi_0^-(\bm\varepsilon)\right)\mathrm{d}\Omega+G_c\int_\Omega\left[\frac 1 {4\epsilon}(1-s)^2+\epsilon|\nabla s|^2\right]\mathrm{d}\Omega
	\label{amor energy functional}
	\end{equation}

\citet{amor2009regularized} applied an alternate minimization algorithm to solve for the energy functional \eqref{amor energy functional} and achieved local minimization. They solved a series of minimization sub-problems on $\bm u$ at a fixed $s$, and vice versa on $s$ at a fixed $\bm u$ until convergence. \citet{ambati2015review} pointed out that a good outcome of the energy split of $\Psi_0$ is the resulting stress-strain relation:
	\begin{equation}
	\begin{aligned}
	\bm\sigma(\bm u,s)=&(s^2+\eta)\frac{\partial\Psi_0^+}{\partial\bm\varepsilon}+\frac{\partial\Psi_0^-}{\partial\bm\varepsilon}\\=&(s^2+\eta)\left[K_n\langle\mathrm{tr}(\bm\varepsilon)\rangle_+\bm I+2\mu\bm\varepsilon^D\right]+K_n\langle\mathrm{tr}(\bm\varepsilon)\rangle_-
	\end{aligned}
	\end{equation}

However, the phase evolution equation suggested by \citet{ambati2015review} is not the variational outcome of the energy functional \eqref{amor energy functional}. Instead, the Ginzburg-Landau type equation is still used and $\Psi_0$ in Eq. \eqref{evolution equation of kuhn} is replaced by $\Psi_0^+$:

	\begin{equation}
	\dot s =-\dot M\left[2s\Psi_0^+(\bm\varepsilon)-G_c\left(2\epsilon\Delta s +\frac{1-s}{2\epsilon}\right)\right]
	\label{evolution equation of suggest}
	\end{equation}

Thus, the evolution of $s$ in Eq. \eqref{evolution equation of suggest} is driven only by the positive parts of the elastic energy. That is, only volumetric dilatation and shear deformation can produce cracks. It should be noted here that the Amor, Marigo, Maurini model does not obey the $\Gamma$-convergence although the energy decomposition model \citep{amor2009regularized} produces adequate numerical results. The reason is that it is not clear what kind of functional and physical process are to be recovered when $\epsilon\rightarrow\infty$.

\subsubsection{Miehe et al. model, 2010}

\citet{miehe2010phase,miehe2010thermodynamically} presented a thermodynamically consistent phase field model of brittle fracture. The model of \citet{miehe2010phase,miehe2010thermodynamically} is another important development after the  variational framework of fracture was presented by \citet{francfort1998revisiting} and \citet{bourdin2000numerical}. This model  is based on continuum mechanics and uses an auxiliary scalar field $\phi\in[0,1]$ (the so-called phase field). The additional phase field is used to smear the sharp crack shape. We can also regard $\phi = 1-s$ here with relation to the former phase field models such as the Amor, Marigo, Maurini model. $\phi=0$ and $\phi =1$ represent the intact and fully broken states, respectively. A length scale parameter $l_0$ is used in Miehe et al. model and $l_0$ controls the transition region between the fully broken and intact bodies. Another elastic energy decomposition method is used with $\Psi_0=\Psi_0^++\Psi_0^-$ based on the spectral decomposition of strain tensor $\bm\varepsilon$: $\bm\varepsilon=\sum_{I=1}^3 \varepsilon_I\bm n_I\otimes \bm n_I$. $\{\varepsilon\}_{I=1}^3$ and $\{\bm n\}_{I=1}^3$ are the principal strains and their directions. The decomposed elastic energy reads
	\begin{equation}
	\Psi_0^\pm(\bm\varepsilon)=\frac1 2\lambda\langle\mathrm{tr}(\bm\varepsilon)\rangle_\pm^2+\mu\mathrm{tr}(\bm\varepsilon_\pm^2)
	\end{equation}
\noindent where $\bm\varepsilon_\pm=\sum_{I=1}^3\langle \varepsilon_I\rangle_\pm\bm n_I\otimes \bm n_I$.

The modified elastic energy $\Psi_e$ used in the total energy functional is expressed as follows,
	\begin{equation}
	\Psi_e=\left[(1-\phi)^2+\eta)\right]\Psi_0^++\Psi_0^-
	\end{equation}

The governing equation of the displacement is $\nabla\cdot\bm\sigma=\bm 0$ and the stress tensor is expressed as
	\begin{equation}
	\bm\sigma(\bm u,\phi)=\left[(1-\phi)^2+\eta\right]\frac{\partial\Psi_0^+}{\partial\bm\varepsilon}+\frac{\partial\Psi_0^-}{\partial\bm\varepsilon}
	\label{miehe sigma}
	\end{equation}

The evolution equation of phase field in the model of \citet{miehe2010phase,miehe2010thermodynamically} is
	\begin{equation}
	\xi\dot\phi=2(1-\phi)\Psi_0^+(\bm\varepsilon)+\frac {G_c}{l_0}\left(\phi-l_0^2\Delta\phi\right)
	\label{miehe phase field}
	\end{equation}
\noindent where $\xi$ is a viscosity parameter.

It should be noted that if $l_0=2\epsilon$ and $\xi=\frac 1 M$, the structure of Eqs. \eqref{miehe sigma} and \eqref{miehe phase field} resembles those equations in the Amor, Marigo, Maurini model \citep{amor2009regularized}. The energy decomposition of \citet{miehe2010phase,miehe2010thermodynamically} is different from that of \citet{amor2009regularized}. \citet{miehe2010phase,miehe2010thermodynamically} only distinguishes compressive and tensile parts and imposed a degradation of the compressive energy component. However, the deviatoric stress is considered in \citet{amor2009regularized}. Another feature of Miehe's model is that Eq. \eqref{miehe sigma} results in a highly non-linear stress-strain relationship. Therefore, a much higher computational cost is needed compared with former physical and mechanical phase field methods. In order to ensure a monotonically increasing phase-field, the irreversibility condition must be imposed as a constraint during compression or unloading. \citet{miehe2010phase,miehe2010thermodynamically} proposed a new approach where a history-field $H$ is defined in a loading process:
	\begin{equation}
	H(\bm x,t) = \max \limits_{\tau\in[0,t]}\psi_0^+\left(\bm\varepsilon(\bm x,\tau)\right)
	\end{equation}

The introduction of the history field enhances the phase-field formulations and overcomes some implementation difficulties. Replacing $\Psi_0^+$ in Eq. \eqref{miehe phase field} by $H$, the finally used evolution equation reads
	\begin{equation}
	\xi\dot\phi=2(1-\phi)H+\frac {G_c}{l_0}\left(\phi-l_0^2\Delta\phi\right)
	\end{equation}

The common isotropic, anisotropic, and hybrid phase field models in the mechanics community are summarized in Table \ref{Different forms of phase field methods}. Thus, the differences in governing equations and driving forces for different PFMs can be easily identified by readers.

	\begin{table}[htbp]
	\caption{Different forms of phase field methods in mechanics community}
	%\footnotesize
	\scriptsize
	\label{Different forms of phase field methods}
	\centering
	%\begin{tabular}{p{20pt}p{20pt}p{200pt}p{200pt}}%|p{2cm}<{\centering}|
	\begin{tabular}{m{3cm}<{\centering} m{8cm}<{\centering} m{5cm}<{\centering}}
		%\hline
		\toprule
		Type &  Governing equations &  Definition\\
		%\hline
		\midrule
		Isotropic &\begin{equation*}
			\left\{\begin{aligned}
				&\nabla\cdot\bm\sigma+\bm b =0\\
				&-l_0^2\Delta \phi+\phi=\frac {2l_0}{G_c}(1-\phi)H\\
				&\bm\sigma(\bm u,\phi)=(1-\phi)^2\frac {\partial\psi_0}{\partial\bm\varepsilon}\\
				&\mathbb{D}=\frac{\partial\bm\sigma}{\partial\bm\varepsilon}\\
			\end{aligned}\right.
		\end{equation*}&  $H = \max \limits_{\tau\in[0,t]}\psi_0\left(\bm\varepsilon(\bm x,\tau)\right)$ with $\psi_0=\frac \lambda 2 (\mathrm{tr}(\bm\varepsilon))^2+\mu \mathrm{tr}(\bm\varepsilon^2)$\\
		Anisotropic \citep{miehe2010phase} &\begin{equation*}
			\left\{\begin{aligned}
				&\nabla\cdot\bm\sigma+\bm b =0\\
				&\left(\frac{2l_0 H}{G_c}+1\right)\phi-l_0^2\nabla^2\phi=\frac{2l_0 H}{G_c}\\
				&\bm\sigma(\bm u,\phi)=(1-\phi)^2\frac {\partial\psi_\varepsilon^+}{\partial\bm\varepsilon}+\frac {\partial\psi_\varepsilon^-}{\partial\bm\varepsilon}\\
				&\mathbb{D}=\frac{\partial\bm\sigma^+}{\partial\bm\varepsilon}+\frac{\partial\bm\sigma^-}{\partial\bm\varepsilon}\\
			\end{aligned}\right.
		\end{equation*}& $H = \max \limits_{\tau\in[0,t]}\psi_\varepsilon^+\left(\bm\varepsilon(\bm x,\tau)\right)$, $\bm\sigma^+=(1-\phi)^2\frac {\partial\psi_\varepsilon^+}{\partial\bm\varepsilon}$, $\bm\sigma^-=\frac {\partial\psi_\varepsilon^-}{\partial\bm\varepsilon}$, $\psi_{\varepsilon}^{\pm}(\bm \varepsilon) = \frac{\lambda}{2}\langle \mathrm{tr}(\bm\varepsilon)\rangle_{\pm}^2+\mu \mathrm{tr} \left(\bm\varepsilon_{\pm}^2\right)$\\
		Anisotropic \citep{amor2009regularized} &\begin{equation*}
			\left\{\begin{aligned}
				&\nabla\cdot\bm\sigma+\bm b =0\\
				&-l_0^2\Delta \phi+\phi=\frac {2l_0}{G_c}(1-\phi)H\\
				&\bm\sigma(\bm u,\phi)=(1-\phi)^2\frac {\partial\psi_0^+}{\partial\bm\varepsilon}+\frac {\partial\psi_0^-}{\partial\bm\varepsilon}\\
				&\mathbb{D}=\frac{\partial\bm\sigma^+}{\partial\bm\varepsilon}+\frac{\partial\bm\sigma^-}{\partial\bm\varepsilon}\\
			\end{aligned}\right.
		\end{equation*}& $H = \max \limits_{\tau\in[0,t]}\psi_0^+\left(\bm\varepsilon(\bm x,\tau)\right)$, $\bm\sigma^+=(1-\phi)^2\frac {\partial\psi_0^+}{\partial\bm\varepsilon}$, $\bm\sigma^-=\frac {\partial\psi_0^-}{\partial\bm\varepsilon}$, $\psi_0^+=\frac 1 2 K_n\langle(\mathrm{tr}(\bm\varepsilon)\rangle_+^2+\mu(\bm\varepsilon^{\mathrm{dev}}:\bm\varepsilon^{\mathrm{dev}})$,  $\psi_0^-=\frac 1 2 K_n\langle(\mathrm{tr}(\bm\varepsilon)\rangle_-^2$ with $K_n$ a physical parameter\\
		Hybrid \citep{ambati2015review} &\begin{equation*}
			\left\{\begin{aligned}
				&\nabla\cdot\bm\sigma+\bm b =0\\
				&-l_0^2\Delta \phi+\phi=\frac {2l_0}{G_c}(1-\phi)H\\
				&\bm\sigma(\bm u,\phi)=(1-\phi)^2\frac {\partial\psi_0}{\partial\bm\varepsilon}\\
				&\mathbb{D}=\frac{\partial\bm\sigma}{\partial\bm\varepsilon}\\
			\end{aligned}\right.
		\end{equation*}& $H = \max \limits_{\tau\in[0,t]}\psi_\varepsilon^+\left(\bm\varepsilon(\bm x,\tau)\right)$, $\psi_0=\frac \lambda 2 (\mathrm{tr}(\bm\varepsilon))^2+\mu \mathrm{tr}(\bm\varepsilon^2)$, $\psi_{\varepsilon}^{+}(\bm \varepsilon) = \frac{\lambda}{2}\langle \mathrm{tr}(\bm\varepsilon)\rangle_{+}^2+\mu \mathrm{tr} \left(\bm\varepsilon_{+}^2\right)$\\
		Hybrid \citep{zhang2017modification} &\begin{equation*}
			\left\{\begin{aligned}
				&\nabla\cdot\bm\sigma+\bm b =0\\
				&(1-\phi)\left(\frac {H_1}{G_{cI}}+\frac {H_2}{G_{cII}}\right)-\frac 1 {2l_0}\phi + \frac {l_0} 2 \Delta\phi =0\\
				&\bm\sigma(\bm u,\phi)=(1-\phi)^2\frac {\partial\psi_0}{\partial\bm\varepsilon}\\
				&\mathbb{D}=\frac{\partial\bm\sigma}{\partial\bm\varepsilon}\\
			\end{aligned}\right.
		\end{equation*}& $H_1 = \max \limits_{\tau\in[0,t]}\lambda\langle\mathrm{tr}(\bm\varepsilon)\rangle_+^2$, $H_2 = \max \limits_{\tau\in[0,t]}\mu\mathrm{tr}[\langle\bm\varepsilon\rangle_+^2]$, $G_{cI}$ and $G_{cII}$ are the critical energy release rate of modes I and II\\
		%\hline
			Hybrid \citep{zhou2019phase2} &\begin{equation*}
			\left\{\begin{aligned}
				&\nabla\cdot\bm\sigma+\bm b =0\\
				&-l_0^2\Delta \phi+\phi=\frac {2l_0}{G_c}(1-\phi)H\\
				&\bm\sigma(\bm u,\phi)=(1-\phi)^2\frac {\partial\psi_0}{\partial\bm\varepsilon}\\
				&\mathbb{D}=\frac{\partial\bm\sigma}{\partial\bm\varepsilon}\\
			\end{aligned}\right.
		\end{equation*}& $H = \max \limits_{\tau\in[0,t]}\psi_p\left(\bm\varepsilon(\bm x,\tau)\right)$, $\psi_p=\sum_{i=j+1}^{3}\sum_{j=1}^{2}\frac 1 {2G} \left\langle \frac {\mu(\varepsilon_{ip}-\varepsilon_{jp})}{\mathrm{cos}\varphi}+\left[\lambda(\varepsilon_{1p}+\right.\right.$ $\left.\left.+\varepsilon_{2p}+\varepsilon_{3p})+\mu(\varepsilon_{ip}+\varepsilon_{jp})\right]\mathrm{tan}{\varphi}-c \right\rangle_+^2$\\
		\bottomrule
	\end{tabular} 
\end{table}

\subsection{Phase field approximation of the sharp crack topology}
	\subsubsection{Classic crack surface density function}

For the crack phase field $s(\bm x, t)$ or $\phi(\bm x,t)$, let us follow \citet{miehe2010phase,miehe2010thermodynamically} and introduce a regularized functional $A(\phi)$ or $A(s)$ such that
	\begin{equation}
	\left\{
	\begin{aligned}
	&A_d(\phi)=\int_\Omega\gamma(\phi,\nabla\phi)\mathrm{d}\Omega\approx\int_\Omega\delta_s\mathrm{d}\Omega=\int_\Gamma\mathrm{d}S=A_s\\
	&A_d(s)=\int_\Omega\gamma(s,\nabla s)\mathrm{d}\Omega\approx\int_\Omega\delta_s\mathrm{d}\Omega=\int_\Gamma\mathrm{d}S=A_s
	\end{aligned}
	\right.
	\end{equation}
\noindent where $\gamma(\phi,\nabla\phi)$ or $\gamma(s,\nabla s)$ is the crack surface density function, and $A_s$ is the crack area. The crack surface density function approximates the Dirac-delta $\delta_s$ along the crack $\Gamma$, and is composed of the crack phase-field and its spatial gradient.

\citet{bourdin2000numerical} proposed the original form of the crack surface density function as
	\begin{equation}
	\gamma(s,\nabla s)=\frac 1 {4\epsilon}(1-s)^2+\epsilon|\nabla s|^2
	\end{equation}

\citet{miehe2010phase,miehe2010thermodynamically} followed the original form of \citet{bourdin2000numerical} but used different parameters and phase field:
	\begin{equation}
	\gamma(\phi,\nabla\phi) =\frac 1 {2l_0}\phi^2+\frac {l_0} 2 |\nabla \phi|^2
	\end{equation}

In the physical phase field models, the crack surface density function follows a Ginzburg-Landau double-well potential \citep{karma2001phase, hakim2009laws}:
	\begin{equation}
	\gamma(s,\nabla s) =\frac {s^2(1-s^2)} 4 +\frac 1 2D_s|\nabla s|^2
	\end{equation}

	\subsubsection{Generic geometric crack function}
		\paragraph{Kuhn, Schl\"uter, M\"uller model, 2015}

\citet{kuhn2015degradation} proposed a generic energy density functional $\Psi$ of a phase field fracture
model:
	\begin{equation}
	\Psi=\left(g(s)+\eta\right)\Psi_0+\frac {G_c} {2c_w}\left(\frac {w(s)}{4\epsilon}+\epsilon|\nabla s|^2\right)
	\end{equation}

\noindent where $g(s)$ is the energetic degradation function with $g(1)=1$ and $g(0)=0$. The function $w(s)$ models the local fracture energy and $c_w$ is a normalization constant that results in the integral of $ \left(\frac {w(s)}{4\epsilon}+\epsilon|\nabla s|^2\right)/(2c_w)$ over the fractured domain converges to the surface measure of the crack set when $\epsilon\rightarrow0$ \citep{kuhn2015degradation}. \citet{kuhn2015degradation} concluded two types of $w(s)$, namely, the double well function $w(s)=16s^2(1-s^2)$ and monotonous functions $w(s)=(1+\beta s)(1-s)$ with $\beta\in[-1,1]$.  The convex quadratic function with $\beta=-1$ is mostly used, such as in the model of \citet{miehe2010phase}. In addition, models with $\beta=0$ can be found in \citet{HOSSAIN201415}.

\begin{figure}[htbp]
	\centering
	\includegraphics[width = 8cm]{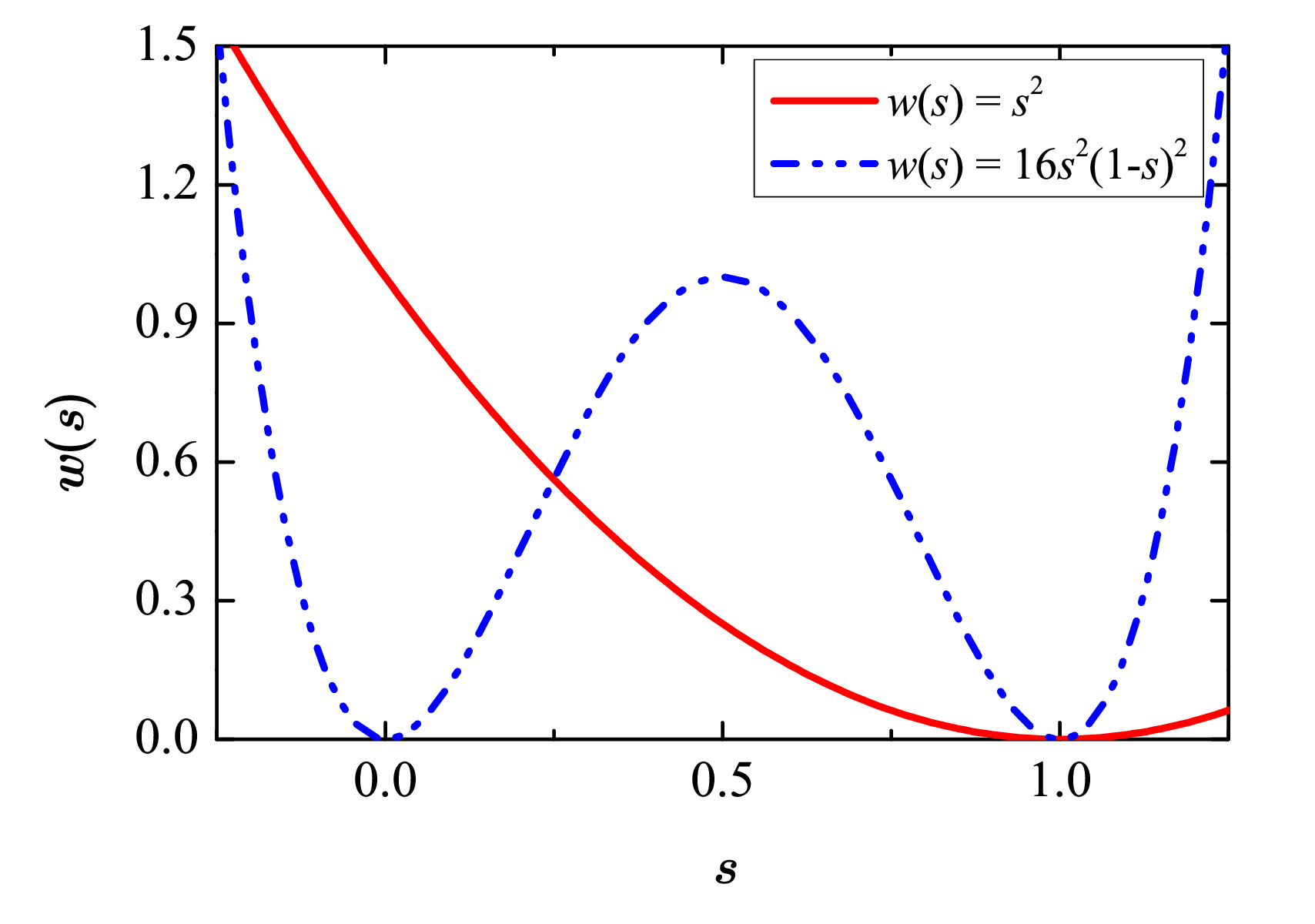}
	\caption{Different degradation functions}
	\label{Different degradation functions}
	\end{figure}

\citet{kuhn2013numerical} compared the function of $w(s)=16s^2(1-s^2)$ and $w(s)=(1-s)^2$. The function $w(s)$ of these two forms is shown in Fig. \ref{Different degradation functions}. Their observations proved that the double well potential naturally models the irreversibility of fracture processes to a certain extend because of local maximum acting as an energy barrier between the broken and undamaged phase. In contrast, the crack phase field needs additional irreversibility to prevent crack healing for $w(s)=(1-s)^2$ due to lack of energy barrier between the broken and undamaged states.

\paragraph{Wu's model, 2017}
\citet{wu2017unified} proposed another generic form for the crack surface density function $\gamma(\phi,\nabla\phi)$:
	\begin{equation}
	\gamma(\phi,\nabla\phi)=\frac 1 {c_0}\left(\frac 1 b \kappa(\phi)+b|\nabla\phi|^2\right)
	\end{equation}
\noindent with 
	\begin{equation}
	c_0=4\int_0^1\sqrt{\kappa(\beta)}\mathrm{d}\beta
	\end{equation}
\noindent where $\kappa(\phi)$ is the geometric function that characterizes homogeneous evolution of the crack phase-field, $b$ is an internal length scale parameter that regularizes the sharp crack, $c_0 > 0$ is a scaling parameter by which the regularized functional $A_d(\phi)$ can be recovered to the
crack surface $A_s$ when $b\rightarrow 0$. The proposed crack geometric function $\kappa(\phi)$ must obey
	\begin{equation}
	\kappa(0)=0,\hspace{2cm}\mathrm{and}\hspace{0.5cm}\kappa(1)=1
	\end{equation}

\subsection{Energetic degradation function}
\subsubsection{Classic forms}
As aforementioned, $g(s)$ is a degradation function that models the degradation of the elastic energy when a fracture initiates. Seen from a simple stress-strain relation $\bm\sigma=(g(s)+\eta)\mathbb c\bm \varepsilon$, the function $g(s)$ also models the decrease in the stiffness in a broken material when the phase field $s\rightarrow0$. In addition, the energetic degradation function $g(s)$ must monotonically increase with $g(1)=1$ and $g(0)=0$.  Because the derivative $g'(s)$ enters the evolution equation of the phase field, $g'(0)=0$ must be imposed to eliminate the elastic driving term $\frac{\partial \Psi}{\partial s}=\frac 1 2 g'(s)\bm\varepsilon{\mathbb c:\bm\varepsilon}$ when $s=0$.

Some of the classic degradation functions have been mentioned in the overview of the most important phase field models such as the quadratic function $g(s)=s^2$ in \citet{bourdin2000numerical}, and the quartic function $ g(s)=4s^3-3s^4$ in the KKL model \citep{karma2001phase}. In addition, \citet{kuhn2015degradation} adopted a cubic function:
	\begin{equation}
	g(s)=3s^2-2s^3
	\end{equation}

\subsubsection{Wu's model, 2017}
\citet{wu2017unified} proposed a unified phase field damage model for brittle fracture. The elastic energy density $\Psi$ is modeled as
	\begin{equation}
	\Psi=v(\phi)\Psi_0
	\end{equation}

The monotonically decreasing energetic function $v(\phi)$ describes degradation of the initial energy $\Psi_0$ with the crack phase-field evolution. The function $v(\phi)$ obeys \citep{miehe2010phase}
	\begin{equation}
	v'(\phi)<0,\hspace{1cm}\mathrm{and}\hspace{0.5cm}v(0)=1,\hspace{0.5cm}v(1)=0,\hspace{0.5cm}v'(1)=0
	\end{equation}

The generic form of the energetic degradation function proposed by \citet{wu2017unified} is given by
	\begin{equation}
	v(\phi)=\frac{(1-\phi)^{p_1}}{(1-\phi)^{p_1}+Q(\phi)}
	\end{equation}
\noindent with the exponent $p_1 > 0$ and continuous function $Q(\phi) > 0$.

For a strictly positive function $Q(\phi)$, the following polynomial is considered
	\begin{equation}
	Q(\phi)=\alpha_1\phi+\alpha_1\alpha_2\phi^2+\alpha_1\alpha_2\alpha_3\phi^3+\cdots=\alpha_1\phi P(\phi)
	\end{equation}
	\begin{equation}
	P(\phi)=1+\alpha_2\phi+\alpha_2\alpha_3\phi^2+\cdots
	\end{equation}

More details about the determination of the coefficients $\alpha_1$, $\alpha_2$, $\cdots$ can be referred to \citet{wu2017unified}.

\subsubsection{Sargado et al. model, 2017}
\citet{sargado2017high} proposed a new parametric family of degradation functions to increase the accuracy of phase field models in predicting critical loads. Expect a better prediction of the crack initiation and propagation, their additional goal is to preserve linear elastic response in the bulk material prior to fracture. Their numerical examples indicated the superiority of the proposed family of functions to the classical quadratic degradation function. 

The family of degradation functions proposed by \citet{sargado2017high} are based on three parameters:
	\begin{equation}
	g_e(\phi)=(1-w)\frac{1-e^{-k(1-\phi)^n}}{1-e^{-k}}+wf_c(\phi)
	\end{equation}

\noindent where $k$, $n$ and $w$ are parameters and $k > 0$, $n \ge 2$ and $w \in [0, 1]$. The function $f_c$ acts as a corrector term. If $w=0$, the proposed function has only two parameters and follows (a) $g_e(\phi)$ is monotonically decreasing, (b) $g_e(0)=1$, $g_e(1)$=0, and (c) $g'_e<0$, $g'_e(1)=0$.

\subsection{On constitutive assumptions}

As aforementioned, the model of \citet{amor2009regularized} and the model of \citet{miehe2010phase,miehe2010thermodynamically} achieve variationally consistent constitutive model for stress. In addition, the crack driving force $D_s$ in the model of \citet{amor2009regularized} is derived by consistent variation:
	\begin{equation}
	D_s= \frac 1 2 K_n\langle\mathrm{tr}(\bm\varepsilon)\rangle_+^2+\mu(\bm\varepsilon^D:\bm\varepsilon^D)
	\end{equation}

Consequently, only positive volume changes and distortion of the shape can produce crack evolution. For the model of \citet{miehe2010phase,miehe2010thermodynamically}, the crack driving force $D_s$:
	\begin{equation}
	D_s=\frac1 2\lambda\langle\mathrm{tr}(\bm\varepsilon)\rangle_+^2+\mu\mathrm{tr}(\bm\varepsilon_+^2)
	\end{equation}

Both volumetric-deviatoric splits \citep{amor2009regularized} and spectral decomposition \citep{miehe2010phase, miehe2010thermodynamically} fully preserve the
variational character of the phase field method. The energy split scheme ensures the energy contribution in the evolution of phase field and the non-interpenetration condition on crack surfaces. This means that the normal stress on the crack should be non-positive and the shear stresses along a frictionless crack should vanish \citep{strobl2016constitutive}. On this, some variationally non-consistent constitutive equations for the stress are developed. A constitutive model based on a crack orientation dependent decomposition of $\bm\sigma$ is mentioned in \citet{strobl2016constitutive} where $\bm\sigma_a$ and $
\bm\sigma_p$ represent the stress affected and unaffected by the phase field respectively. The stress decomposition depends on the normal strain $\varepsilon_{nn}$:
	\begin{equation}
	\left\{\begin{aligned}
	&\bm\sigma_a=g(s)(\lambda\mathrm{tr}(\bm\varepsilon)\bm 1+2\mu\bm\varepsilon),\hspace{3cm}&\mathrm{if}\hspace{0.2cm}\varepsilon_{nn}>0\\
	&\bm\sigma_p=g(s)(\lambda\mathrm{tr}(\bm\varepsilon)\bm 1+2\mu\bm\varepsilon)+(1-g(s))(\lambda+2\mu)(\bm\varepsilon:\bm N)\bm N,&\mathrm{if}\hspace{0.2cm}\varepsilon_{nn}>0
	\end{aligned}\right.
	\end{equation}

\noindent where $\bm N=\bm n_s\otimes\bm n_s$ with $\bm n_s$ being the normal vector of the crack surface. 

A more sophisticated stress model can be seen in \citet{strobl2015novel} and only the stiffness normal to the crack is degraded. Tensile and shear stresses vanish on the fully developed crack surfaces. This results in the following stress:
	\begin{multline}
	\bm\sigma_a=\left(\lambda+(g(s)-1)\frac {\lambda^2}{\lambda+2\mu}\right)\mathrm{tr}(\bm\varepsilon)\bm 1+2\mu\bm\varepsilon+(g(s)-1)\left(\lambda+\frac {\lambda^2}{\lambda+2\mu}\right)(\mathrm{tr}(\bm\varepsilon)\bm N+(\bm\varepsilon:\bm N)\bm 1)\\+4(1-g(s))\left(\lambda+2\mu-\frac {\lambda^2}{\lambda+2\mu}\right)(\bm\varepsilon:\bm N)\bm N+\mu(g(s)-1)(\bm N\cdot \bm \varepsilon+\bm \varepsilon\cdot\bm N)
	\end{multline}

For compression, only shear stresses parallel to the crack surfaces are degraded. The modified tensor is expressed as
	\begin{equation}
	\bm\sigma_p = \lambda \mathrm{tr}(\bm\varepsilon)\bm 1 +2\mu\bm\varepsilon +4\mu(1-g(s))(\bm\varepsilon:\bm N)\bm N+\mu(g(s)-1)(\bm N\cdot \bm \varepsilon+\bm \varepsilon\cdot\bm N)
	\label{strobl stress}
	\end{equation}

Equation \eqref{strobl stress} shows a transverse isotropy with symmetry about the crack normal and the stiffness degradation is described by $\lambda$, $\mu$ and the phase field.

\subsection{Governing equations of PFM in strong form}

In summary, the quasi-static phase-field models, which are developed from the regularized variational formulation of \citet{bourdin2000numerical}, can be classified into three basic types, namely, the isotropic, anisotropic, and hybrid (isotropic-anisotropic) types. The isotropic and anisotropic formulations are the natural results of the variational principle and the basic structures can be expressed as follows.

(1) Isotropic formulation
\begin{equation}
\left\{\begin{aligned}
&\nabla\cdot\bm\sigma+\bm b =0\\
&-l_0^2\phi\Delta \phi+\phi=\frac {2l_0}{G_c}(1-d)H
\end{aligned}\right.
\end{equation}

\noindent where $\bm\sigma(\bm u,\phi)=(1-\phi)^2\frac {\partial\Psi_0}{\partial\bm\varepsilon}$ and $H = \max \limits_{\tau\in[0,t]}\psi_0\left(\bm\varepsilon(\bm x,\tau)\right)$.

(2) Anisotropic formulation
\begin{equation}
\left\{\begin{aligned}
&\nabla\cdot\bm\sigma+\bm b =0\\
&-l_0^2\phi\Delta \phi+\phi=\frac {2l_0}{G_c}(1-d)H^+
\end{aligned}\right.
\end{equation}

\noindent where $\bm\sigma(\bm u,\phi)=(1-\phi)^2\frac {\partial\Psi_0^+}{\partial\bm\varepsilon}+\frac{\partial\Psi_0^-}{\partial\bm\varepsilon}$ and $H^+ = \max \limits_{\tau\in[0,t]}\psi_0^+\left(\bm\varepsilon(\bm x,\tau)\right)$.

It should be noted again that the isotropic formulation is easier to implement and has lower computational cost than the anisotropic formulation because the stress-strain constitutive relationship is linear in the isotropic approach. However, non-linear stress-strain relation exists in the anisotropic formulation. Despite of the lower implementation cost, the isotropic formulation can be used only in some very simple problems because fractures in compression and interpenetration of the crack faces are allowed. The anisotropic formulation which uses energy split technology, naturally overcomes the drawbacks emanating from the isotropic formulation. However, because of non-linearity from strain decomposition, the numerical implementation of the anisotropic models is laborious.

\citet{ambati2015review} developed the hybrid phase field method, which retains a linear momentum balance equation. Thus, a favorable computational cost of the isotropic methods is retained. However, the evolution equation of the anisotropic methods is also retained to avoid fractures in compression. The hybrid model of \citet{ambati2015review} reads

\begin{equation}
\left\{\begin{aligned}
&\nabla\cdot\bm\sigma+\bm b =0\\
&-l_0^2\phi\Delta \phi+\phi=\frac {2l_0}{G_c}(1-d)H^+\\
&\bm\sigma(\bm u,\phi)=(1-\phi)^2\frac {\partial\Psi_0}{\partial\bm\varepsilon}\\
&\forall \bm x:\hspace{0.1cm}\Psi_0^+<\Psi_0^-\rightarrow\phi=0
\end{aligned}\right.
\end{equation}

\citet{wu2017unified} also used the hybrid formulation but the driving force of the crack field is modified as
	\begin{equation}
	D_s=\frac 1 {2E_0} \bar{\sigma}_{eq}
	\end{equation}

\noindent where $E_0$ is the undamaged Young's modulus and $\bar{\sigma}_{eq}$ is the equivalent stress. More detailed descriptions about the equivalent stress can be seen in \citet{wu2017unified}.

Another variationally non-consistent phase field method can be seen in \citet{zhang2017modification}. The driving term $H/G_c$ in the evolution equation is decomposed into two parts: $H_1/G_{cI}$ and $H_2/G_{cII}$ with
	\begin{equation}
	H_1 = \max \limits_{\tau\in[0,t]}\lambda\langle\mathrm{tr}(\bm\varepsilon)\rangle_+^2
	\end{equation}

\noindent and
	\begin{equation}
	H_2 = \max \limits_{\tau\in[0,t]}\mu\mathrm{tr}[\langle\bm\varepsilon\rangle_+^2]
	\end{equation}

\citet{zhang2017modification} regarded the parameters $G_{cI}$ and $G_{cII}$ as the respective strain energy release rates due to mode I and mode II deformations
at a crack tip. However, their definition cannot practically distinguish between the tension and shear fractures. The governing equations of \citet{zhang2017modification} are express as follows
	\begin{equation}
	\left\{\begin{aligned}
	&\nabla\cdot\bm\sigma+\bm b =0\\
	&(1-\phi)\left(\frac {H_1}{G_{cI}}+\frac {H_2}{G_{cII}}\right)-\frac 1 {4\epsilon}\phi + \epsilon \Delta\phi =0
	\end{aligned}\right.
	\end{equation}

\subsection{Similarities and differences between gradient-damage and phase field methods}

In mathematics, the damage-based gradient-damage method \citep{de2016gradient} and the phase-field methods for fractures are similar and these two types of methods have almost the same structures. Therefore, the difference between gradient-damage models and phase-field models is mainly in their interpretation and the interior length scale. Both damage-based gradient-damage and phase field methods introduce an intrinsic length scale into the discretization to smear the fracture over a localization band of finite width \citep{zhou2018phase2}. However, in more detail, gradient-damage models used a spatial averaging operator and a local damage concept, whereas the phase field models follow the regularized energy variation due to fracture evolution and use a thermodynamic driving force for the smeared fracture. In addition, the vanishing derivative of the degradation function in the phase field ensures that the crack cannot broaden once the crack forms. Another difference is that crack broadening can be not predicted by the gradient-damage models \citep{de2016gradient}.

\section{PFMs coupled with different discretization methods}\label{PFMs coupled with different discretization methods}

\subsection{Finite element method}
Most of the phase field methods are solved within the framework of the finite element methods (FEM) because the governing equations in the phase field models are commonly seen partial difference equations that can be solved for by using FEM. The space domain can be easily discretized by finite elements and suitable decoupling technology can be used to solve the fully coupled phase field fracture problems. Some FE implementations of the phase field models of fracture can be seen in \citet{kuhn2008phase, amor2009regularized, miehe2010phase, miehe2010thermodynamically, bourdin2012variational, hesch2014thermodynamically, wheeler2014augmented, miehe2015minimization, mikelic2015quasi, mikelic2015phase, heister2015primal, lee2016pressure, wick2016fluid, yoshioka2016variational, miehe2016phase, liu2016abaqus, wu2017unified, ehlers2017phase, santillan2017phase, sargado2017high, zhou2018phase, zhou2018phase2, zhou2018phase3}. In the following section, more details about the implementation by using the finite element methods will be demonstrated.

Although FEM has good capacity of implementing the phase field models, it is still difficult to solve high-order phase field equations by FEM. Therefore, some researchers tried to couple the phase field methods with other discretization technologies such as the isogeometric analysis and meshfree technology. More details about the isogeometric and meshfree methods will be presented in the following two subsections.

\subsection{Isogeometric method}
\citet{borden2014higher} proposed a fourth-order model for the phase-field approximation of the variational formulation for brittle fracture. Based on energy balance, thermodynamically consistent governing equations are derived for the fourth-order phase-field model by using the variational fracture approach. The higher-order model has higher regularity in solving for the exact phase field. To implement the higher-order model, \citet{borden2014higher} employed the isogeometric analysis framework and the smooth spline function spaces are used. The proposed higher-order model improves the convergence rate of the numerical solution and complex 3D crack propagation can be captured by the model of \citet{borden2014higher}.

The modified energy functional in the model of \citet{borden2014higher} is expressed as
	\begin{equation}
	E(\bm u, s)=\int_\Omega\left((s^2+\eta)\Psi_0^+(\bm\varepsilon)+\Psi_0^-(\bm\varepsilon)\right)\mathrm{d}\Omega+G_c\int_\Omega\left[\frac 1 {4\epsilon}(1-s)^2+\frac 1 2 \epsilon|\nabla s|^2+\frac 1 4\epsilon^3(\Delta s)^2    \right]\mathrm{d}\Omega
	\end{equation}

By replacing $\Psi_0^+$ by the history field $H$, the resulting evolution equation of the higher-order phase field model is
	\begin{equation}
	\frac {4\epsilon s H} {G_c} +s -2\epsilon^2\Delta s+\epsilon^4\Delta(\Delta s) =0
	\end{equation}

\citet{schillinger2015isogeometric} tried to use the isogeometric collocation methods for the discretization of the second-order and fourth-order phase-field fracture models. The used isogeometric collocation methods are shown to speed up the phase-field fracture computations over general isogeometric Galerkin method because point evaluations are reduced. In addition, a hybrid collocation-Galerkin formulation is recommended by \citet{schillinger2015isogeometric} because it provides consistent weakly enforcing Neumann boundary conditions and multi-patch interface constraints. The hybrid collocation–Galerkin formulation can also deal with multiple boundary integral terms that arise from the weighted residual formulation and improve the phase field resolution \citep{schillinger2015isogeometric}.

\citet{ambati2016phase} applied a phase field model to investigate fracture in shells. The shell is modeled based on solid-shell kinematics with small rotations and displacements. The displacement and phase fields are discretized by using quadratic Non-Uniform Rational B-Spline basis functions. \citet{goswami2020adaptive} proposed a fourth-order phase field method with cubic-stress degradation function. Hybrid-staggered approach is used to solve the model, and PHT-splines are used to implement the adaptive refinement.

\subsection{Meshfree method}

\citet{amiri2016fourth} applied a fourth order phase-field model for fracture based on local maximum entropy (LME) approximants. Fourth order phase-field equations can be directly solved without splitting the fourth order differential equation into two second order differential equations due to the higher order continuity of the meshfree LME approximants.

The higher order crack surface density function used in \citet{amiri2016fourth} is identical to that in \citet{borden2014higher}. The meshfree discretization form of the displacement $\bm u(\bm x)$ and phase field $s(\bm x)$ is shown as follows,
	\begin{equation}
	\left\{\begin{aligned}
	\bm u(\bm x)=\sum_a p_a\bm u_a\\
	s(\bm x)=\sum_a p_a s_a
	\end{aligned}\right.
	\end{equation}

\noindent where $p_a$ are LME basis functions, and $u_a$ and $s_a$ are nodal values of the displacement and phase fields. The same discretization is used on virtual displacements and virtual phase field parameters. In addition, the LME basis functions are non-negative and must satisfy the $C^0$ and $C^1$ consistency:
	\begin{equation}
	p(\bm x)\ge 0
	\end{equation}
	\begin{equation}
	\sum_{a=1}^N p_a(\bm x) = 1
	\end{equation}
	\begin{equation}
	\sum_{a=1}^N p_a(\bm x) \bm x_a = \bm x
	\end{equation}

\noindent where the vector $\bm x_a$ denotes the positions of the nodes associated with each basis function. \citet{amiri2016fourth} used the local maximum entropy basis functions as follows,
	\begin{equation}
	p_a(\bm x)=\frac{1}{Z(\bm x, \bm \lambda^*(\bm x))}\mathrm{exp}\left[-\beta_a|\bm x-\bm x_a|^2+\bm \lambda^*(\bm x-\bm x^*)\right ]
	\end{equation}
\noindent where 
	\begin{equation}
	Z(\bm x, \bm \lambda(\bm x)) = \sum_{b=1}^N \mathrm{exp}\left[-\beta_b|\bm x-\bm x_b|^2+\bm \lambda(\bm x-\bm x_b)\right ]
	\end{equation}

\noindent is a function associated with a set of nodes $X=\left\{\bm x_a\right\}_{a=1,\cdots,N}$ and $\bm \lambda^*$ is defined by
	\begin{equation}
	\bm \lambda^*=\mathrm{arg} \min\limits_{\bm \lambda\in \mathbb{R}^d} \mathrm {log} Z(\bm x, \bm \lambda)
	\end{equation}

\noindent with $d$ the dimension of the crack problem. The discrete stiffness and force matrices are then obtained by using the meshfree discretization to solve the crack problems. Note that the crack can propagate, branch and merge but cannot be recovered because a strong constraint $s_i\le s_{i-1}$ is imposed with $s_i$ and $s_{i-1}$ are the phase fields at step $i$ and $i-1$.

\subsection{Physics informed neural network}
\citet{goswami2020transfer} proposed the first physic informed neural network for phase field modeling of fracture. It seems that the PINN model treats the fracture problems one level higher on the model hierarchy than FEM. The proposed machine learning method is easy to implement and only a few lines of code are required. The approach can have a good computational savings compared with FEM after the network being trained.

\subsection{FFT solver for phase field modeling}
\citet{chen2019fft} introduced an FFT algorithm to solve the phase-field model of brittle fracture. By using a staggered update scheme, the FFT algorithm solves the fracture and mechanical problems separately. The proposed method inherits the advantages of classical FFT methods in terms of simplicity of mesh generation and parallel implementation. In addition, \citet{ma2020fft} proposed an FFT-based solver for higher-order and multi-phase-field fracture, while the model was applied to strongly anisotropic brittle materials.

\section{Finite element implementation of phase field methods}\label{Finite element implementation of phase field methods}
In this section, more details about the FE implementation of the phase field models are presented.
\subsection{FE approximation}

To show better of the FE approximation, we start from the governing equations in \citet{zhou2018phase} for example:
	\begin{equation}
	  \left\{
	   \begin{aligned}
	\frac {\partial {\sigma_{ij}}}{\partial x_j}+b_i=\rho{\ddot u_i}
	\\ \left[\frac{2l_0(1-k)H}{G_c}+1\right]\phi-l_0^2\frac{\partial^2 \phi}{\partial {x_i^2}}=\frac{2l_0(1-k)H}{G_c}
	   \end{aligned}\right.
	\label{Strong form}
	\end{equation}

Note that these governing equations are used for dynamic problems. For quasi-static problems, the initial term vanishes. The weak form of the governing equations are subsequently given  by
	\begin{equation}
	\int_{\Omega}\left(-\rho \bm{\ddot u} \cdot \delta \bm u -\bm\sigma:\delta \bm {\varepsilon}\right) \mathrm{d}\Omega +\int_{\Omega}\bm b \cdot \delta \bm u  \mathrm{d}\Omega +\int_{\Omega_{h}}\bm f \cdot \delta \bm u  \mathrm{d}S=0
	\label{weak form 1}
	\end{equation}

\noindent and
	\begin{equation}
	\int_{\Omega}-2(1-k)H(1-\phi)\delta\phi\mathrm{d}\Omega+\int_{\Omega}G_c\left(l_0\nabla\phi\cdot\nabla\delta\phi+\frac{1}{l_0}\phi\delta\phi\right)\mathrm{d}\Omega=0
	\label{weak form 2}
	\end{equation}

The standard vector-matrix notation is used and the discretization of the displacement and phase field can be expressed as

	\begin{equation}
	\bm u = \bm N_u \bm d,\hspace{0.5cm} \phi = \bm N_{\phi} \hat{\bm\phi}
	\end{equation}

\noindent where $\bm u_i$ and $\phi_i$ are the nodal values of the displacement and phase field. $\bm d$ and $\hat{\bm\phi}$ are the vectors consisting of node values $\bm u_i$ and $\phi_i$. $\bm N_u$ and $\bm N_{\phi}$ are shape function matrices:
	\begin{equation}		
			\bm N_u = \left[ \begin{array}{ccccccc}
			N_{1}&0&0&\dots&N_{n}&0&0\\
			0&N_{1}&0&\dots&0&N_{n}&0\\
			0&0&N_{1}&\dots&0&0&N_{n}
			\end{array}\right], \hspace{0.5cm}
			\bm N_\phi = \left[ \begin{array}{cccc}
			N_{1}&N_{2}&\dots&N_{n}
			\end{array}\right]
	\end{equation}

\noindent where $n$ is the node number in one element and $N_i$ is the corresponding shape function. The same discretization is assumed on the test functions:
	\begin{equation}
	\delta \bm u = \bm N_u \delta \bm d,\hspace{0.5cm} \delta \phi = \bm N_{\phi} \delta \hat{\bm\phi}
	\end{equation}

\noindent where $\delta \bm d$ and $\delta \hat{\bm\phi}$ are the vectors consisting of node values of the test functions.

These gradients are used:
	\begin{equation}
	\bm \varepsilon =  \bm B_u \bm d,\hspace{0.5cm} \nabla\phi = \bm B_\phi \hat{\bm\phi}, \hspace{0.5cm}\bm \delta \varepsilon =  \bm B_u \delta \bm d,\hspace{0.5cm} \nabla\phi = \bm B_\phi \delta \hat{\bm\phi}
	\end{equation}

\noindent where $\bm B_u$ and $\bm B_\phi$ are the derivatives of the shape functions defined by
	\begin{equation}
	\bm B_u=\left[
		\begin{array}{ccccccc}
		N_{1,x}&0&0&\dots&N_{n,x}&0&0\\
		0&N_{1,y}&0&\dots&0&N_{n,y}&0\\
		0&0&N_{1,z}&\dots&0&0&N_{n,z}\\
		N_{1,y}&N_{1,x}&0&\dots&N_{n,y}&N_{n,x}&0\\
		0&N_{1,z}&N_{1,y}&\dots&0&N_{n,z}&N_{n,y}\\
		N_{1,z}&0&N_{1,x}&\dots&N_{n,z}&0&N_{n,x}
		\end{array}\right],\hspace{0.2cm}
		\bm B_\phi=\left[
		\begin{array}{cccc}
		N_{1,x}&N_{2,x}&\dots&N_{n,x}\\
		N_{1,y}&N_{2,y}&\dots&N_{n,y}\\
		N_{1,z}&N_{2,z}&\dots&N_{n,z}
		\end{array}\right]
		\label{BiBu}
	\end{equation}

The equations of weak form \eqref{weak form 1} and \eqref{weak form 2} are subsequently written as
	\begin{equation}
	-(\delta\bm d)^\mathrm{T} \left[\int_{\Omega}\rho\bm N_u^\mathrm{T}\bm N_u \mathrm{d}\Omega \ddot{\bm d}+\int_{\Omega} \bm B_u^\mathrm{T} \bm D_e \bm B_u \mathrm{d}\Omega \bm d \right]+ 
	(\delta\bm d)^\mathrm{T} \left[\int_{\Omega}\bm N_u^\mathrm{T}\bm b \mathrm{d}\Omega+\int_{\Omega_{h}} \bm N_u^\mathrm{T} \bm f\mathrm{d}S \right]=0
	\label{discrete equation 1}
	\end{equation}
	\begin{equation}
	-(\delta\hat{\bm \phi})^{\mathrm{T}} \int_{\Omega}\left\{\bm B_\phi^{\mathrm{T}} G_c l_0 \bm B_\phi +\bm N_\phi^{\mathrm{T}} \left [ \frac{G_c}{l_0} + 2(1-k)H \right ]  \bm N_\phi \right \} \mathrm{d}\Omega \hat{\bm \phi}+ (\delta\hat{\bm \phi})^\mathrm{T} \int_{\Omega}2(1-k)H\bm N_\phi^{\mathrm{T}} \mathrm{d}\Omega = 0
	\label{discrete equation 2}      
	\end{equation}

Because of admissible arbitrary test functions, Eqs. \eqref{discrete equation 1} and \eqref{discrete equation 2} results in the discretized weak form equations:
	\begin{equation}
	-\underbrace{\int_{\Omega}\rho\bm N_u^\mathrm{T}\bm N \mathrm{d}\Omega \ddot{\bm d}}_{\bm F_u^{ine}=\bm M \ddot{\bm d}}-\underbrace{\int_{\Omega} \bm B_u^\mathrm{T} \bm D_e \bm B_u \mathrm{d}\Omega \bm d}_{\bm F_u^{int}=\bm K_u \bm d} + \underbrace{\int_{\Omega}\bm N_u^\mathrm{T}\bm b \mathrm{d}\Omega+\int_{\Omega_{h}} \bm N_u^\mathrm{T} \bm f\mathrm{d}S}_{\bm F_u^{ext}}=0
\label{discrete equation 3}
	\end{equation}
	\begin{equation}
	-\underbrace{ \int_{\Omega}\left\{\bm B_\phi^{\mathrm{T}} G_c l_0 \bm B_\phi +\bm N_\phi^{\mathrm{T}} \left [ \frac{G_c}{l_0} + 2(1-k)H \right ] \bm N_\phi \right \} \mathrm{d}\Omega \hat{\bm \phi}}_{\bm F_\phi^{int}=\bm K_\phi \hat{\bm \phi}}+ \underbrace{\int_{\Omega}2(1-k)H\bm N_\phi^{\mathrm{T}} \mathrm{d}\Omega}_{\bm F_\phi^{ext}} = 0 
\label{discrete equation 4}
	\end{equation}

\noindent where $\bm F_u^{ine}$, $\bm F_u^{int}$, and $\bm F_u^{ext}$ are the inertial, internal, and external forces for the displacement field and $\bm F_\phi^{int}$ and $\bm F_\phi^{ext}$ are the internal and external force terms of the phase field \citep{zhou2018phase}.  Additionally, the mass and stiffness matrices read
	\begin{equation}
	\left\{\begin{aligned}\bm M &= \int_{\Omega}\rho\bm N_u^\mathrm{T}\bm N \mathrm{d}\Omega\\
\bm K_u &= \int_{\Omega} \bm B_u^\mathrm{T} \bm D_e \bm B_u \mathrm{d}\Omega\\
	\bm K_\phi &= \int_{\Omega}\left\{\bm B_\phi^{\mathrm{T}} G_c l_0 \bm B_\phi +\bm N_\phi^{\mathrm{T}} \left [ \frac{G_c}{l_0} + 2(1-k)H \right ] \bm N_\phi \right \} \mathrm{d}\Omega
	\end{aligned}\right .
	\end{equation}

\subsection{Solution schemes}
There are two basic approaches to solve the coupling discretized equations \eqref{discrete equation 3} and \eqref{discrete equation 4}: monolithic and staggered schemes. In a monolithic scheme, the displacement and phase field are solved simultaneously. Whereas, the phase field and displacement are solved independently for a staggered scheme. At a given time step, the displacement is solved first at a fixed phase field. By using the updated displacement, the phase-field equation is subsequently solved.

\subsubsection{Monolithic scheme}

For a monolithic scheme, the implicit Hughes-Hibert-Taylor (HHT) time integration can be adopted in the dynamic calculations (see \citet{liu2016abaqus}). According to our experience, other integration methods such as the implicit BDF (backward differentiation formulas) and the generalized-$\alpha$ methods can also be successfully used in phase field modeling. As an example, the implicit HHT method is presented herein. Letting $i$ be the Newton iteration, the residual force vectors now are re-written as \citep{liu2016abaqus}:
	\begin{equation}
	(\bm R^u)_{i}^{t+\Delta t}=(\bm F_u^{ext})_{i}^{t+\Delta t}-(\bm F_u^{ine})_{i}^{t+\Delta t}-(1+\alpha)(\bm F_u^{int})_{i}^{t+\Delta t}+\alpha (\bm F_u^{int})_{i}^{t}
	\end{equation}
	\begin{equation}
	(\bm R^\phi)_{i}^{t+\Delta t}=(\bm F_\phi^{ext})_{i}^{t+\Delta t}-(1+\alpha)(\bm F_\phi^{int})_{i}^{t+\Delta t}+\alpha (\bm F_\phi^{int})_{i}^{t}
	\end{equation}

\noindent where $\alpha\in[\frac 1 3, 0]$ is the HHT integration operator and $\Delta t = t_{n+1}-t_{n}$ is the time step size. If the material damping effect is neglected, the tangent matrix in one element is given by \citep{liu2016abaqus}:
	\begin{equation}
	\bm S_{ij}=\bm M_{ij}\frac{\mathrm{d}\ddot{\bm u}}{\mathrm{d}{\bm u}}+(1+\alpha)\bm K_{ij}
	\end{equation}
	\begin{equation}
	\frac{\mathrm{d}\ddot{\bm u}}{\mathrm{d}{\bm u}}=\frac 1 {\beta \Delta t^2},\hspace{1cm}\beta=\frac {(1-\alpha)^2} 4
	\end{equation}

\subsubsection{Staggered scheme}
In a staggered scheme, greater flexibility exists in solving the displacement and phase field. This means both implicit and explicit time schemes can be used. In an implicit approach, the procedure of solving displacement field is similar to those in a monolithic scheme. Whereas for an explicit method, e.g. the explicit central-difference time-integration method, some modification is needed \citep{liu2016abaqus}:
	\begin{equation}
	(\bm u)_{i}^{t+\Delta t}=(\bm u)_{i}^{t}+(\dot{\bm u})_{i}^{t+0.5\Delta t}\Delta t
	\end{equation}
	\begin{equation}
	(\dot{\bm u})_{i}^{t+0.5\Delta t}=(\dot{\bm u})_{i}^{t-0.5\Delta t}+(\ddot{\bm u})_{i}^{t}\Delta t
	\end{equation}
\noindent and
	\begin{equation}
	(\ddot{\bm u})_{i}=(\bm M_{ij})^{-1}\left( (\bm F_u^{ext})_{i}-(\bm F_u^{int})_{i} \right)
	\end{equation}

\citet{selevs2019residual} discussed the details of a residual control staggered solution scheme for the phase-field modeling of brittle fracture. The use of a stopping criterion within the broadly used staggered algorithm was discussed. The methods of stopping criterion of the iterative scheme are based on the residual norm and implemented in commercial software ABAQUS. The proposed method can avoid fine loading incrementation to produce an accurate solution in a common single iteration staggered algorithm.

\subsection{Implementation codes}

To reduce the implementation effort on the phase field methods and facilitate new learners, some phase field models are developed and implemented with commonly used commercial software. Commonly seen are the Abaqus, FEniCS and COMSOL implementations.

	\subsubsection{Abaqus implementation}

\citet{msekh2015abaqus, liu2016abaqus} implement the isotropic and anisotropic types of phase field methods in Abaqus. For the monolithic scheme, a UEL subroutine is used. The time integration and element discretization are implemented in the Abaqus standard and the UEL subroutine calls for each iteration in a given increment. Because Abaqus itself does not support a plot of the results on the used elements, external visualization software is required for better exporting the calculated results \citep{liu2016abaqus}. 

For a staggered scheme, A UMAT or VUMAT subroutine is used. If an implicit time integration scheme is used, a UMAT subroutine is employed to exchange the local history field $H$ and the phase field $\phi$ on integration points. Subsequently, the stress $\bm\sigma$ and its tangent modulus with $\bm\varepsilon$ must be renewed in every iteration.  Whereas, if an explicit time integration scheme is used, a VUMAT subroutine is employed to exchange the local history field $H$ and the phase field $\phi$ instead, and only the stress $\bm \sigma$ needs to be renewed during each increment.

\citet{zhang2018iteration} implemented an iterative scheme in Abaqus for cohesive fractures. \citet{fang2019phase} presented Abaqus implementation procedures of phase field fracture of elasto-plastic solids in a staggered manner. The subroutines UEL and UMAT are also used. The UMAT describes the constitutive behavior of elasto-plastic solids, while the UEL is designed for the phase field fracture. The authors solved the phase field and displacement field separately using the Newton-Raphson iteration method.

	\subsubsection{COMSOL implementation}

\begin{figure}[htbp]
	\centering
	\includegraphics[width = 10cm]{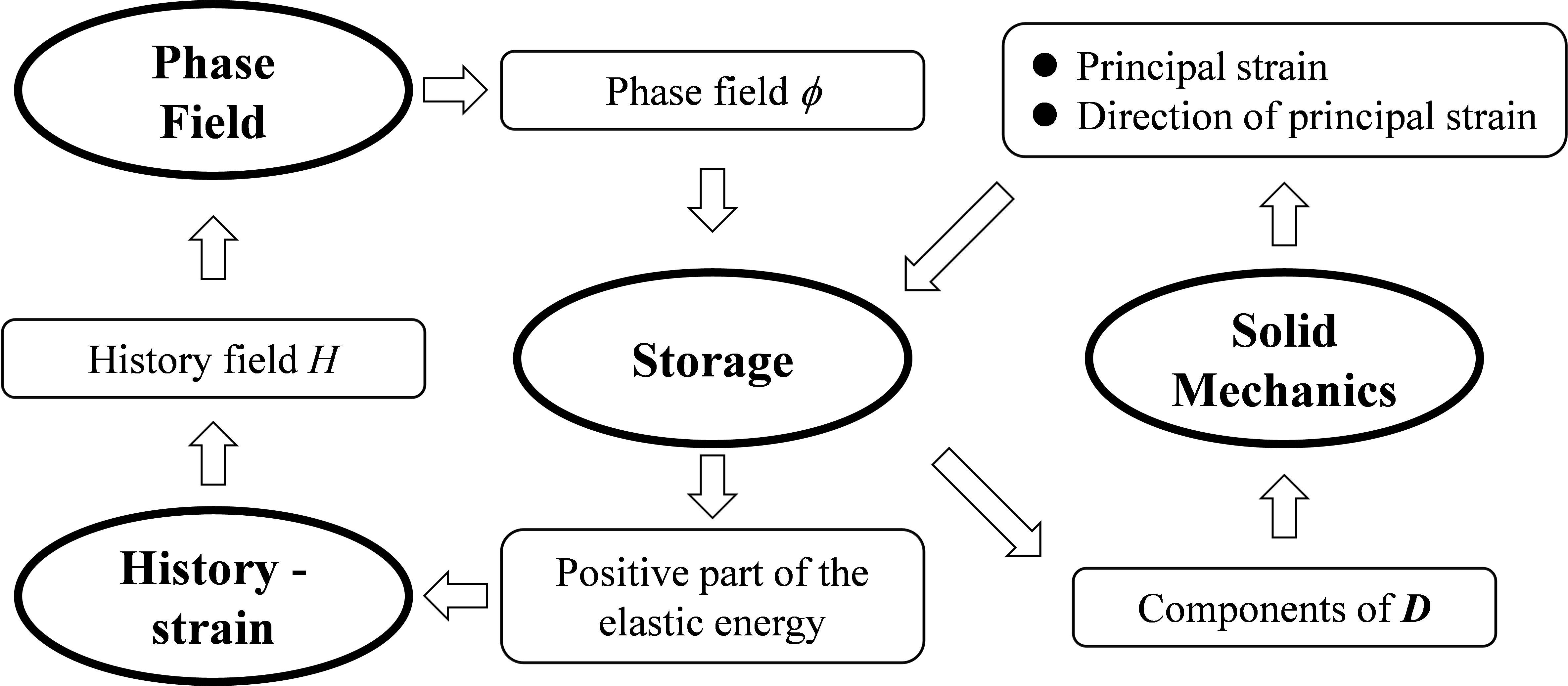}
	\caption{Established modules in COMSOL for the phase field modeling}
	\label{Established modules in COMSOL for the phase field modeling}
	\end{figure}

\begin{figure}[htbp]
	\centering
	\includegraphics[width = 8cm]{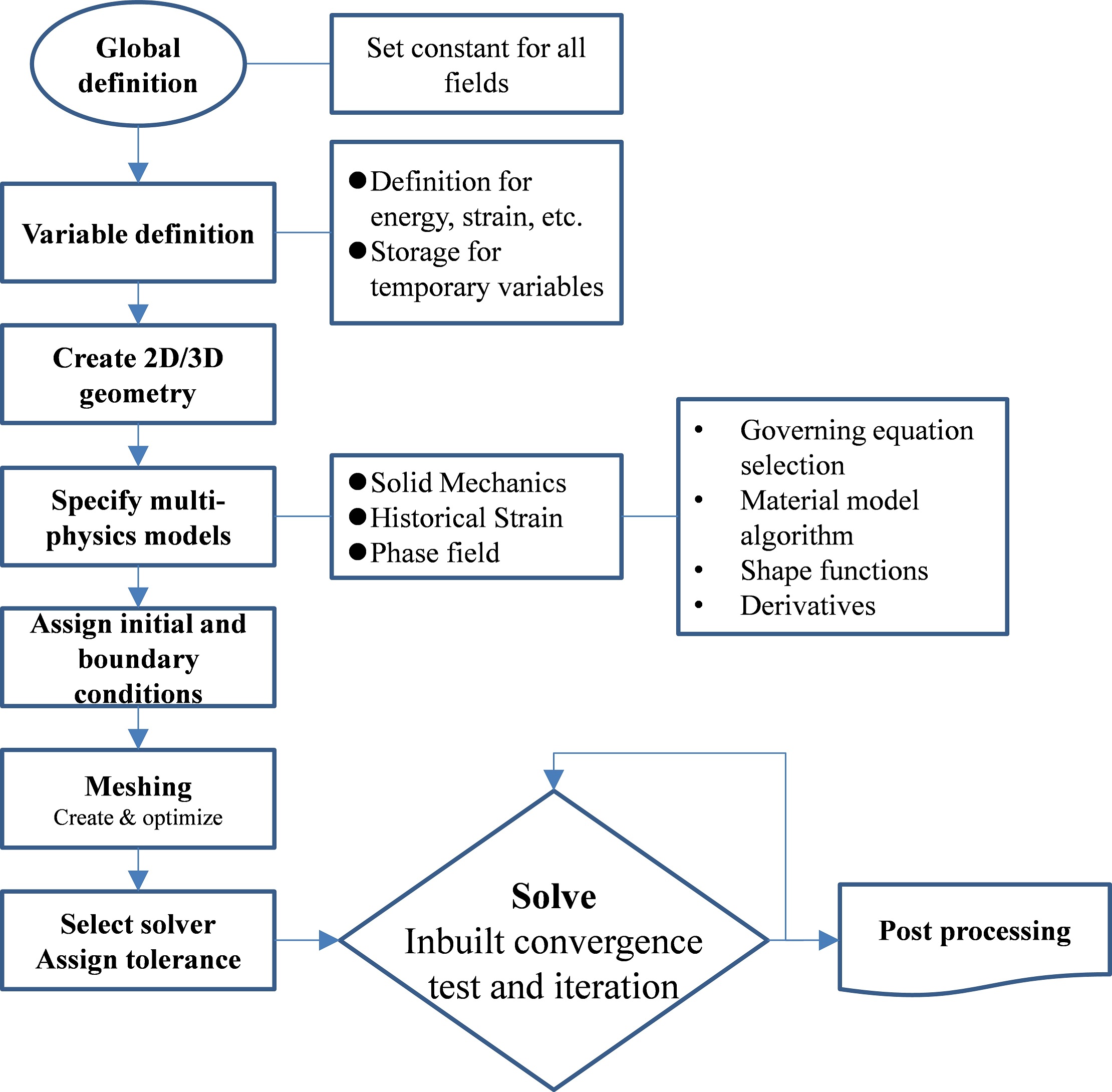}
	\caption{Implementation procedure in COMSOL for the phase field modeling \citep{zhou2018phase}}
	\label{Implementation procedure in COMSOL for the phase field modeling}
	\end{figure}

\citet{zhou2018phase, zhou2018phase2, zhou2018phase3} implemented the anisotropic type of phase field models in the multi-field simulator--COMSOL Multiphysics. The implementations in COMSOL are much easier to be extended to problems with more fields than other software such as Abaqus. The modules established in COMSOL is shown in Fig. \ref{Established modules in COMSOL for the phase field modeling} and the implementation procedure in COMSOL is shown in Fig. \ref{Implementation procedure in COMSOL for the phase field modeling}. 

As shown in Fig. \ref{Established modules in COMSOL for the phase field modeling}, the ``Storage Module'' stores the results obtained from the ``Solid Mechanics Module'', such as the principal strains and their directions. The positive part of the elastic energy is then calculated to update the local history strain field. Subsequently, the updated history strain is used to solve the phase-field. Because of the highly nonlinearity, the elasticity matrix in the ``Solid Mechanics Module'' must be modified by the updated phase-field solution and varying principal strains and their corresponding directions. Note that only staggered schemes are used in \citet{zhou2018phase, zhou2018phase2, zhou2018phase3}. In addition, the implicit Generalized-$\alpha$ method is used for time integration. For convergence issues, Anderson acceleration technology is used to increase the convergence rate in COMSOL and the open-access codes of the phase field modeling in COMSOL can be found in ``https://sourceforge.net/projects/phasefieldmodelingcomsol/".

\subsubsection{FEniCS implementation}
\citet{natarajan2019fenics} presented a FEniCS implementation scheme of the phase field method for quasi-static brittle fracture. The FEniCS provides a framework for the automated solutions of the partial differential equations. Therefore, the phase field and displacement field can be solved, respectively. However, in recent days, the performance of FEniCS implementation has not been verified in comparison with Abaqus and COMSOL implementations.

\subsection{Computational cost}

To obtain accurate results predicted from the phase field models of fracture, it is heavily time-costing even in 2D problems when the finite element method is used. One reason is that the resolution of the small length-scale requires extremely fine meshes, at least near the region with high phase field $\phi$. That is, if no adaptive mesh refinement technology is used, the phase field models are calculated on fine fixed meshes, which requires high computational cost. Another reason is that energy split technology is used to obtain realistic crack patters under tension and compression. The special split such as the strain decomposition \citep{miehe2010phase} results in a highly non-linear constitutive model of stress to strain. Rather many iterations are needed to solve the highly non-linear constitutive model. In addition, to obtain a better crack resolution, the loading and displacement increments must be small enough. Too small increments make the computational cost expensive even for an simple isotropic phase field model. 

\subsection{Element technologies}

The phase field methods are in general laborious approaches to fracture and thereby some element technologies are developed to improve the efficiency of the phase field methods within the framework of FEM.

	\subsubsection{Adaptive mesh refinement}

To develop a robust and efficient numerical scheme for the phase field modeling, a primal-dual active set method and predictor-corrector mesh adaptivity technology were used in \citet{heister2015primal}. The primal-dual active set method is a semi-smooth Newton method, which maintains the crack irreversibility as a constraint and achieved increasing converging rate. The predictor-corrector mesh adaptivity technology is also developed by \citet{heister2015primal} to reduce the computational cost. This adaptive mesh refinement technology has following features:

	\begin{itemize}
	\begin{spacing}{0.25}
	\item Keep a single fixed, small strain $\bm\varepsilon$ during the entire computation.
	\item Ensure $h < \epsilon$ inside crack region.
	\item Error is controlled by $\epsilon$, not the element size $h$.
	\item No requirement of prior knowledge about crack paths.
	\item Handling fast growing cracks.
	\end{spacing}
	\end{itemize}

\citet{wick2016goal} developed a posteriori error estimation and goal-oriented adaptive mesh refinement technology for phase field crack propagation. Goal functionals and dual-weighted residual (DWR) methods are used simultaneously. Their approach is based on a partition-of-unity (PU) and does not require strong residuals nor jumps over element edges.

In addition, \citet{badnava2018h, Zhou2018} used another adaptive mesh refinement method to couple the phase field method (so-called h-adaptive phase field method). The adaptive mesh refinement does not require the mesh refinement around a pre-fixed crack path. Instead, a predictor-corrector scheme for mesh adaptivity is used. First, the coupled system with coarser elements is solved and predicts an initial crack path. Subsequently, a refinement threshold such as $\phi>0.85$ is used to judge if the mesh requires refinement. Finally, all the fields are solved again based on the refined mesh \citep{Zhou2018}. The predictor-corrector mesh refinement is continuously imposed on the elements in one time step unless the elements do not need refinement. \citet{tian2019hybrid} proposed a novel hybrid adaptive finite element phase‐field method (ha‐PFM) for fractures under quasi‐static and dynamic loading. The method refines adaptively the meshes based on a crack tip identification strategy while the refined meshes in the noncrack progression region are reset as coarse meshes. The proposed method prominently reduces the CPU time and memory usage. \citet{jansari2019adaptive} used a recovery based error indicator and quadtree decomposition to establish an adaptive phase field method for quasi-static brittle fracture. \citet{kristensen2020phase}
investigated the potential of quasi-Newton methods in facilitating convergence of monolithic solution schemes for phase field modelling. A new adaptive time increment scheme is proposed to reduce the computational cost. The study indicates computation times can be reduced by several orders of magnitude. On the other hand, the number of load increments required by the staggered solution will be up to 3000 times higher. \citet{noii2020adaptive} proposed an adaptive global--local approach for phase-field modeling of anisotropic brittle fracture.

	\subsubsection{Multi-scale phase field method}

\citet{PATIL2018254} coupled a multi-scale finite element method (MsFEM) with the hybrid type of phase field model for brittle fracture problems. This multi-scale method is also coupled with adaptive remeshing technology and named the adaptive multi-scale phase field method (AMPFM). An important feature of this phase field method is that the degrees of freedom of coarse-mesh and fine-mesh are linked together using multi-scale basis functions during mesh refinement \citep{PATIL2018254}. The crack propagation path is automatically tracked and refined around the crack by using the current phase field and its increment. Another benefit of AMPFM is that memory and CPU time are dramatically reduced when AMPFM is used to simulate the brittle fracture in heterogeneous structures with uniformly distributed small-size discontinuities. \citet{gerasimov2018non}  adopted non-intrusive global/local approaches when the fracture was modeled by using the phase-field framework. 

	\subsubsection{New element shape function}

In another effort to reduce computational cost of the phase field, \citet{kuhn2011new} developed special engineered FEM shape functions to discretize the phase field. The shape functions have an exponential nature and their forms are as follows,
	\begin{equation}
	N_{1}^e(\xi,\delta)=1-\frac{\mathrm{exp}(-\delta(1+\xi)/4)-1}{\mathrm{exp}(\delta/2)-1}
	\label{function 1}
	\end{equation}
	\begin{equation}
	N_{2}^e(\xi,\delta)=\frac{\mathrm{exp}(-\delta(1+\xi)/4)-1}{\mathrm{exp}(\delta/2)-1}
	\label{function 2}
	\end{equation}

\noindent where $\delta$ is the ration between the element size and the scale parameter of the phase field method. 

Note that Eqs. \eqref{function 1} and \eqref{function 2} are only available in 1D 2-node element. For 2D, the shape functions for a 4-node element are obtained as the tensor products of the above Eqs.  \eqref{function 1} and \eqref{function 2} with some changes. The results of \citet{kuhn2011new} showed that accurate prediction of the surface energy can be achieved in a much lower element refinement level by using the special shape functions. A drawback of the approach of \citet{kuhn2011new} is that some prior information about the fracture direction is needed to construct the exponential shape functions.

\subsubsection{Virtual element and smooth finite element}
\citet{aldakheel2018phase} proposed an efficient low order virtual element scheme for phase-field modeling of brittle fracture. The virtual element formulation has flexible choice of node number in an element that can be changed in a simulation. The potential density is written in terms of polynomial functions rather than the unknown shape functions for complicated element geometries. \citet{bhowmick2018phase} developed a phase field model in the framework of cell-based smoothed finite element method (CS-FEM). The CS-FEM is softer than the standard FEM, and CS-FEM is used to solve the equations that govern the continuum mechanics of solids. The computational cost of CS-FEM is slightly lower than the finite element counterpart.

\subsection{Special treatments}
	\subsubsection{Modelling pre-existing cracks}

How to model pre-existing cracks is another important issue in phase field modelling. A straightforward manner is to model initial cracks as discrete cracks using double nodes in the element mesh. However, the phase field can also be used to model the initial pre-existing cracks. The first way is to constrain the phase field by the Dirichlet boundary condition $\phi=1$ at the pre-existing crack. The nodes with $\phi =1$ can be placed on a single line or a small region. But the latter produces larger error on calculating the crack surface energy. Another way to model pre-existing crack is the approach proposed by \citet{borden2012phase} where an initial history energy field is introduced. The history field has the following form:
	\begin{equation}
	H_0(\bm x)=B\left\{\begin{aligned}
	&\frac {G_c}{4\epsilon}\left(1-\frac{d(\bm x, l)}{\epsilon}\right)\hspace{1cm}&d(\bm x, l)\le\epsilon\\&0&d(\bm x, l)>\epsilon
	\end{aligned}\right.
	\end{equation}

\noindent where $B=1\times 10^3$ and $d(\bm x, l)$ is the closest distance from $\bm x$ to the line $l$.

	\subsubsection{Determining crack tip position}

Because the phase field method applies diffusive description of the sharp crack, there is no obvious sharp crack tip in the modelling. But for post-processing, the crack tip position is sometimes needed especially when the crack velocity and accumulated crack length are calculated. For 2D problems, the crack tip is commonly defined by using the iso-curves of the phase-field \citep{borden2012phase, liu2016abaqus, zhou2018phase}. For example, \citet{borden2012phase} used the iso-curves of $\phi=0.75$ to fix the crack tip position. In addition, \citet{nguyenamodeling} adopted another approach. They selected all nodes that have a phase field larger than 0.8, while 
the crack tip is fixed on the node that has the maximum horizontal coordinate among those selected nodes.

\subsubsection{Hierarchical meshes}
\citet{goswami2019adaptive} proposed a novel dual-mesh based adaptive phase field method for solving fracture problems. The implementation of the phase field model is based two sets of  meshes with different characteristic element sizes. A coarser mesh for the displacement field and a finer mesh is for the phase field. To facilitate the exchange of information between the meshes, the authors also proposed an efficient data transfer algorithm.

\section{Extensions and applications of the PFMs}\label{Extensions and applications of the phase field methods}

\subsection{Ductile fracture}

Extensions of the phase field methods to ductile fractures are exclusively based on the variational approach to fracture. These extensions are mostly based on the coupling of phase field methods with models of elasto-plasticity \citep{miehe2016phase1}. \citet{duda2015phase} examined a series of brittle fractures in elastic-plastic solids. Other variational approaches of modelling combined brittle-ductile fractures can be seen in \citet{ulmer2013phase, alessi2015gradient}. The model of \citet{ambati2015phase} suggested a characteristic degradation function. By using this function, the damage is coupled to plasticity in a multiplicative format \citep{miehe2016phase1}. However, combination of the local plasticity models and the gradient-damage-type phase field modeling of fracture in \citet{ambati2015phase} do not meet the constraints on the plastic and damage length scales \citep{miehe2016phase1}. The model of \citet{ambati2015phase} also lacks a canonical structure based on variational principles. This drawback is overcome by the following work of \citet{miehe2015phase}, which couples gradient plasticity to gradient damage at finite strains.

\citet{miehe2016phase1} proposed a consistent variational framework for the phase field modelling of ductile fractures in elastic-plastic solids at large strains. The bases of the model in \citet{miehe2016phase1} are the formulation of variational gradient plasticity in \citet{miehe2014variational} and the original phase field model for brittle fractures in \citet{miehe2010phase, miehe2010thermodynamically}. \citet{ambati2016phase} applied a phase field model to investigate fracture in shells. The solid-shell formulation is distinguished between elastic and plastic materials. A brittle phase-field model is used for elastic materials, while a ductile fracture model for elasto-plastic materials with $J_2$ plasticity and isotropic hardening.

In \citet{ambati2016phase}, the total free energy functional $E_l$ for the ductile fracture is the sum of elastic, plastic, and fracture energy contributions:
	\begin{equation}
	E_l(\bm\varepsilon_e,\bm\varepsilon_p,h,s)=\int_\Omega\left[g(s,p)\Psi_e^+(\bm\varepsilon)+\Psi_e^-(\bm\varepsilon)+\Psi_p(h)+\gamma(s,\nabla s)\right]\mathrm{d}\Omega
	\end{equation}

\noindent where $\Psi_p(h)$ is the plastic strain energy density function with isotropic hardening variable $h$. $\bm\varepsilon_e$ and $\bm\varepsilon_p$ are the respective elastic and the plastic strain tensors with the total strain $\bm\varepsilon=\bm\varepsilon_e+\bm\varepsilon_p$. The similar decomposition of the energy to Eq. \eqref{armor decomposition} is used
	\begin{equation}
	\left\{\begin{aligned}
	&\Psi_e^+(\bm\varepsilon)=\frac 1 2 K_n\langle\mathrm{tr}(\bm\varepsilon)\rangle_+^2+\mu(\bm\varepsilon^D:\bm\varepsilon^D)\\
	&\Psi_e^+(\bm\varepsilon)=\frac 1 2 K_n\langle\mathrm{tr}(\bm\varepsilon)\rangle_-^2
	\end{aligned}\right.
	\end{equation}

\noindent where $K_n=\lambda+2/3\mu$ is the bulk modulus of the material.

The stress degradation function $g(s, p)$ is chosen as follows:
\begin{equation}
g(s, p)=s^{2p^m}+\eta
\end{equation}

\noindent where $m$ and $p$ are two parameters with
	\begin{equation}
	p=\frac{\varepsilon_p^{eq}}{\varepsilon_{p,crit}^{eq}},\hspace{1cm}\varepsilon_p^{eq}(t)=\sqrt{\frac 2 3}\int_0^t \sqrt{\dot{\bm\varepsilon_p}:\dot{\bm\varepsilon_p}}
	\end{equation}

\noindent and $\varepsilon_{p,crit}^{eq}$ is a threshold value, and $\varepsilon_{p}^{eq}$ is the von Mises equivalent plastic strain.

In addition, \citet{borden2016phase} presented a phase field formulation for fractures in ductile materials. They introduced a cubic degradation function, which produces a more accurate stress-strain response prior to crack initiation. A microforce-driven governing equation is used to replace the general energy potential for finite deformation problems. \citet{borden2016phase} also introduced a yield surface degradation function that accounts for plastic softening and non-physical elastic deformations after crack initiation.

\citet{shanthraj2016phase} applied a obstacle phase field energy model to formulate the fracture behavior in a finite strain elasto-viscoplastic material. The obstacle energy model can produce physically realistic fracture behaviors at the vicinity of the crack tip. The resulting variational inequality is discretized by a finite element method, and is solved by using a reduced space Newton method. \citet{shanthraj2016phase} also emphasized a significant decrease in the computational cost by using their method. \citet{alessi2018coupling} also proposed a phase field model for plasticity. Based on a minimization algorithm, the coupled elasto-damage-plasticity can be solved by using the proposed method. \citet{dittmann2018variational} proposed a higher order phase-field model for non-linear ductile fracture problems. The approach can easily account for the entire range of ductile fracture in the framework of non-linear elastoplasticity. A novel multiplicative triple split of the deformation gradient and a novel critical fracture energy are involved in the proposed model.

\subsection{Cohesive fracture}

For cohesive fracture, \citet{verhoosel2013phase} developed a phase field model for straight crack propagation and the numerical modeling is implemented within the framework of the finite element method. In addition to the displacement and phase field fields in a general phase field method for brittle fractures, an auxiliary field was used to represent the displacement jump across the crack. This third field is kept constant orthogonal to the crack. \citet{vignollet2014phase} extended the work of \citet{verhoosel2013phase} and an arc length method and staggered scheme were used. However, the model of \citet{vignollet2014phase} requires a pre-defined path for the fracture propagation.

\subsection{Dynamic fracture}

Numerical modelling of dynamic fractures in solids based on sharp crack discontinuities (e.g. XFEM) is difficult in dealing with complex crack topologies and requires special branching criteria such as XFEM. The phase field based dynamic modelling can overcome these drawbacks. Following the static phase field model of \citet{miehe2010phase, miehe2010thermodynamically}, \citet{hofacker2012continuum, hofacker2013phase} proposed a computational framework of phase field modelling of diffusive dynamic fractures. The proposed modelling method allows complex crack patterns. The dynamic approach follows the history energy field introduced by \citet{miehe2010phase}. This auxiliary field contains a maximum reference energy obtained in the deformation history and drives the diffusive crack evolution. In addition, \citet{hofacker2012continuum, hofacker2013phase} applied the energy split scheme introduced in \citet{miehe2010phase}. Some representative examples of complex crack patterns are also presented by \citet{hofacker2012continuum, hofacker2013phase}.

\citet{borden2012phase} proposed a compact phase field description of dynamic fractures. The presented method is similar to that proposed by \citet{hofacker2012continuum, hofacker2013phase}. Both  monolithic and staggered time integration schemes are presented by \citet{borden2012phase}. For the dynamic description, the kinetic energy $\Psi_{kin}$ is considered:
	\begin{equation}
	\Psi_{kin}(\dot{\bm u})=\frac 1 2\int_\Omega \rho \dot{u}_i \dot{u}_i \mathrm{d}\Omega
	\end{equation}

\noindent where $\dot{\bm u}=\frac {\partial \bm u}{\partial t}$ and $\rho$ is the mass density of the material.

The final energy functional is modified as
	\begin{equation}
	E = \int_\Omega \left\{ \frac 1 2 \rho \dot{u}_i \dot{u}_i-\Psi_e-G_c\left[\frac {(1-s)^2} {4\epsilon}+\epsilon\frac{\partial s}{\partial x_i}\frac{\partial s}{\partial x_i}\right]\right\}\mathrm{d}\Omega
	\end{equation}

\noindent with 
	\begin{equation}
	\Psi_e=\left[(1-k)s^2+k\right]\Psi_0^++\Psi_0^-
	\end{equation}

The resulting governing equations of the dynamic problem are expressed as follows,
	\begin{equation}
	\left\{\begin{aligned}
	&\nabla\cdot\bm\sigma=\rho\ddot{\bm u}\\
	&\left[\frac{4\epsilon(1-k)H}{G_c}+1\right]s-4\epsilon^2\Delta s =1
	\end{aligned}\right.
	\end{equation}
	
\citet{nguyen2018modeling} extended the phase-field cohesive zone model for static fracture to dynamic fracture in brittle and quasi-brittle solids. The performance of the dynamic model is tested on several benchmarks for both dynamic and cohesive brittle fractures. \citet{geelen2019phase} also extended the phase field formulation to dynamic cohesive fracture. The model is characterized by a regularized fracture energy that is linear in the phase field, as well as non-polynomial degradation functions. The authors examined two categories of degradation functions. \citet{ren2019explicit} proposed an explicit phase field model for dynamic brittle fracture. The mechanical field is integrated with a Verlet-velocity scheme, and the phase field is incremented with sub-steps at each step. Adaptive sub-stepping is applied by using the phase field residual, and the explicit scheme avoids the numerical difficulty in convergence and the calculation of anisotropic stiffness tensor. In addition, the phase field modulus is used rather than conventional phase field viscosity.

\subsection{Finite deformation fracture}

The variational approach for brittle fracture such as the original regularized formulation of \citet{bourdin2000numerical} can be extended to the problems of finite deformation fracture.  \citet{kuhn2013numerical} modified the original phase field model to describe fractures in Neo-Hookean materials and the elastic energy density becomes:
	\begin{equation}
	\Psi_e=(s^2+\eta)\left[\frac \lambda 4\left(J^2-1\right)-(\frac \lambda 2+\mu)\mathrm{ln}J+\frac \mu 2\left(\mathrm{tr}(\bm C -3) \right)\right]
	\end{equation}

\noindent where $J$ is the determinant of the deformation gradient, and $\bm C$ is the right Cauchy-Green tensor. If the surface energy density $\gamma(s,\nabla s)$ is unaltered, the thermodynamical restriction yields the constitutive relation:
	\begin{equation}
	\bm S=2\frac{\partial \Psi_e}{\partial \bm C}=(s^2+\eta)\left[\frac \lambda 2(J^2-1)\bm C^{-1}+\mu(\bm 1-\bm C^{-1})\right]
	\end{equation}

\noindent where $\bm S$ is the second Piola-Kirchhoff stress tensor, and $\bm S = \bm F^{-1} \bm P$ with $\bm P$ being the first Piola-Kirchhoff stress tensor. Being constructed in the reference configuration, the balance equation of the phase field modelling reads
	\begin{equation}
	\nabla\cdot\bm P+\bm f_0=\bm 0
	\end{equation}

\noindent where $\bm f_0$ is the body force vector.

Finally, the modified evolution equation of the crack phase field is expressed as
	\begin{equation}
	-\frac{\dot{s}} M=s\left[\frac \lambda 2 (J^2-1)-(\lambda +2\mu)\mathrm{ln} J+\mu\mathrm{tr}({\bm C})-3)\right]-G_c\left(2\epsilon\Delta s+\frac{1-s}{2\epsilon}\right)
	\label{kuhn finite deformation}
	\end{equation}

Note that all the parameters used in Eq. \eqref{kuhn finite deformation} are consistent with those in the model of \citet{kuhn2008phase}. \citet{hesch2014thermodynamically} established another phase-field method for finite deformations and general nonlinear materials. A multiplicative split of the principal stretches is used to account for the fracture behaviors under tension and compression. \citet{hesch2014thermodynamically} also used an energy-momentum consistent integrator and their phase field model is thermodynamically consistent.

\subsection{Anisotropic fracture}

Anisotropy is inherent to crystalline materials (among others) due to the symmetry of the atomic lattice \citep{nguyen2017phase}. Failure anisotropy seemingly conflicts with the local symmetry and maximum energy release rate used in commonly used phase field model such as the model of \citet{miehe2010phase,miehe2010thermodynamically,zhou2019phase2}. Therefore, \citet{li2015phase} proposed a variational
phase-field model for strongly anisotropic fracture. In their model, higher-order phase-field description is implemented in a direct Galerkin way with smooth local maximum entropy approximants.

\citet{nguyen2017phase} proposed a phase field model that could simulate non-free anisotropic crack bifurcations in a robust and fast implementation framework. The key in the model of \citet{nguyen2017phase} is an introduction of the new crack surface density function:
	\begin{equation}
	\gamma(\phi,\nabla\phi,\bm w)=\frac 1{2l_0}\phi^2+\frac{l_0}2\bm w:(\nabla\phi\otimes\nabla\phi)
	\end{equation}

\noindent where $\bm w$ is a second-order structural tensor that characterizes the type of anisotropy and $\bm w$ is an invariant with respect to rotations. $\bm w$ can be expressed as follows,
	\begin{equation}
	\bm w =\bm 1+\beta_2(1-\bm M\otimes\bm M)
	\end{equation}

\noindent where $\bm M$ is the unit vector normal to the preferential cleavage plane, and parameter $\beta_2\ge0$ is used to penalize the damage on planes not normal to $\bm M$. In the case of isotropic material, $\beta_2$ is naturally equal to 0.

\citet{bryant2018mixed} proposed a mixed-mode phase field fracture model in anisotropic rocks with consistent kinematics. The mixed-mode driving force of the phase field is obtained by balancing the microforce. In the method, local fracture dissipation determines the crack propagation and kinematics modes. \citet{pillai2020anisotropic} proposed an anisotropic cohesive phase field model for quasi-brittle fractures in thin fibre-reinforced composites.

\subsection{Plate and shell fractures}

In this section, we provide an overview of phase field approaches applied to dimensionally reduced continua according to their complexity from linear to nonlinear regimes. Generally, phase field models applied in solid bodies can be applied to thin structures, however, the most difficulty is how to make quantities in phase field modeling consistent with ones in shell kinematics which are often described on local frames, i.e, curvilinear coordinates. On the other hand,  what makes it different between published works until now is related to  types of shell models, of phase field models and of discretization methods used, which are summarized in Table \ref{Table1}. Here, we adopt the notion that classifies types of phase field model according to  \citet{miehe2010phase}, in which an approach that splits strain energy and stress tensor into negative and positive parts is regarded as anisotropic phase field model. While a model whose entire elastic energy is degraded in fracture zones is called isotropic, one that still keeps linear in the momentum balance equation but performs a decomposition on the elastic strain energy is referred as a hybrid phase field model \cite{ambati2015review}. We also note that details of related shell models are not emphasized in this section, instead, we refer to other review papers of shell models, see e.g \citet{Bischoff2018}.

	\begin{table}[h]
	\caption{Overview of phase field approaches in thin structures}
	\footnotesize
	\centering
	  \begin{tabular}{c c c c}
	  \hline
	  Model name                                  & Shell model             & Phase field model & Discretization method \\
	  \hline 
	  \citet{Ulmer2012}       & Mindlin-Reissner    & anisotropic           & standard FEM \\
	  \citet{amiri2014phase}          & Kirchhoff-Love      & isotropic               & Local Maximum-Entropy meshfree method\\  
	  \citet{Kiendl2016}       & Kirchhoff-Love      & anisotropic           &  NURBS-based isogeometric analysis        \\
	  \citet{ambati2016phase}    & Solid-shell             & anisotropic           & NURBS-based isogeometric analysis      \\
	  \citet{Areias2016}      & Corotational shell & hybrid                  & standard FEM \\
	  \citet{Reinoso2017} & Solid-shell             & isotropic              & standard FEM \\
	  \hline
	  \end{tabular}
	\label{Table1}
	\end{table}

The first work that applied phase field model in plates and shells is proposed by \citet{Ulmer2012}, in which the shell model is considered as a combination of standard plates and membranes, which allows them to use  standard Lagrangian polynomials to approximate solution fields as the shell model only requires C0$-$continuity of basis functions. On the other hand, elastic energy is split into bending part and membrane part which further is divided into positive and negative components. Accordingly, full of bending energy and positive membrane energy contribute to fracture progression. The free energy functional of this approach takes the form of

	\begin{equation}
	\label{EnergyFunctional111}
		E(\bm{u}, d) = \int \limits_{V} [ g(d) (\mathcal{E}_m^+(\bm{\varepsilon}_m) + \mathcal{E}_b(\bm{\kappa}) ) + \mathcal{E}_m^-(\bm{\varepsilon}_m)  + \mathcal{F}(d, \nabla d)  ]d \, V ,
	\end{equation}
where $\mathcal{E}_m$ is the membrane energy, $\mathcal{E}_b$ bending energy, $ \mathcal{F}(d, \nabla d)$ fracture energy, and $g(d)$ the degradation function.

In 2014, \citet{amiri2014phase} applied the isotropic phase field model to Kirchhoff -Love thin shell, for which they use Local Maximum-Entropy meshfree approximations to fulfill the requirement of C1$-$continuity of basis functions which arises from the appearance of second derivatives in the  bending strain of the shell model.  \citet{Kiendl2016} introduced an approach that combine the anisotropic phase field model with Kirchhoff-Love thin shell and NURBS basis functions are used to approximate solution fields. They also shown that the two mentioned works had some limitations as they may prevent crack propagation and  be not realistic for fracture phenomenon at some circumstances. Amiri's approach employs the isotropic model that makes it only applicable to cases of tensile stress states, while Ulmer's model leads to delay of cracking in cases of combination between bending and compressive membrane strains or may result in evolving damage under cases of purely compressive strains . Accordingly, Eq. \ref{EnergyFunctional111} is rewritten as

\begin{equation}
\label{EnergyFunctional222}
	E(\bm{u},d) = \int \limits_{V} [g(d) \mathcal{E}^+(\bm{\varepsilon}) +  \mathcal{E}^-(\bm{\varepsilon}) +  \mathcal{F}(d, \nabla d) ] d \, V
\end{equation}
where	total strain $ \bm{\varepsilon} = \bm{\varepsilon}_m + \theta^3 \, \bm{\kappa}$ with $\bm{\varepsilon}_m$ and $\bm{\kappa}$ as membrane strain and curvature change respectively.
In Eq. \ref{EnergyFunctional222}, both membrane and bending strains contribute to total strain tensor which is split into tensile and membrane components, which prohibits the decomposition of elastic energy into membrane and bending terms .
In contrast with the aforementioned works that employed the classical thin shell models, \citet{ambati2016phase} investigated the anisotropic phase field approach to a solid-shell element whose kinematics quantities are defined through the thickness instead of on the mid-surface as in  the Kirchhoff-Love shell. The employed  solid-shell formulation is rotation-free, which motivates  inherenting the same nodes and degrees of freedom of the solid element, allowing to model fracture for elasto-plastic materials by using general three dimensional (3D) elasto-plastic constitutive models, i.e J2 plasticity, see e.g \citet{ambati2015phase} for ductile fracture model in solid.
\\
One thing to note is that all the phase field models in thin structures  above are formulated in linear regime, which makes them very limited , since in reality shell structures are usually undergone  large deformation and rotation. To overcome this, \citet{Areias2016} developed a hybrid phase field approach to the so-called corotational shell at finite strain. Kiendl's work \citep{Kiendl2016} which ensures the irreversibility of crack evolution in the case of elasto-plasticity by the  local history field which is the maximum of the positive elastic energy obtained within each loading step,  Areias's model suggests another criteria for the irreversibility of crack phase field as
	\begin{equation}
	\mathcal{H} = <max(W_p - W_p^*)>_+
	\end{equation}
with $W_p$ as the plastic work and $W_p^*$ as the critical value of the plastic work, which enables to employ the  work of separation in the ductile fracture \citep{Siegmund2000}. Furthermore, to get consistent in the extension of infinitesimal-strain regime for the continuum phase field model to the case of the employed finite strain shell, a consistent updated-Lagrangian algorithm, where a reference configuration is not chosen as the undeformed configuration, is adopted, see \citet{Areias2016} for details. Recently, \citet{Reinoso2017} introduced an approach combining the isotropic phase field model with a six-parameter  solid-shell element which consists of three displacements on the mid-surface, two independent rotations and additional degree of freedom accounting for thickness stretch. Large deformation as well as linear and nonlinear elastic constitutive laws are considered  in this work.  Note that, both Areias' and Reinoso's models describe variations of phase field through the shell thickness that leads to a correct description of fracture in  bending-dominated cases. This is in contrast with previous works that assume invariance of the phase field variable over the shell thickness. Following \citet{Areias2016}, phase field is linearly interpolated between top and bottom faces of the shell as follows
	\begin{equation}
	d(\theta^1, \theta^2, \theta^3) = \frac{1}{2} (1+ \theta^3) d_{top}(\theta^1, \theta^2) +   \frac{1}{2} (1- \theta^3) d_{bottom}(\theta^1, \theta^2) 
	\end{equation}
where $d_{top}$ and $d_{botoom}$ are phase field variables at top and bottom faces respectively, while $\theta^1$, $\theta^2$ and $\theta^3$ are coordinates in the parametric space with $(\theta^1, \theta^2)$ as the in-plane directions and $\theta^3$ as the thickness direction.

%\citet{amiri2014phase} presented a phase field model for fractures in Kirchoff-Love thin shells. Their PFM is coupled with the local maximum-entropy (LME) meshfree method. Statistical learning techniques are used to describe the geometry of the shell and no global parametrization is used because general point set surfaces are allowed to tackle complex surface topology. The presented model shows a good flexibility and robustness in two numerical examples: plate in tension and a set of open connected pipes.
%\citet{ambati2016phase} applied another phase field model to investigate fracture in shells. They examined brittle fractures in single-edge notched specimen tension/shear tests, an annular plate, and a notched cylinder under internal pressure. The ductile fractures in a circumferentially notched cylinder under axial tension, a simply supported plate, and an internally pressurized hexagonal base are also investigated.

\subsection{Thermal fracture}

The phase field approach can be used for thermal fracture modelling. To study thermally induced fractures, the thermal strains are required in the total energy functional. At small strains, the total strain is the sum of elastic and thermal strains. With a thermal expansion tensor $\bm \alpha$, the thermal strain is $\bm\varepsilon^\theta=\theta\bm\alpha$, and the total strain $\bm\varepsilon$ reads
	\begin{equation}
	\bm\varepsilon=\bm\varepsilon^e+\bm\varepsilon^\theta
	\end{equation}

\noindent where $\bm\varepsilon^e$ is the elastic strain and $\theta$ is the temperature variation. Only the elastic part $\bm\varepsilon^e$ of the strain tensor contributes to the elastic energy density \citet{kuhn2013numerical}:
	\begin{equation}
	\Psi=\frac 1 2(s^2+\eta)\bm\varepsilon^e:[\mathbb C:\bm\varepsilon^e]=\frac 1 2(s^2+\eta)(\bm\varepsilon-\bm\varepsilon^\theta):[\mathbb C:(\bm\varepsilon-\bm\varepsilon^\theta)]
	\end{equation}

The Cauchy stress is then expressed in terms of the total and thermal strains:
	\begin{equation}
	\bm\sigma=\frac{\partial\Psi}{\partial\bm\varepsilon}=\frac 1 2(s^2+\eta)\mathbb C:(\bm\varepsilon-\bm\varepsilon^\theta)
	\end{equation}

The crack evolution equation in \citet{kuhn2008phase} then can be modified as
	\begin{equation}
	\dot s = -M\left\{s(\bm\varepsilon-\bm\varepsilon^\theta):[\mathbb C:(\bm\varepsilon-\bm\varepsilon^\theta)]-G_c\left(2\epsilon\Delta s+\frac{1-s}{2\epsilon}\right)\right\}
	\end{equation}

In addition to the balance equation and phase field evolution equation, the governing equation for the temperature field is \citep{kuhn2013numerical}:
	\begin{equation}
	-\nabla\cdot\bm q^\theta=\rho c \dot\theta
	\end{equation}
\noindent where $c$ is the specific heat capacity and $\bm q^\theta$ is the heat flux. \citet{kuhn2009phase} considered the influence of the phase field on the heat conduction behavior and the modified Fourier's law can be expressed as follows
	\begin{equation}
	\bm q^\theta=-\left(\beta(s^2+\eta-1)+1\right)\bm\kappa\nabla\theta
	\end{equation}
\noindent where $\beta\in [0, 1]$ is a parameter that defines the influence of the crack field on the thermal conductivity $\bm\kappa$. If $\beta = 0$, the heat conduction is not affected by the phase field. If $\beta = 1$, the thermal conductivity degrades and tends to zero for $s = 0$. Consequently, there is no heat flux across a crack, i.e. cracks are isolating \citep{kuhn2013numerical}.

The thermally induced phase field patterns can be seen in \citet{bourdin2007variational,corson2009thermal}. \citet{badnava2018h} made a recent contribution. They adopted the similar decomposition technology of \citet{miehe2010phase} to the elastic parts of the strain:
	\begin{equation}
	\bm\varepsilon_\pm^e=\sum_{I=1}^3 \langle\varepsilon_I^e\rangle\bm n_I\otimes \bm n_I
	\end{equation}

The initial elastic energy density is modified as 
	\begin{equation}
	\Psi_0^e=\Psi_0^{e+}+\Psi_0^{e-}
	\end{equation}
\noindent with 
	\begin{equation}
	\Psi_0^{e\pm}=\frac 1 2 \bm\varepsilon_\pm^e:\mathbb C:\bm\varepsilon_\pm^e
	\end{equation}

\citet{dittmann2020phase} introduced a framework to simulate porous-ductile fracture in isotropic thermo-elasto-plastic solids and considered large deformations. In the model, they combined a modified Gurson-Tvergaard-Needleman GTN-type plasticity model with a phase-field fracture approach. Therefore, the temperature-dependent growth of voids on micro-scale followed by crack initiation and propagation on macro-scale can be well modeled. 

\subsection{Hydraulic fracture}

PFMs seem to be a valuable alternative for modeling hydraulic fractures (HF) because all advantages of the phase field methods are appealing for HF. PFMs for HF have been proposed for instance in \citet{bourdin2012variational,wheeler2014augmented,miehe2015minimization,mikelic2015quasi, mikelic2015phase,heister2015primal, lee2016pressure,wick2016fluid, yoshioka2016variational,miehe2016phase,ehlers2017phase,santillan2017phase, zhou2019phase, zhuang2020hydraulic}. \citet{wheeler2014augmented} rewrote the energy functional by including poroelastic terms and succeeded in extending the phase field model to porous media. However, the variation in the reservoir and fracture domains with time is treated as a moving boundary problem and thereby extra work is needed for implementation. Later, \citet{mikelic2015quasi, mikelic2015phase} fully coupled the three fields: elasticity, phase field, and pressure. They modified the energy functional from their previous work,  the flow in their entire system is governed by Biot equations. The permeability tensor was also modified to consider a higher permeability along the fracture. The implementation approaches of \citet{mikelic2015quasi, mikelic2015phase} were then enhanced using adaptive element schemes by \citep{heister2015primal, lee2016pressure}. \citet{wick2016fluid, yoshioka2016variational} proposed approaches to couple the phase field model to reservoir simulators. Miehe et al. \citep{miehe2015minimization,miehe2016phase} proposed new minimization and saddle point principles for Darcy-Biot-type flow in fractured poroelastic media coupled with phase field modeling. The evolution of the phase field was driven by the effective stress in the solid skeleton and a stress threshold was set. Moreover, the flow in the fractures was set as Poiseuille-type by modeling Darcy flow with an anisotropic permeability tensor. Recently, \citet{ehlers2017phase} embedded a phase-field approach in the theory of porous media to model dynamic hydraulic fracturing. \citet{santillan2017phase} proposed an immersed-fracture formulation for impermeable porous media. 

\citet{lee2018optimal} presented a framework that couples the fluid-filled fracture propagation and a genetic inverse algorithm for optimizing hydraulic fracturing scenarios in porous media. \citet{lee2018phase} proposed an immiscible two phase flow fracture model, based on a traditional phase-field model for predicting fracture initiation and propagation in porous media. The multifluid model extends the classical flow models and nonzero capillary pressure is considered. \citet{mikelic2019phase} studied propagation of hydraulic fractures using the fixed stress splitting method. The mechanical step involving displacement and phase field unknowns is studied under a given pressure. However, these recently developed approaches for extending PFM to HF are quite complicated and computationally expensive. 

Another interesting work can be seen in \citet{zhou2018phase2}. Biot poroelasticity theory is applied on the porous medium with a phase field description of the fracture behavior. An additional pressure-related term is added to the original energy functional  presented by \citet{miehe2010phase}. However, different from  \citet{miehe2015minimization,miehe2016phase}, only the elastic energy drives the fracture propagation and no stress threshold is imposed in the formulation. In addition, the phase field is used to construct indicator functions to transit fluid property from the intact medium to the fully broken one. \citet{zhou2018phase2} implemented the phase field modelling by using the aforementioned Comsol Multiphysics and a staggered scheme. The numerical results are also verified by analytical solutions.

The modified energy functional in \citet{zhou2018phase2} reads
	\begin{equation}
	E(\bm u,\Gamma) = \int_{\Omega}\Psi_e(\bm \varepsilon) \mathrm{d}{\Omega}-\int_{\Omega}\alpha p \cdot (\nabla \cdot \bm u) \mathrm{d}{\Omega}+\int_{\Gamma}G_c \mathrm{d}S-\int_{\Omega} \bm b\cdot{\bm u}\mathrm{d}{\Omega} - \int_{\partial\Omega_h} \bm f\cdot{\bm u}\mathrm{d}S
	\end{equation}
\noindent where $p$ is the fluid pressure, $\alpha$ is the Biot coefficient, $\bm b$ is the body force, and $\bm f$ is the surface traction on the Neumann boundaries.

The degradation of the elastic energy is modeled by the following equation:
	\begin{equation}
	\Psi_e(\bm\varepsilon)=\left[(1-k)(1-\phi)^2+k\right]\Psi_0^+(\bm \varepsilon)+\Psi_0^-(\bm \varepsilon)
	\end{equation}

\noindent where $0<k\ll1$ is a model parameter that prevents the tensile part of the elastic energy density from vanishing and avoids numerical singularity when the phase field $\phi$  tends to 1. The same decomposition as \citet{miehe2010phase} is used to obtain $\Psi_0^+$, $\Psi_0^-$, $\bm\varepsilon_+$, and $\bm\varepsilon_-$.

With the first variation of the energy functional and the history reference field $H(\bm x,t) = \max \limits_{\tau\in[0,t]}\psi_0^+\left(\bm\varepsilon(\bm x,\tau)\right)$ being used, the strong form is written as
	\begin{equation}
	\left\{
	\begin{aligned}
	\frac {\partial {\sigma_{ij}^{por}}}{\partial x_i}+b_i=0
	\\ \left[\frac{2l_0(1-k)H}{G_c}+1\right]\phi-l_0^2\frac{\partial^2 \phi}{\partial {x_i^2}}=\frac{2l_0(1-k)H}{G_c}
	\label{governing equation1}
	\end{aligned}\right.
	\end{equation}
\noindent with the Cauchy stress tensor $\bm\sigma^{por}=\bm \sigma(\bm\varepsilon)-\alpha p \bm I$.

Recently, new PFMs are developed for hydraulic fractures. For example, \citet{shiozawa2019effect, zhou2020phase} considered the effect of stress boundary in the phase field models and the correct HF type under stress boundary can be well predicted. \citet{zhou2020phase2} proposed a phase field model for hydraulic fracturing in transversely isotropic porous media.

\subsection{length scale insensitive phase-field model}
\citet{wu2018length} extended their work in quasi-brittle failure and proposed for the first time a length scale insensitive phase-field model for brittle fracture. Their model involves a phase-field regularized cohesive zone model (CZM) with linear softening law and several optimal characteristic functions. As an extension of the common phase field models, the length scale insensitive model has both
failure strength and traction-separation law that are independent of the internal length scale parameter. The best merit of the proposed model is that it gives length scale independent global responses for fracture problems.

\subsection{Rock fracture}
\citet{choo2018coupled} combined a pressure-sensitive plasticity model and a phase field model to model fractures in geological materials. \citet{zhou2019phase2} revisited the formulation of \citet{ambati2015review} and established a new driving force for the phase field evolution to consider compressive-shear fractures in rocks. The phase field model is established in a hybrid framework. To the authors' best knowledge, the model is the first model that can simulate well the compression-induced fracture in rocks and consider the friction effect. \citet{fei2020phase} developed another phase field model for shear fracture in pressure-sensitive geomaterials with emphasis on the effect of friction on fracture evolution. Governing equations for different contact conditions are established while energy is assumed dissipated during slip. The degradation function and threshold energy are fully considered such that the stress responses are insensitive to the length scale. However, the fracture direction is not automatically determined by the evolution equation of the phase field. Instead, the direction is determined using an extra criterion. \citet{wang2020phase} proposed a phase-field model for mixed-mode fracture based on a unified tensile fracture criterion. The proposed method can be also applied to rock fracture. The model involves an additional material parameter. However, the fracture angle is determined by the tensile fracture criterion.

% \subsection{Phase field for optimization}

\section{Representative numerical examples}\label{Representative numerical examples}

In this section, some representative examples are presented to show the capability of the phase field modeling of fracture.

\subsection{Single-phasic problem}
	\subsubsection{Quasi-static fracture}
\paragraph {1. 2D notched square plate subjected to tension}

\begin{figure}[htbp]
	\centering
	\subfigure[Geometry and boundary conditions]{\includegraphics[width = 5cm]{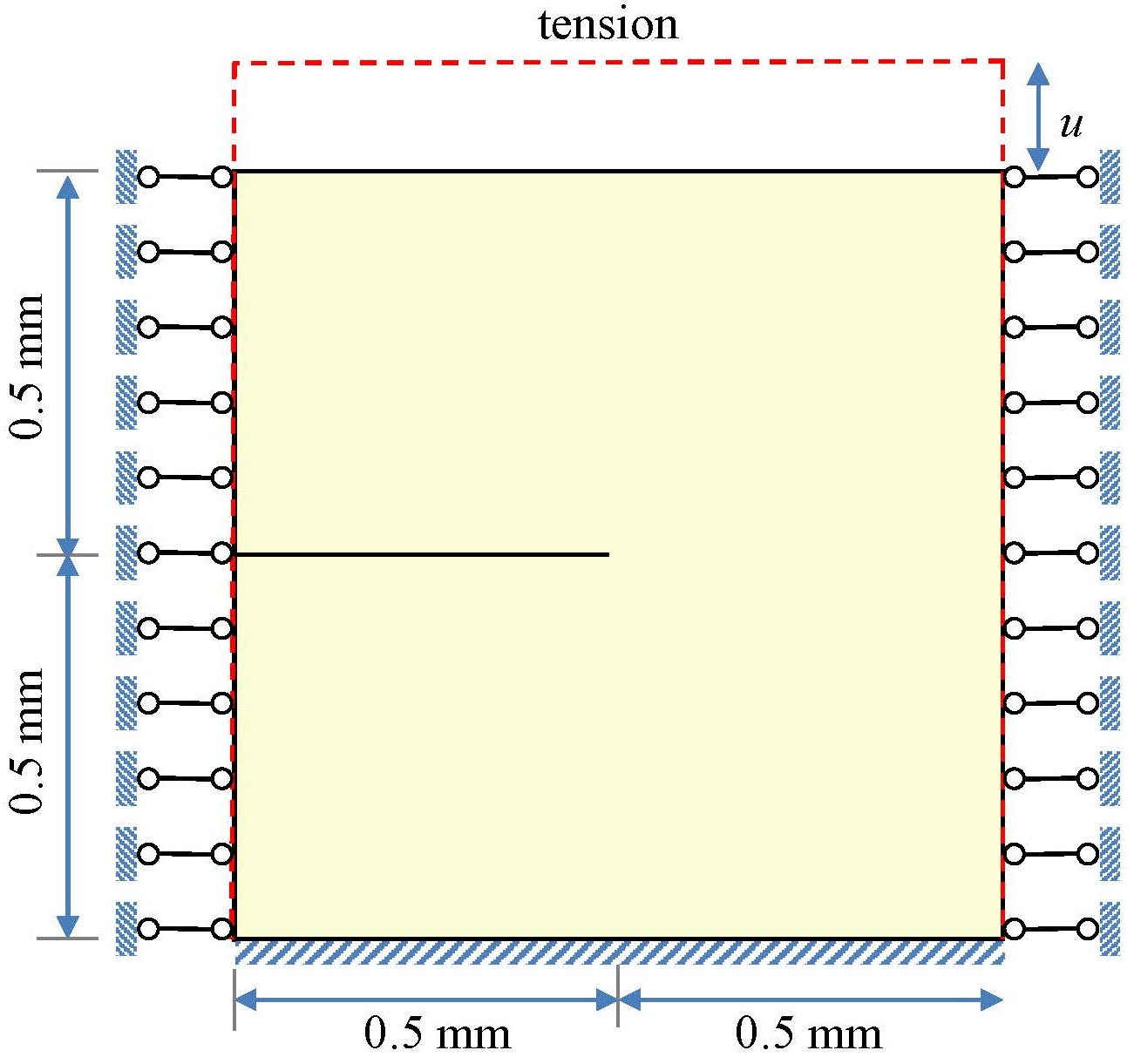}}
	\subfigure[$l_0=1.5\times10^{-2}$ mm]{\includegraphics[width = 5cm]{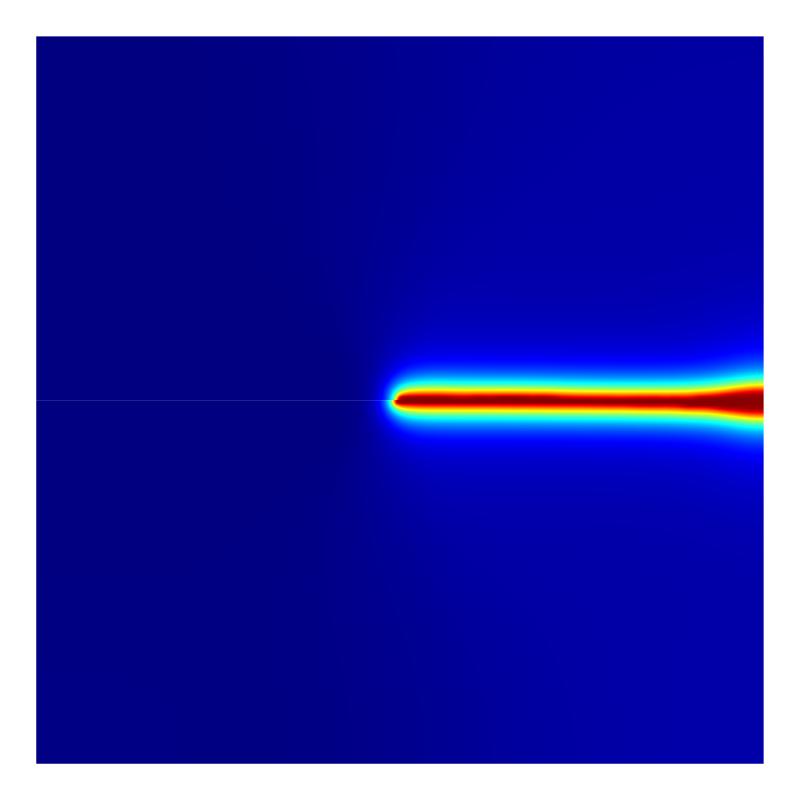}}
	\subfigure[$l_0=7.5\times10^{-3}$ mm]{\includegraphics[width = 5cm]{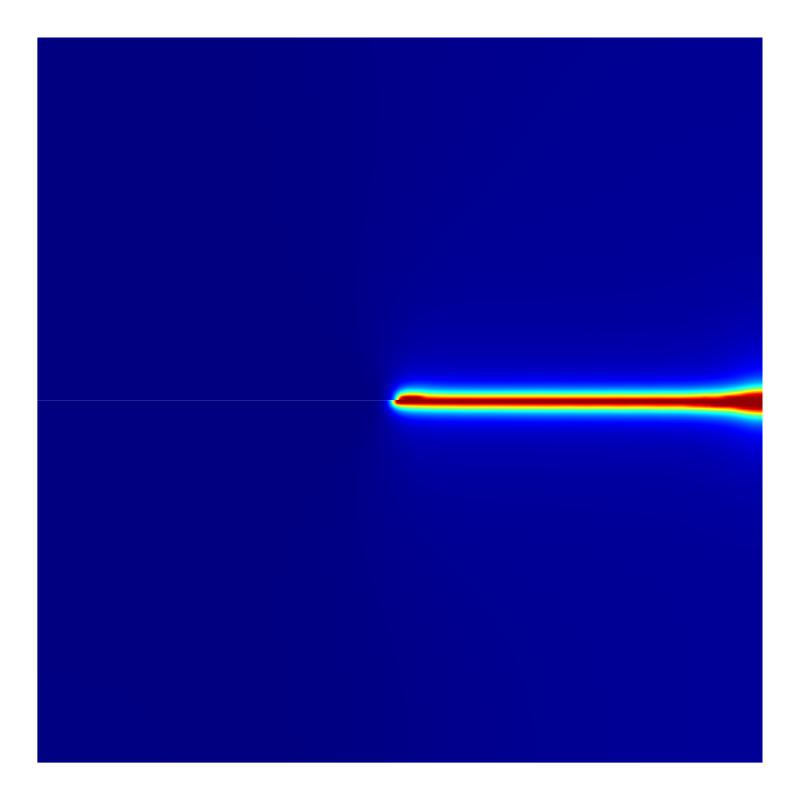}}\\
	\caption{Fracture patterns of a single-edge-notched square plate subjected to tension}\citep{zhou2018phase}
	\label{Fracture patterns of a single-edge-notched square plate subjected to tension}
	\end{figure}

Fracture patterns in a square plate with an initial notch subjected to static tensile loading are presented in Fig. \ref{Fracture patterns of a single-edge-notched square plate subjected to tension} \citep{zhou2018phase}. This benchmark test has been calculated and analyzed by \citet{miehe2010phase, miehe2010thermodynamically, hesch2014thermodynamically, liu2016abaqus}. The geometry and loading condition are shown in Fig. \ref{Fracture patterns of a single-edge-notched square plate subjected to tension}a. The material parameters are: Young's modulus $E$ = 210 GPa, Poisson's ratio $\nu=0.3$, and critical eneryg release rate $G_c$ = 2700 J/m$^2$. Figure \ref{Fracture patterns of a single-edge-notched square plate subjected to tension}b and c shows a horizontal cracks in the middle of the plate.

\begin{figure}[htbp]
	\centering
	\subfigure[isotropic PFM]{\includegraphics[width = 5cm]{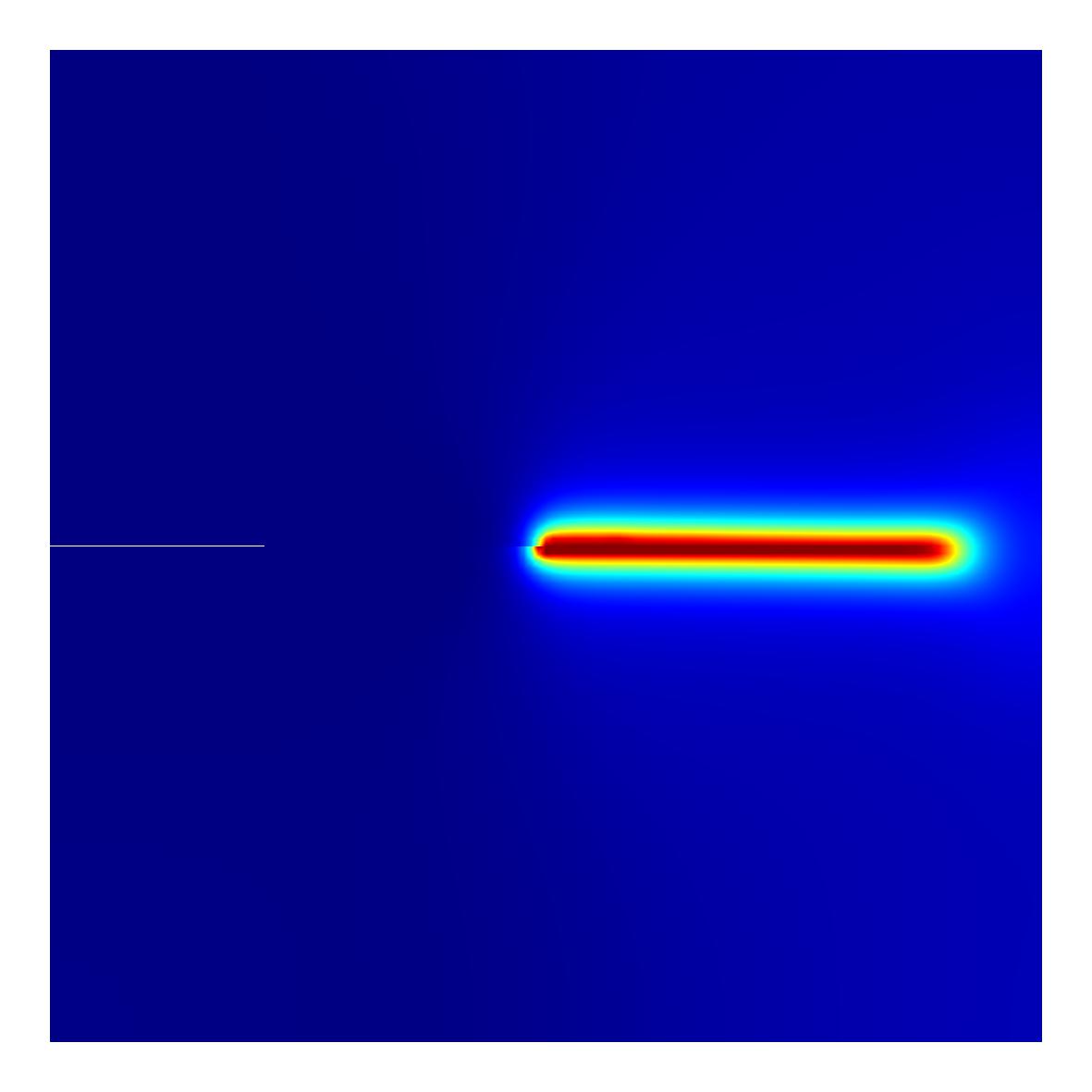}}
	\subfigure[\citet{miehe2010thermodynamically}]{\includegraphics[width = 5cm]{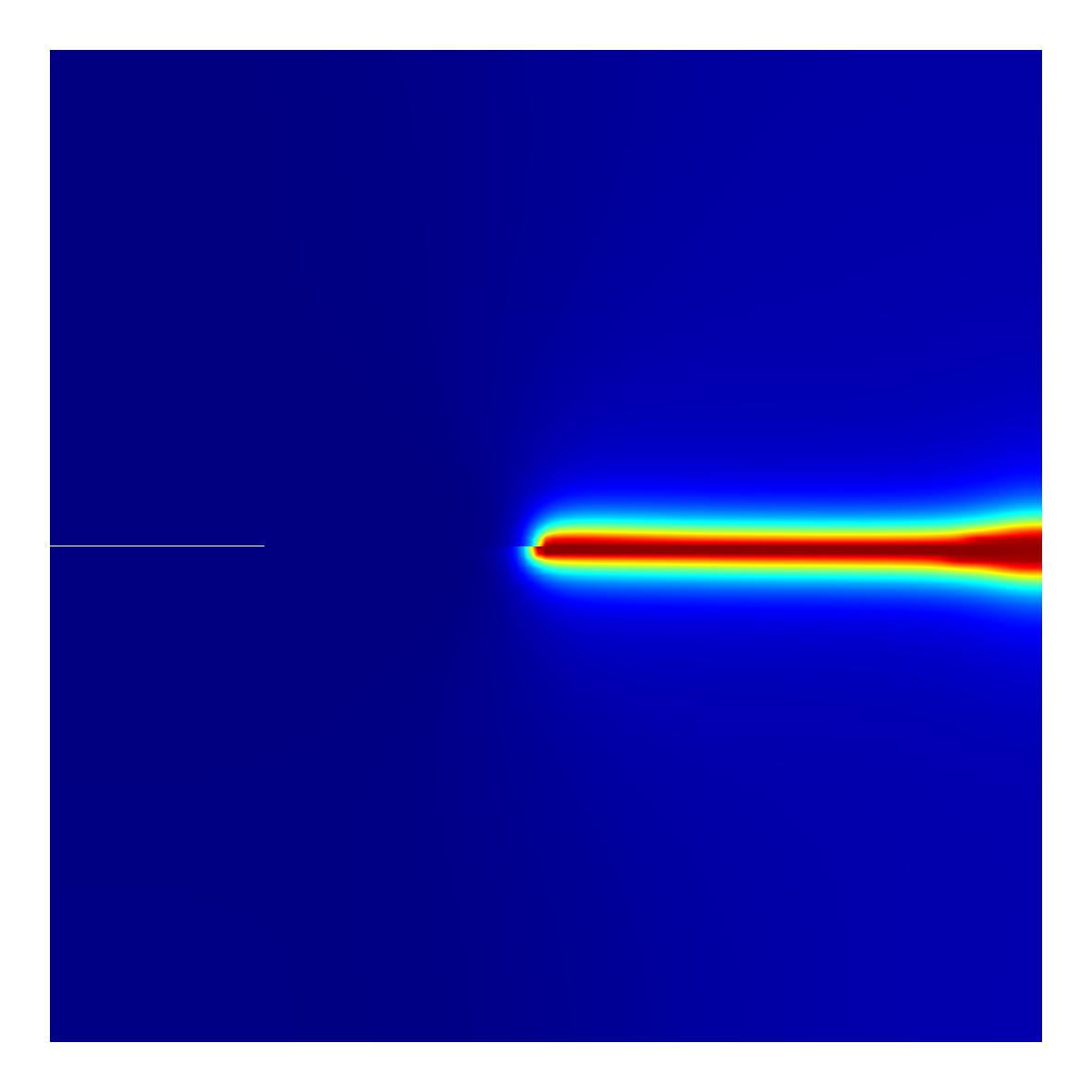}}\\
	\subfigure[\citet{ambati2015review}]{\includegraphics[width = 5cm]{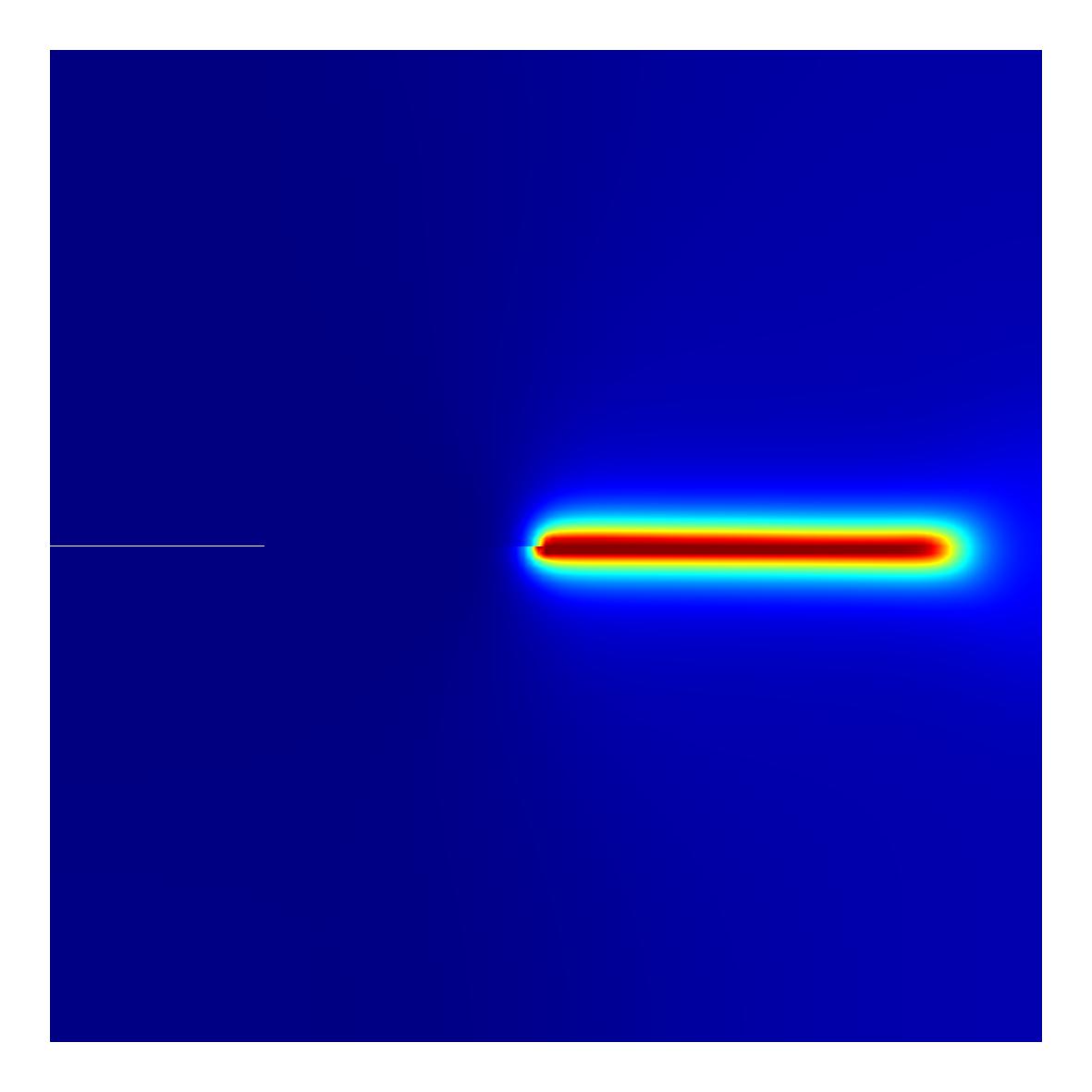}}
	\subfigure[\citet{amor2009regularized}]{\includegraphics[width = 5cm]{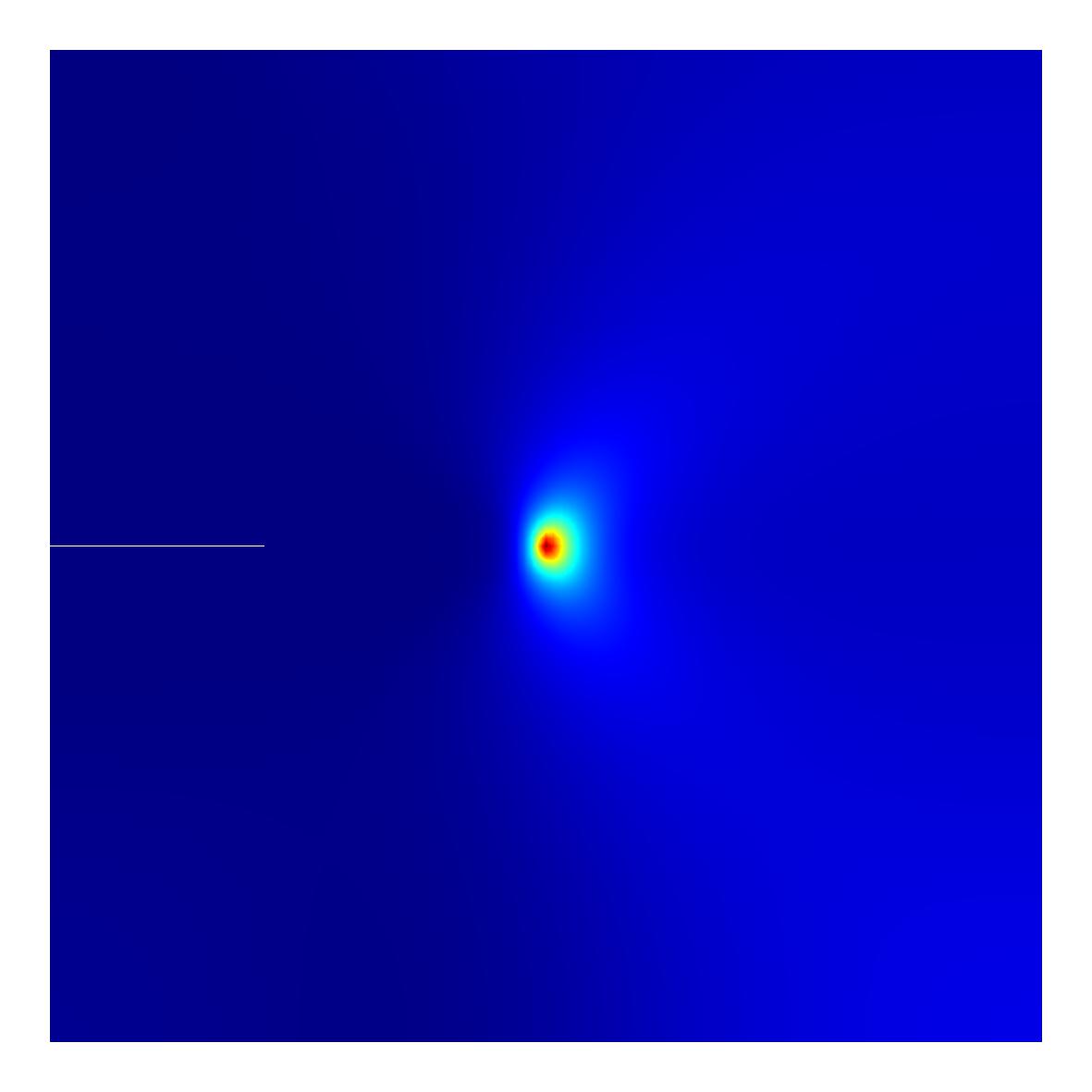}}\\
	\caption{Fracture patterns of a single-edge-notched square plate subjected to tension when $u=6.2\times 10^{-3}$ mm}
	\label{Fracture patterns of a single-edge-notched square plate subjected to tension when u=6.2times 10-3 mm}
	\end{figure}

\begin{figure}[htbp]
	\centering
	\includegraphics[width = 10cm]{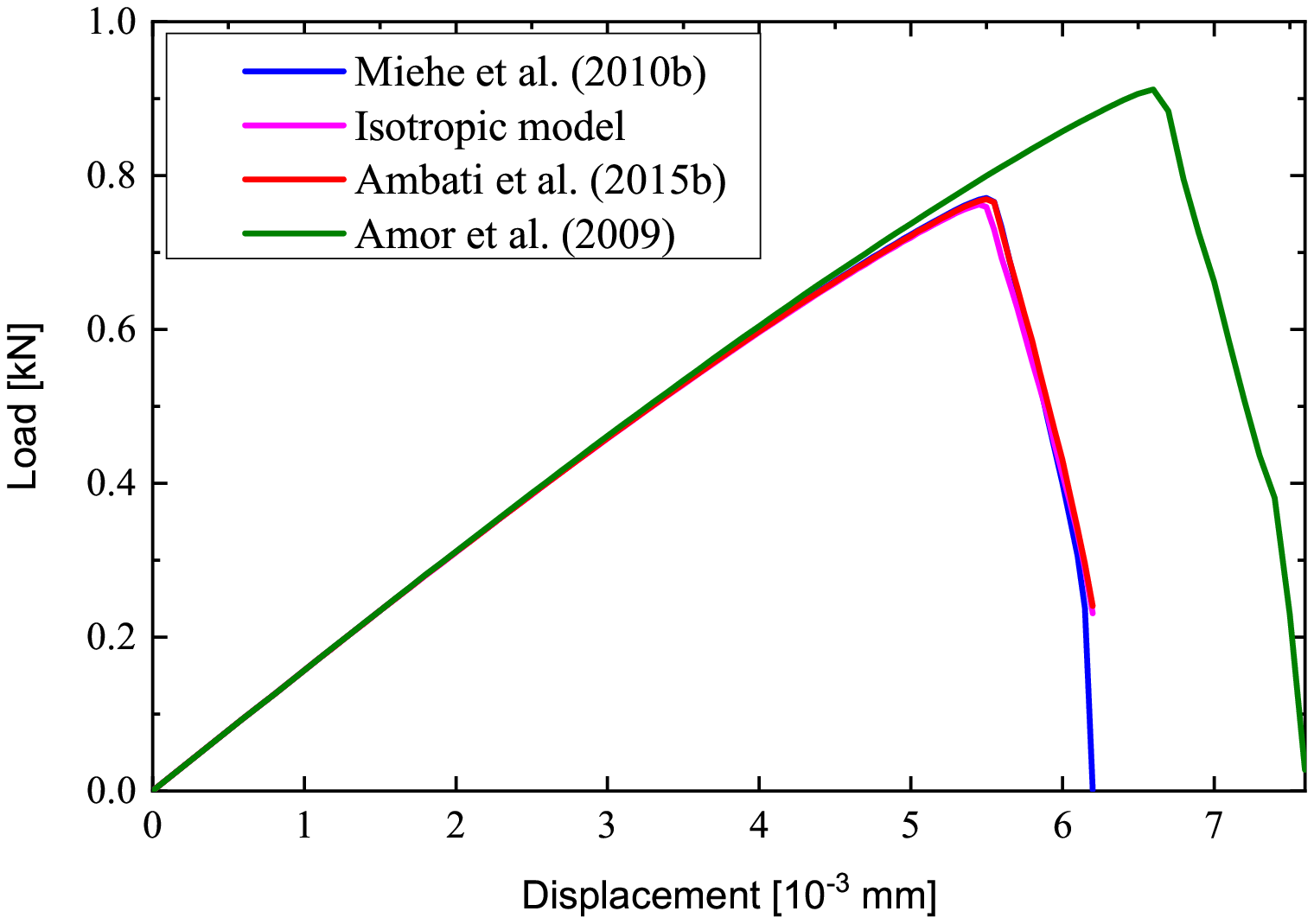}
	\caption{Load-displacement curve of a single-edge-notched square plate subjected to tension for different PFMs}
	\label{Load-displacement curve of a single-edge-notched square plate subjected to tension for different PFMs}
	\end{figure}

The effects of different phase field models on this tension example are also tested. We show the results obtained from \citet{miehe2010thermodynamically}, \citet{ambati2015review}, \citet{amor2009regularized} and the isotropic model under length scale $l_0=1.5\times10^{-2}$ mm and maximum mesh size $h=7.5\times10^{-3}$ mm. The numerical simulations indicate that all the mentioned four PFMs can obtain the same fracture pattern. However, the fracture initiation and propagation differ at the same vertical displacement as shown in Fig. \ref{Fracture patterns of a single-edge-notched square plate subjected to tension when u=6.2times 10-3 mm}. In addition, different PFMs show different load-displacement curves as shown in Fig \ref{Load-displacement curve of a single-edge-notched square plate subjected to tension for different PFMs}. \citet{amor2009regularized} predicts a much higher peak load compared with the other methods.

\paragraph {2. 2D notched square plate subjected to shear}

\begin{figure}[htbp]
	\centering
	\subfigure[Geometry and boundary conditions]{\includegraphics[width = 5cm]{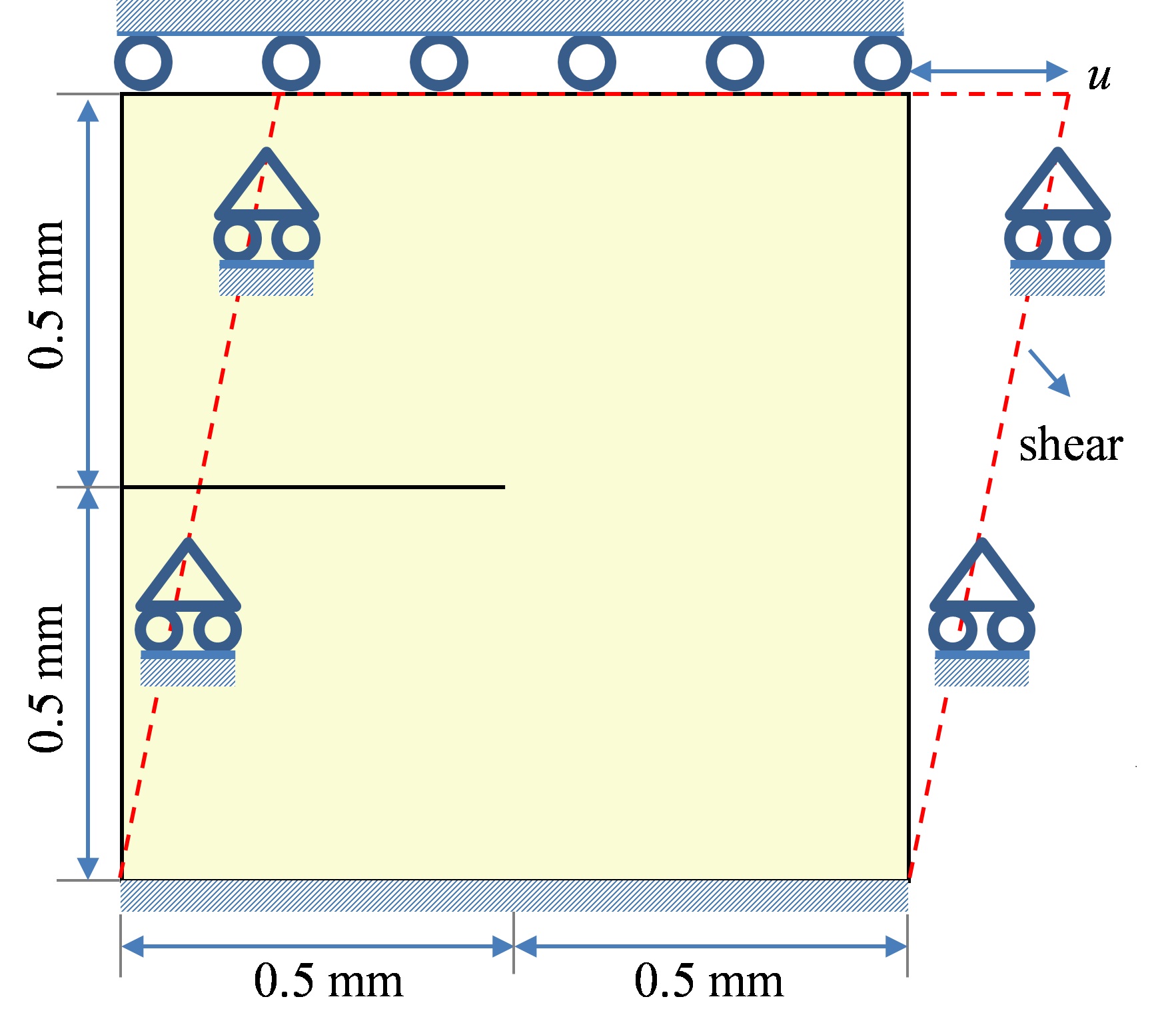}}
	\subfigure[$l_0=1.5\times10^{-2}$ mm]{\includegraphics[width = 5cm]{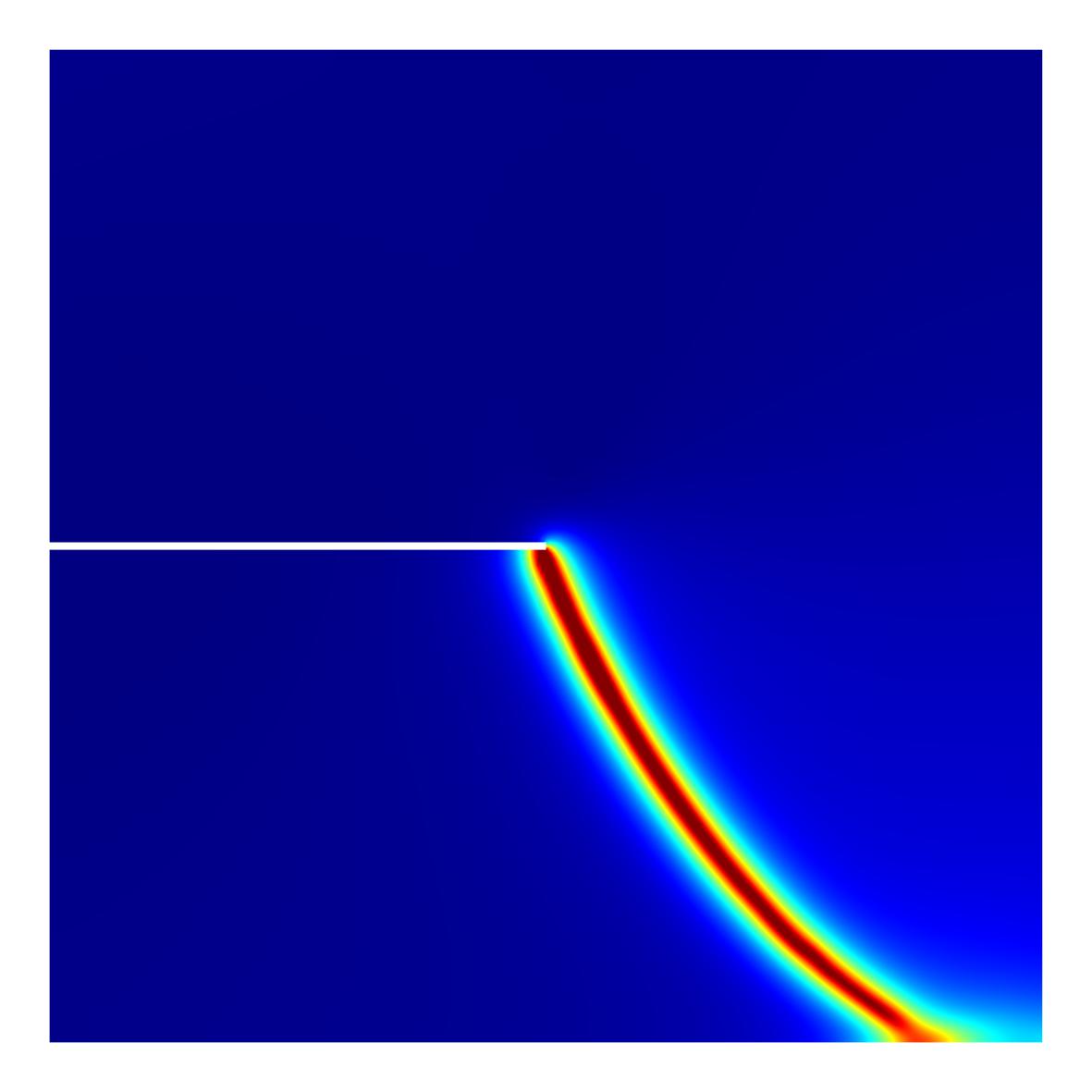}}
	\subfigure[$l_0=7.5\times10^{-3}$ mm]{\includegraphics[width = 5cm]{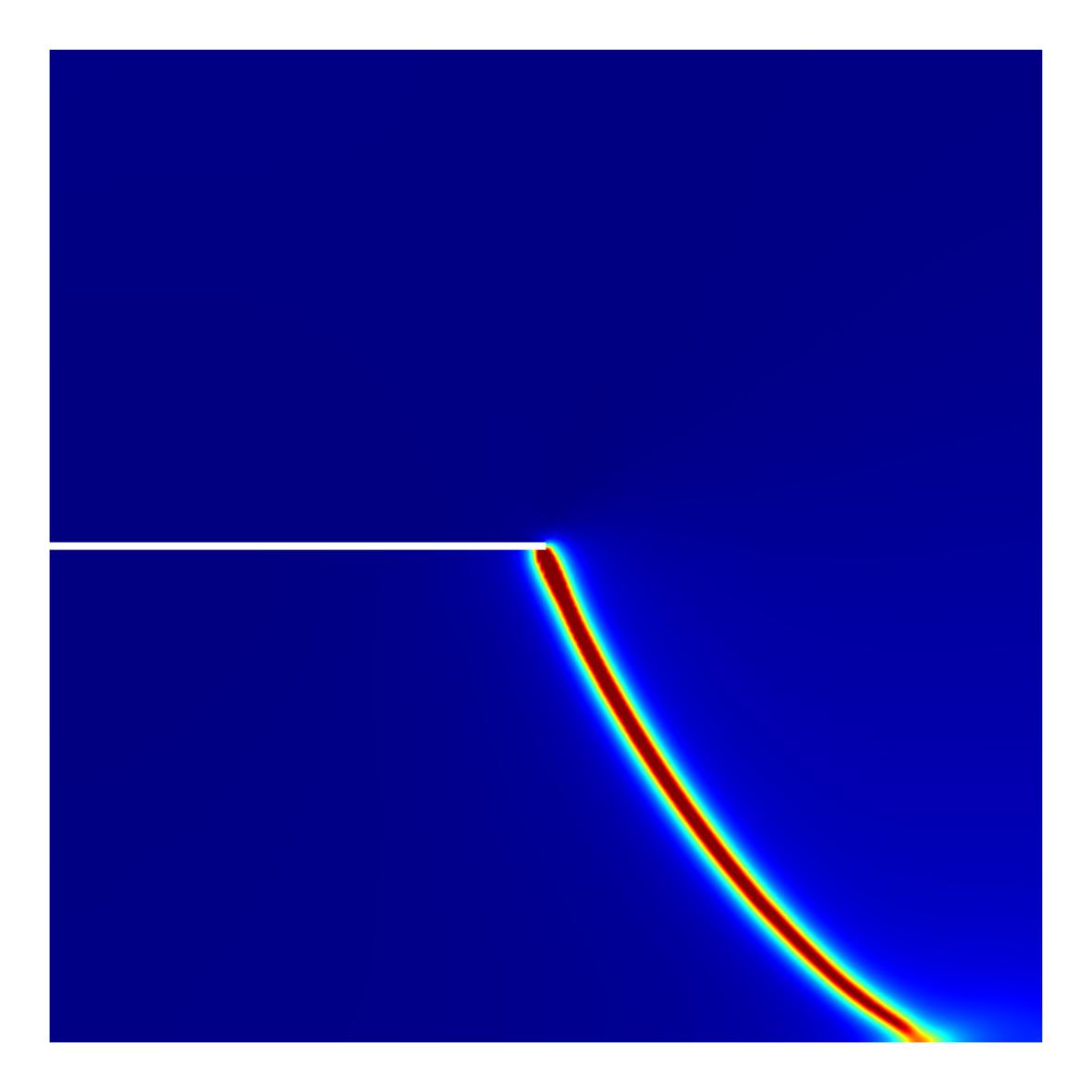}}\\
	\caption{Fracture patterns of a single-edge-notched square plate subjected to shear}\citep{zhou2018phase}
	\label{Fracture patterns of a single-edge-notched square plate subjected to shear}
	\end{figure}

Fracture patterns in a square plate with an initial notch subjected to static shear loading are presented in Fig. \ref{Fracture patterns of a single-edge-notched square plate subjected to shear} \citep{zhou2018phase}. This benchmark test is also shown in  \citet{miehe2010phase, miehe2010thermodynamically, hesch2014thermodynamically, liu2016abaqus}. The geometry and loading condition are shown in Fig. \ref{Fracture patterns of a single-edge-notched square plate subjected to shear}a. Figure \ref{Fracture patterns of a single-edge-notched square plate subjected to shear}b and c shows an inclined crack from the right tip of the pre-existing notch. As expected, the crack has a larger width when $l_0 = 1.5\times10^{-2}$ mm.

\begin{figure}[htbp]
		\centering
		\subfigure[isotropic PFM]{\includegraphics[width = 5cm]{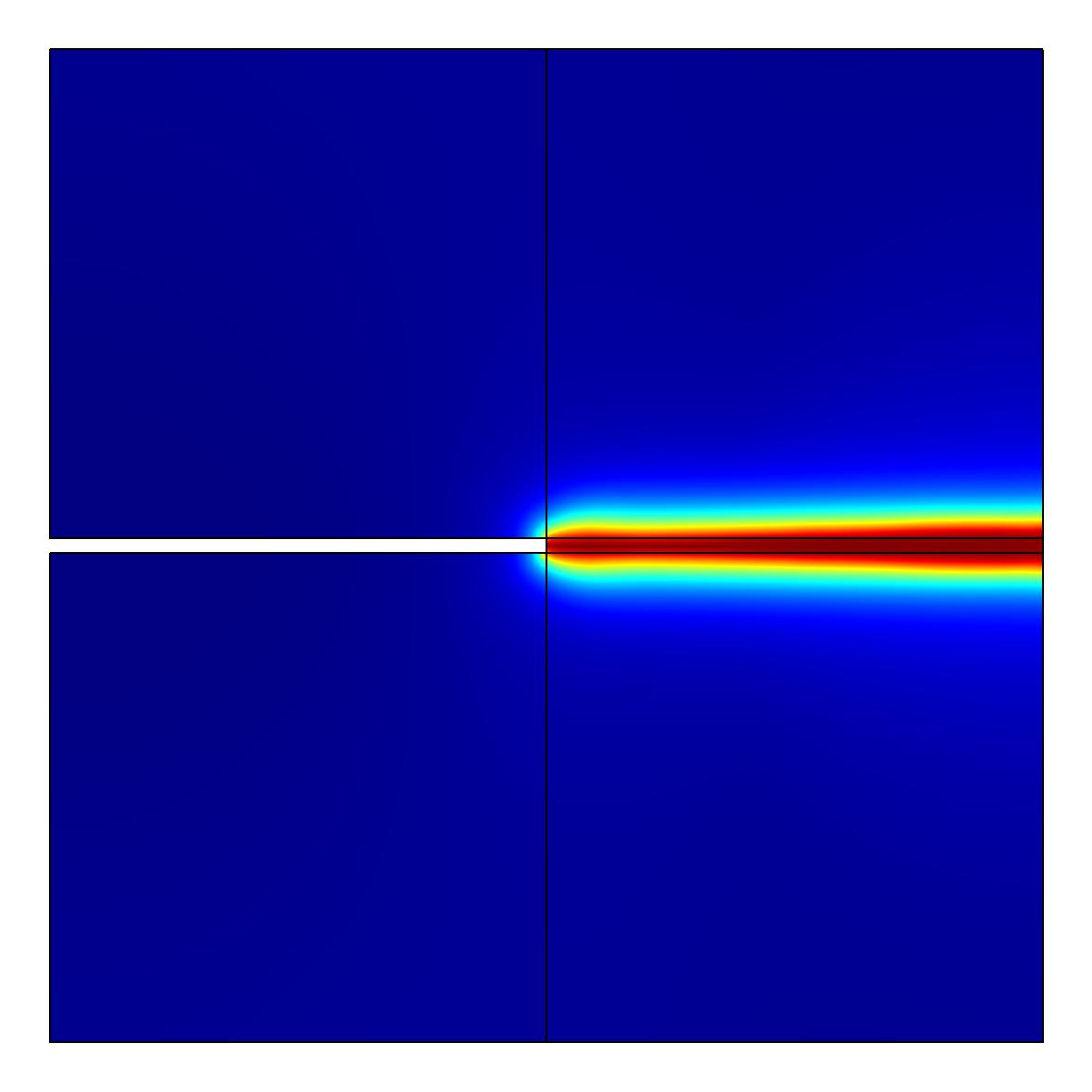}}
		\subfigure[\citet{miehe2010thermodynamically}]{\includegraphics[width = 5cm]{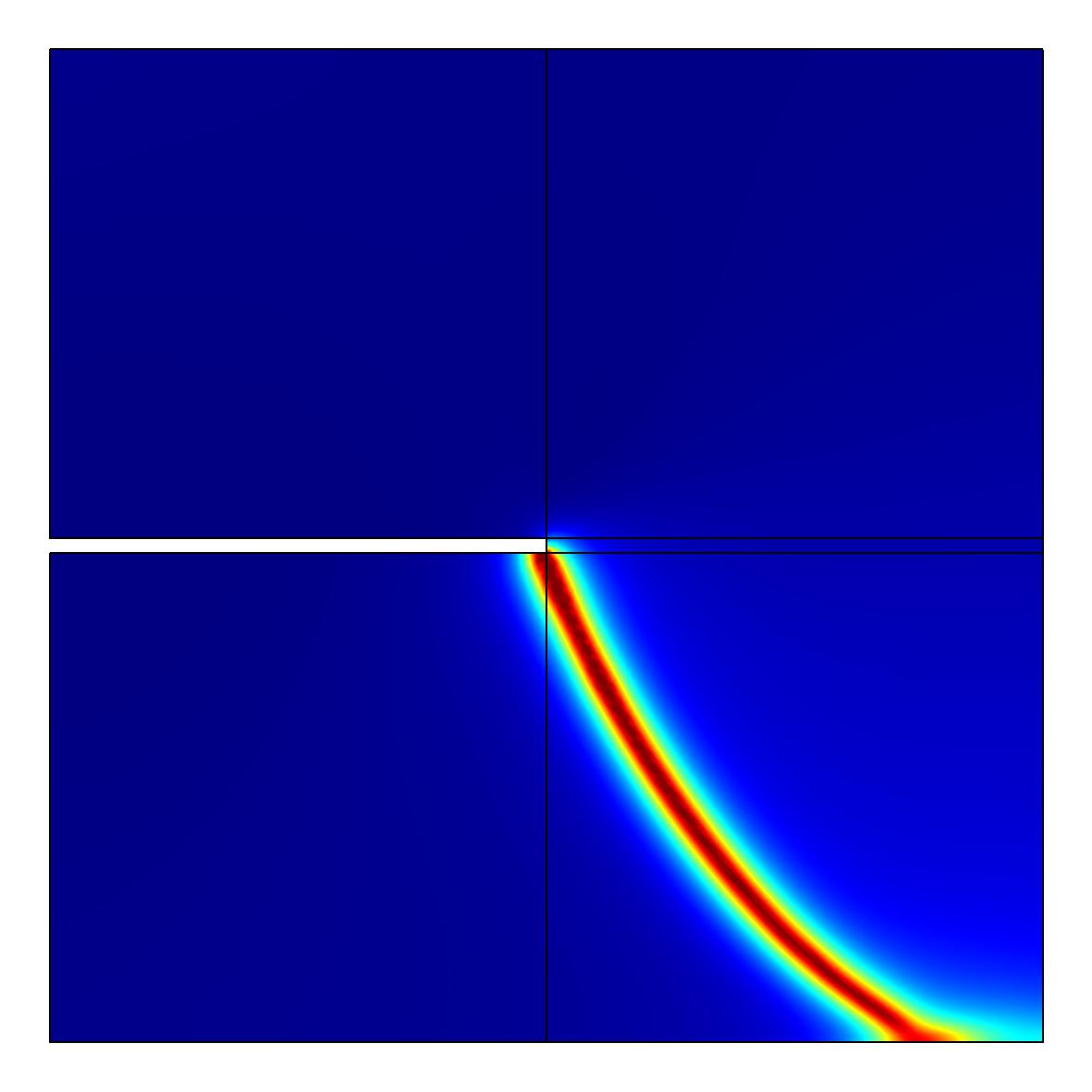}}\\
		\subfigure[\citet{ambati2015review}]{\includegraphics[width = 5cm]{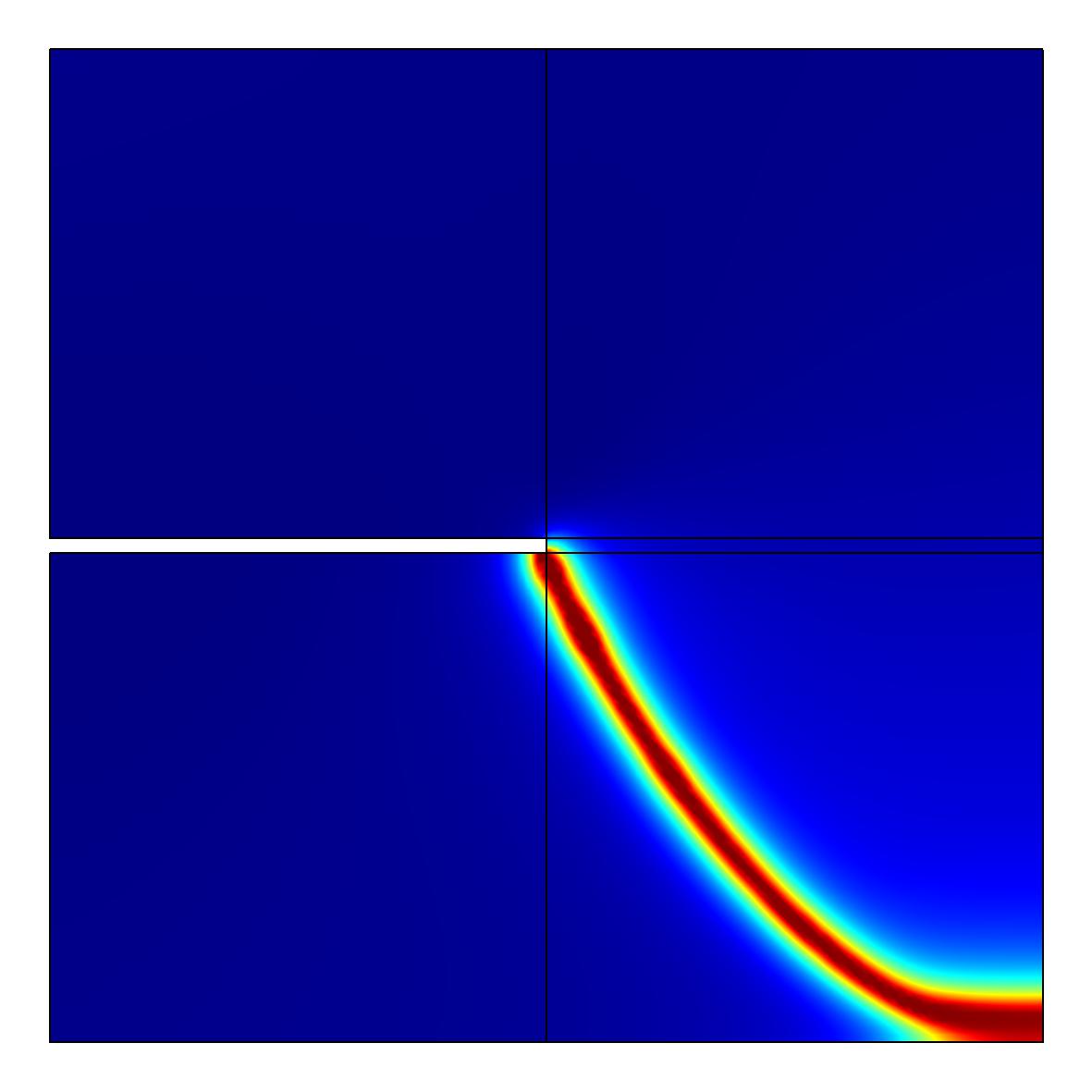}}
		\subfigure[\citet{amor2009regularized}]{\includegraphics[width = 5cm]{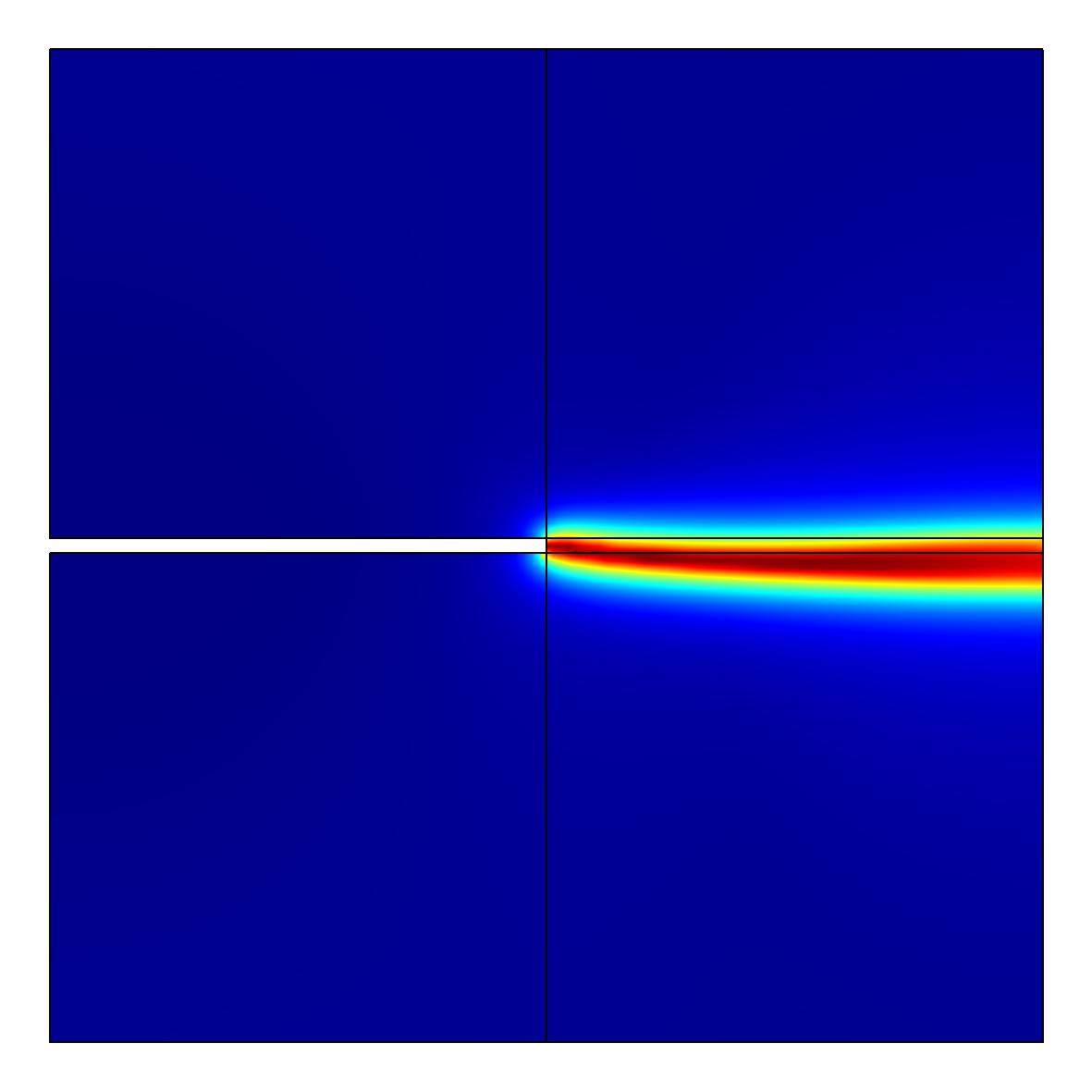}}\\
		\caption{Fracture patterns of a single-edge-notched square plate subjected to shear: different PFMs}
		\label{Fracture patterns of a single-edge-notched square plate subjected to shear different PFms}
	\end{figure}

\begin{figure}[htbp]
		\centering
		\includegraphics[width = 10cm]{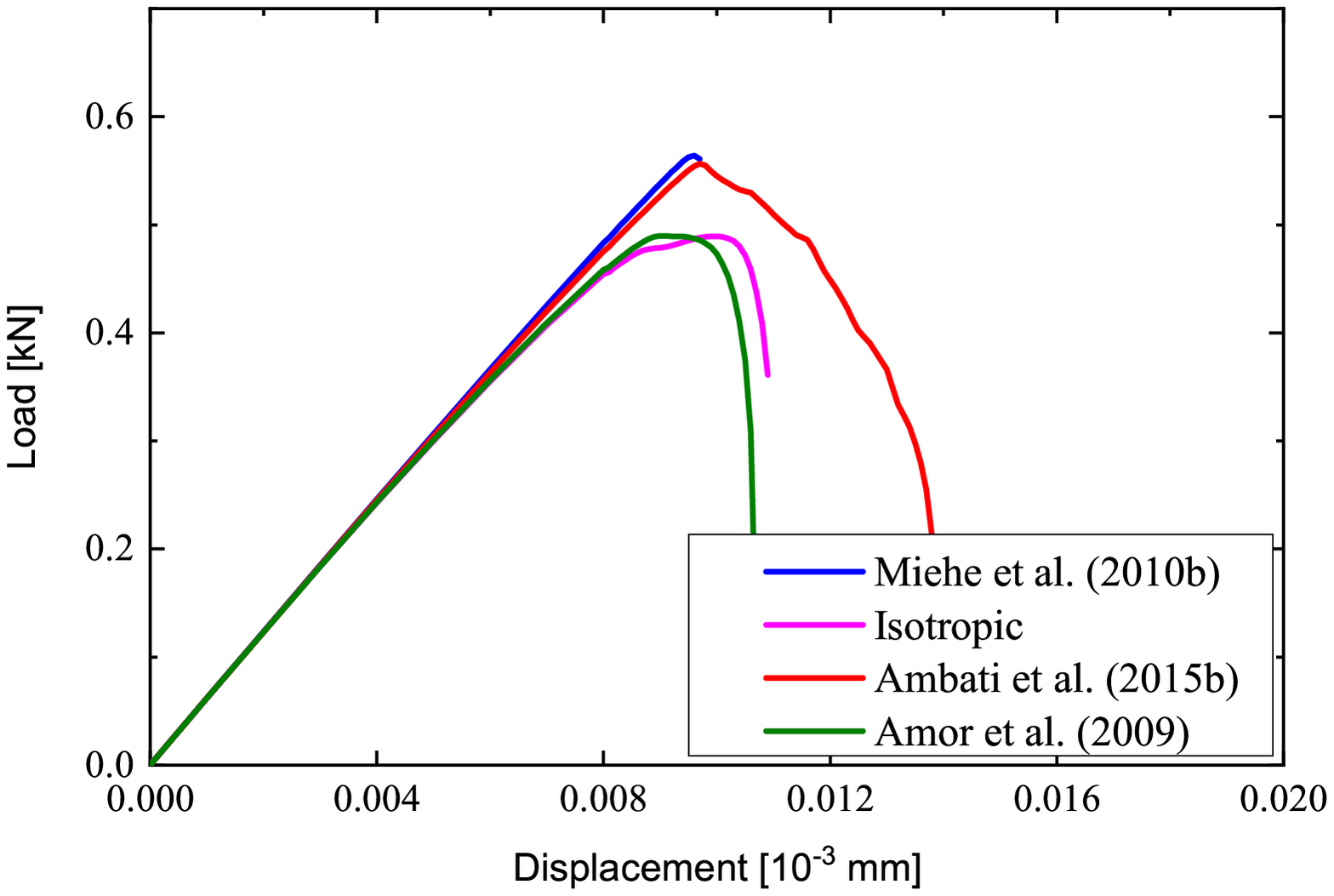}
		\caption{Load-displacement curve of a single-edge-notched square plate subjected to shear for different PFMs}
		\label{Load-displacement curve of a single-edge-notched square plate subjected to shear for different PFMs}
	\end{figure}

The performance of the models of \citet{miehe2010thermodynamically}, \citet{ambati2015review}, \citet{amor2009regularized} and the isotropic model on the shear example is also tested. Note that the same parameters in the tension example are used. Figure \ref{Fracture patterns of a single-edge-notched square plate subjected to shear different PFms} shows the final fracture patterns obtained by different phase field models. \citet{miehe2010thermodynamically} and \citet{ambati2015review} obtain a similar inclined shear fracture while the isotropic model and \citet{amor2009regularized} achieve a horizontal pure mode II fracture. The load-displacement curves for shear are shown in Fig. \ref{Load-displacement curve of a single-edge-notched square plate subjected to shear for different PFMs}. Compared with the isotropic model and \citet{amor2009regularized}, the plate can sustain to a larger shear load when the models of \citet{miehe2010thermodynamically} and \citet{ambati2015review} are applied.

\paragraph {3. 2D notched square plate subjected to tension and shear}

\begin{figure}[htbp]
	\centering
	\subfigure[Geometry and boundary conditions]{\includegraphics[height = 5cm]{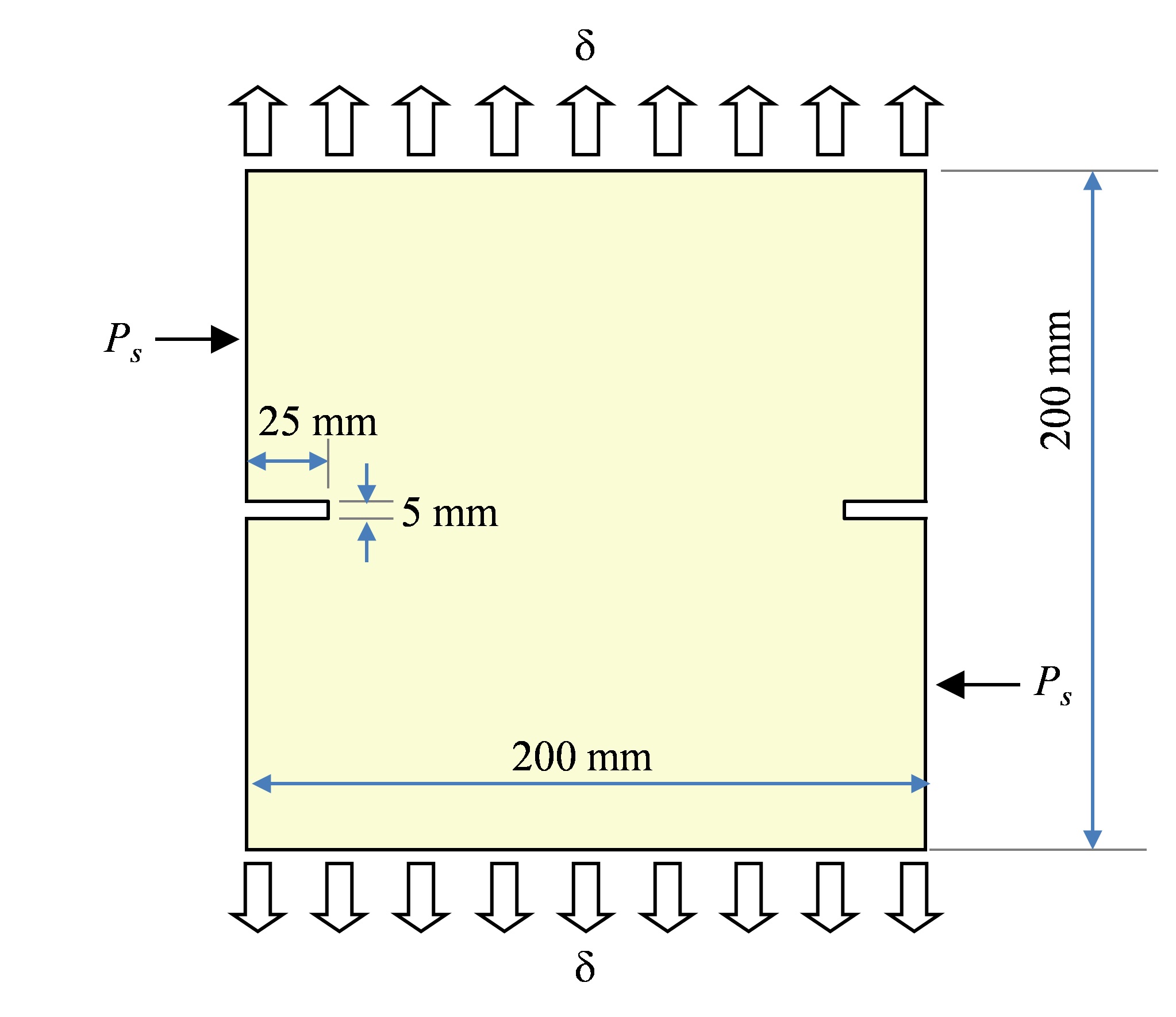}}
	\subfigure[$G_c= 25$ J/m$^2$]{\includegraphics[width = 5cm]{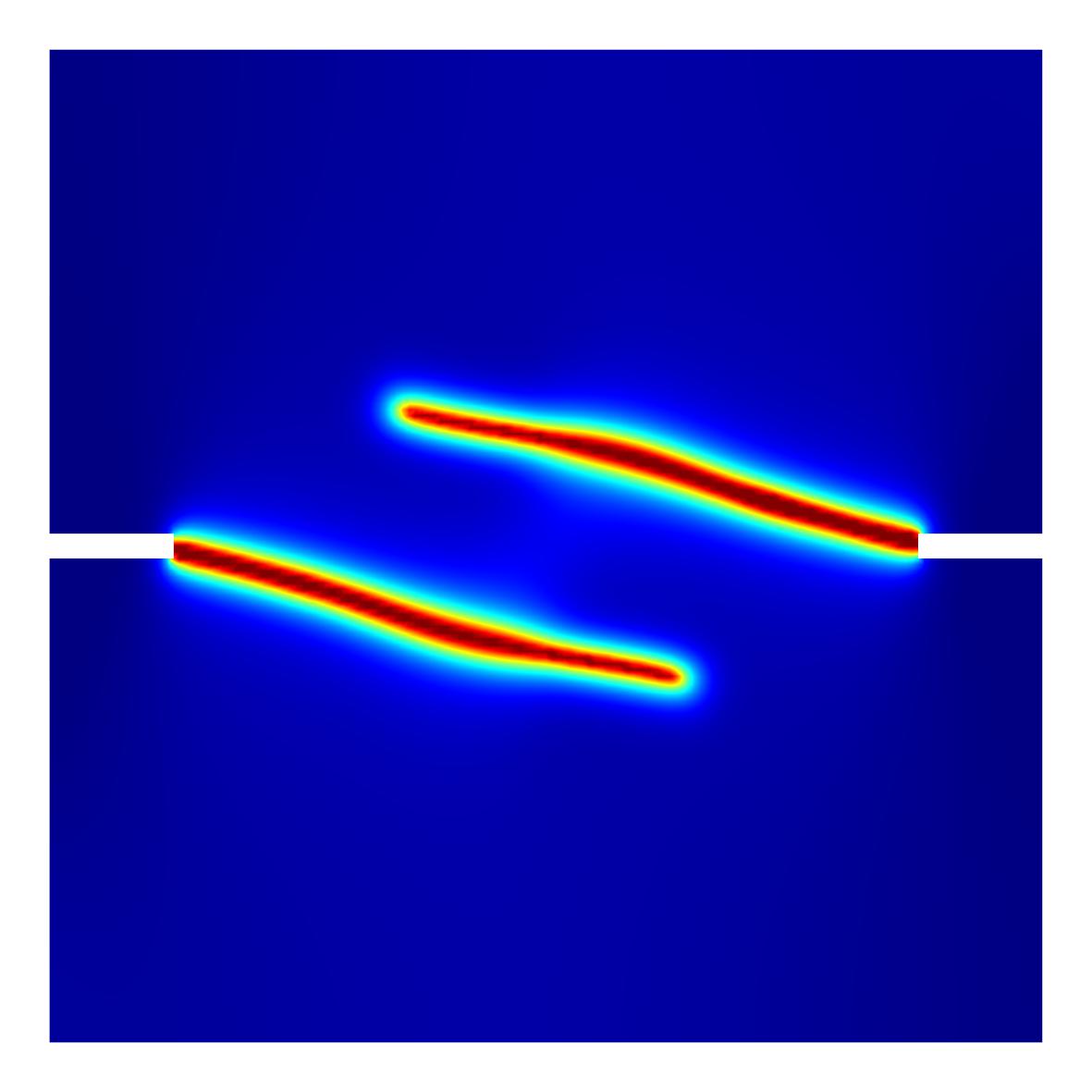}}
	\subfigure[$G_c= 50$ J/m$^2$]{\includegraphics[width = 5cm]{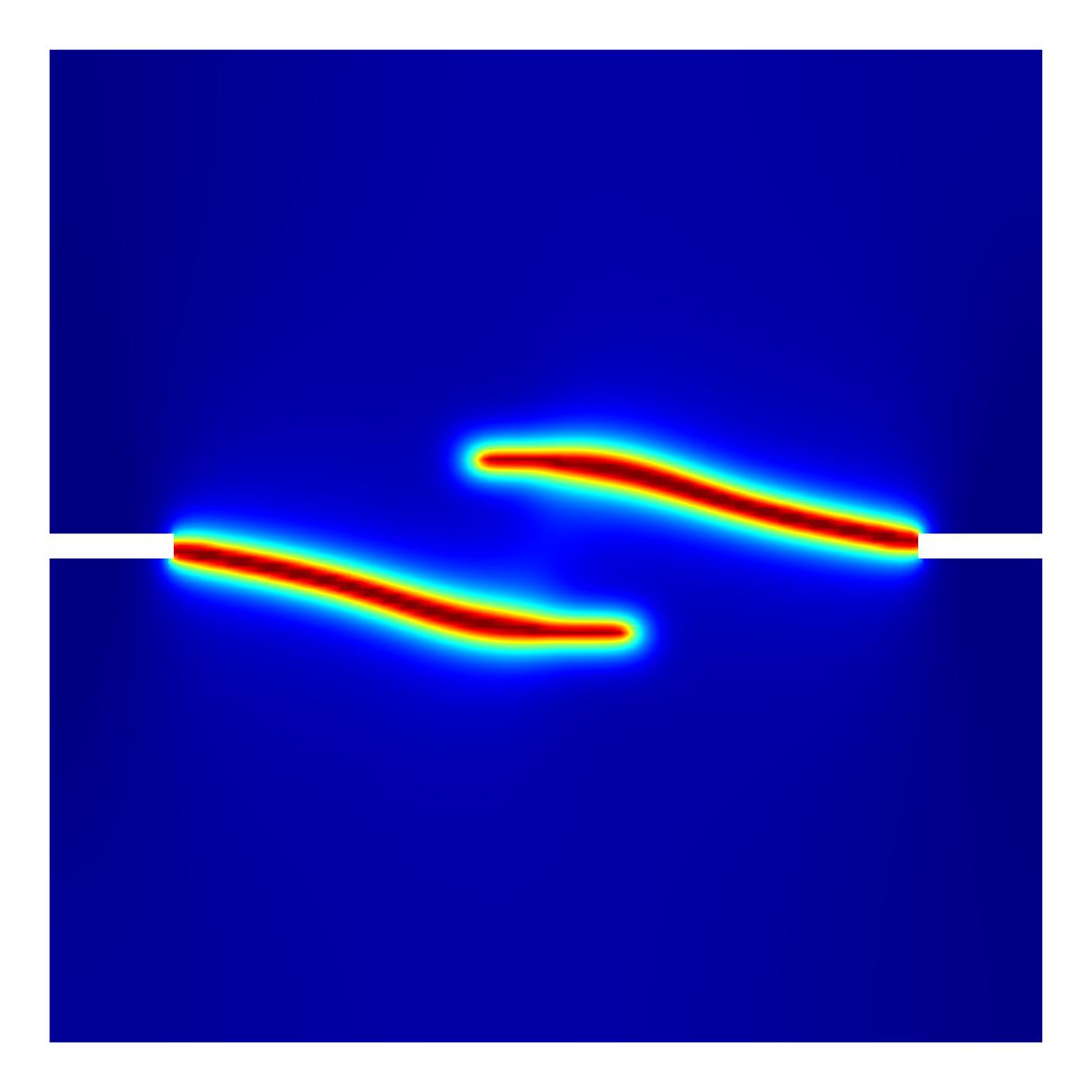}}\\
	\subfigure[$G_c= 75$ J/m$^2$]{\includegraphics[width = 5cm]{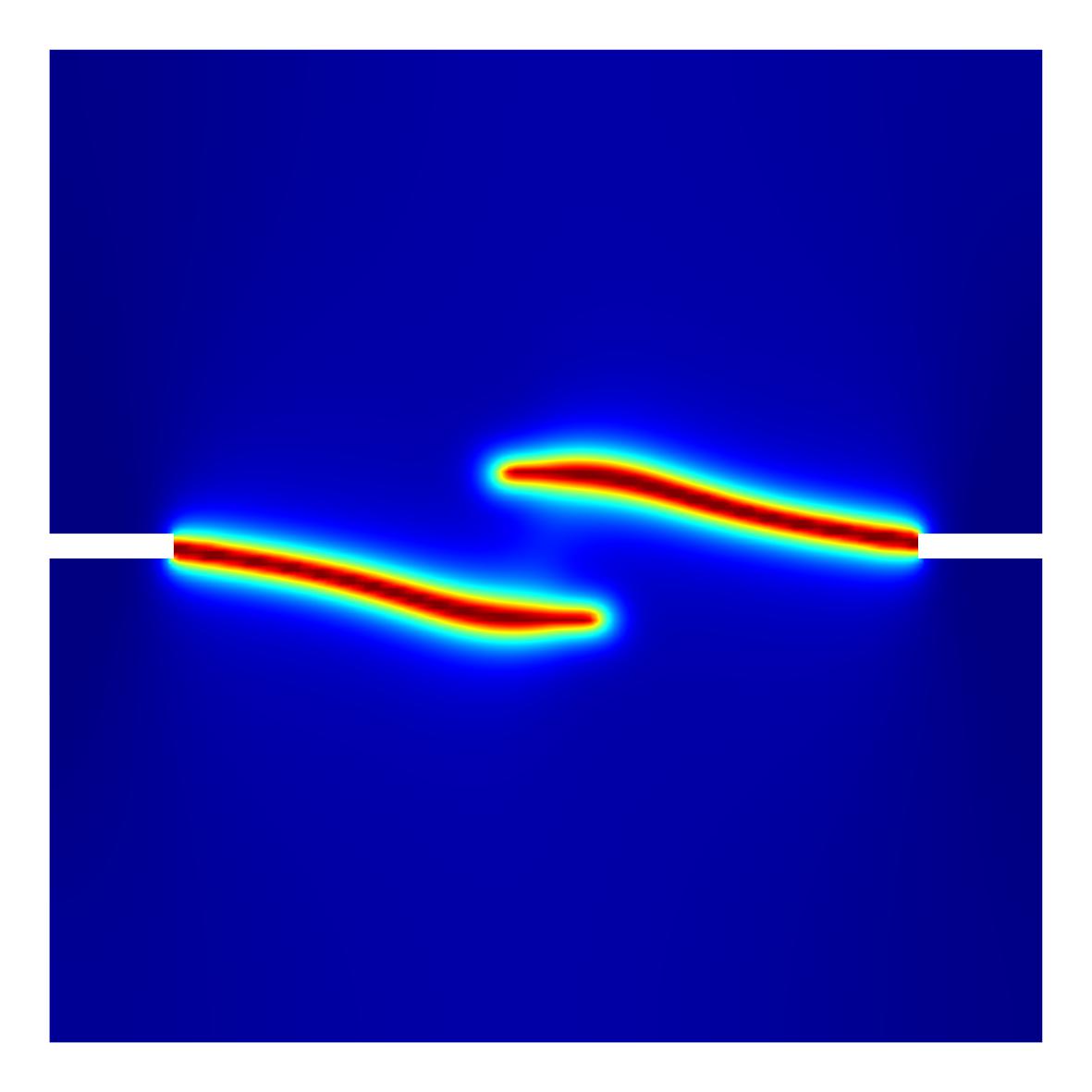}}
	\subfigure[$G_c= 100$ J/m$^2$]{\includegraphics[width = 5cm]{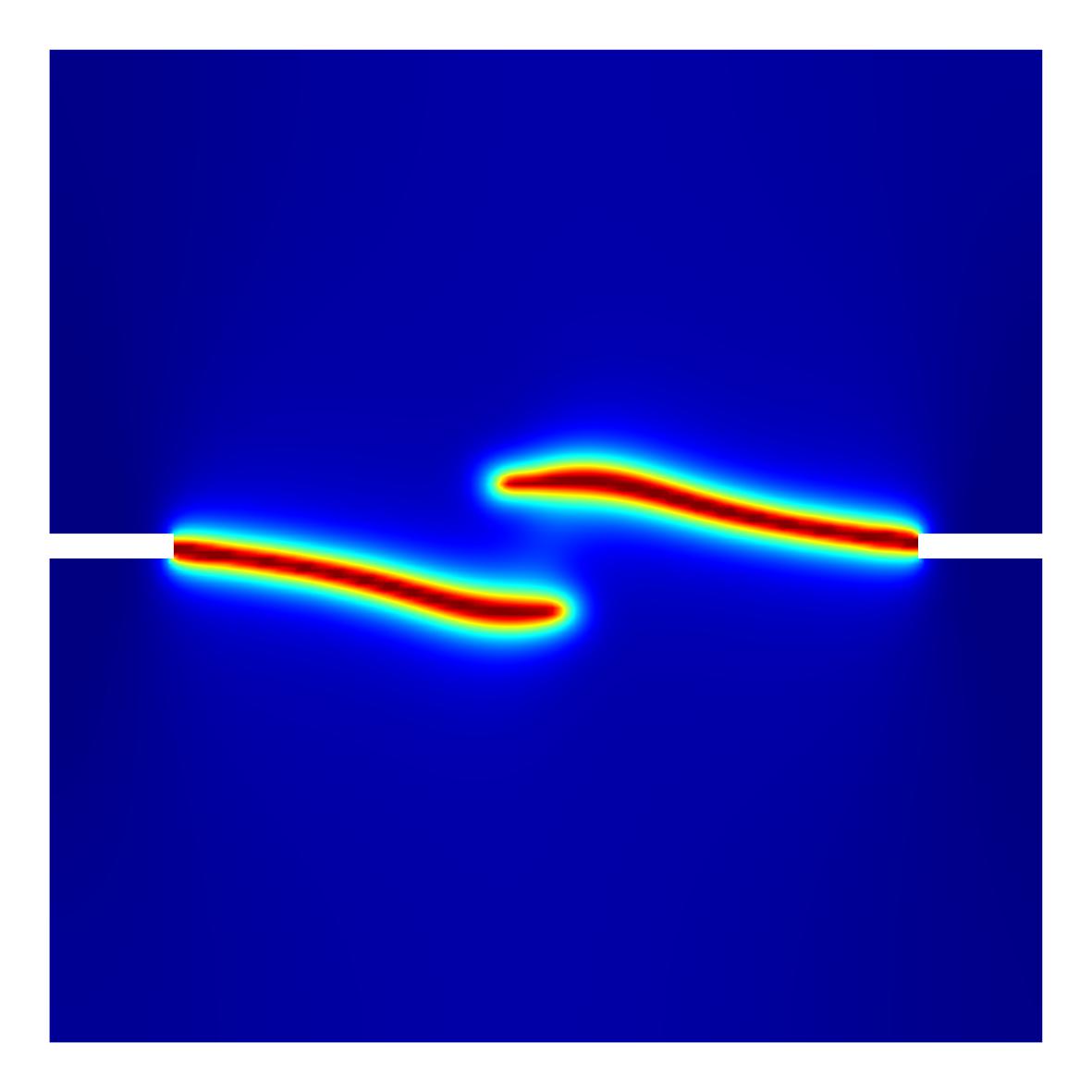}}
	\caption{Fracture patterns of a notched square plate subjected to tension and shear}\citep{zhou2018phase}
	\label{Fracture patterns of a notched square plate subjected to tension and shear}
	\end{figure}

This example shows fractures in a double-edge-notched plate subjected to both tension and shear loading. Figure \ref{Fracture patterns of a notched square plate subjected to tension and shear}a shows the geometry and boundary conditions. The length, height and thickness of the plate are 200, 200 and 50 mm, respectively. Two horizontal notches are 25 mm $\times$ 5 mm. More details can be referred to \citet{zhou2018phase}. Figure \ref{Fracture patterns of a notched square plate subjected to tension and shear}b to \ref{Fracture patterns of a notched square plate subjected to tension and shear}e shows the fracture patterns under different critical energy release rates, i.e. $G_c =$ 25, 50, 75 and 100 J/m$^2$, respectively.

\paragraph {4. 2D notched semi-circular bend (NSCB) test}

\begin{figure}[htbp]
	\centering
	\subfigure[Geometry and boundary conditions]{\includegraphics[height = 4cm]{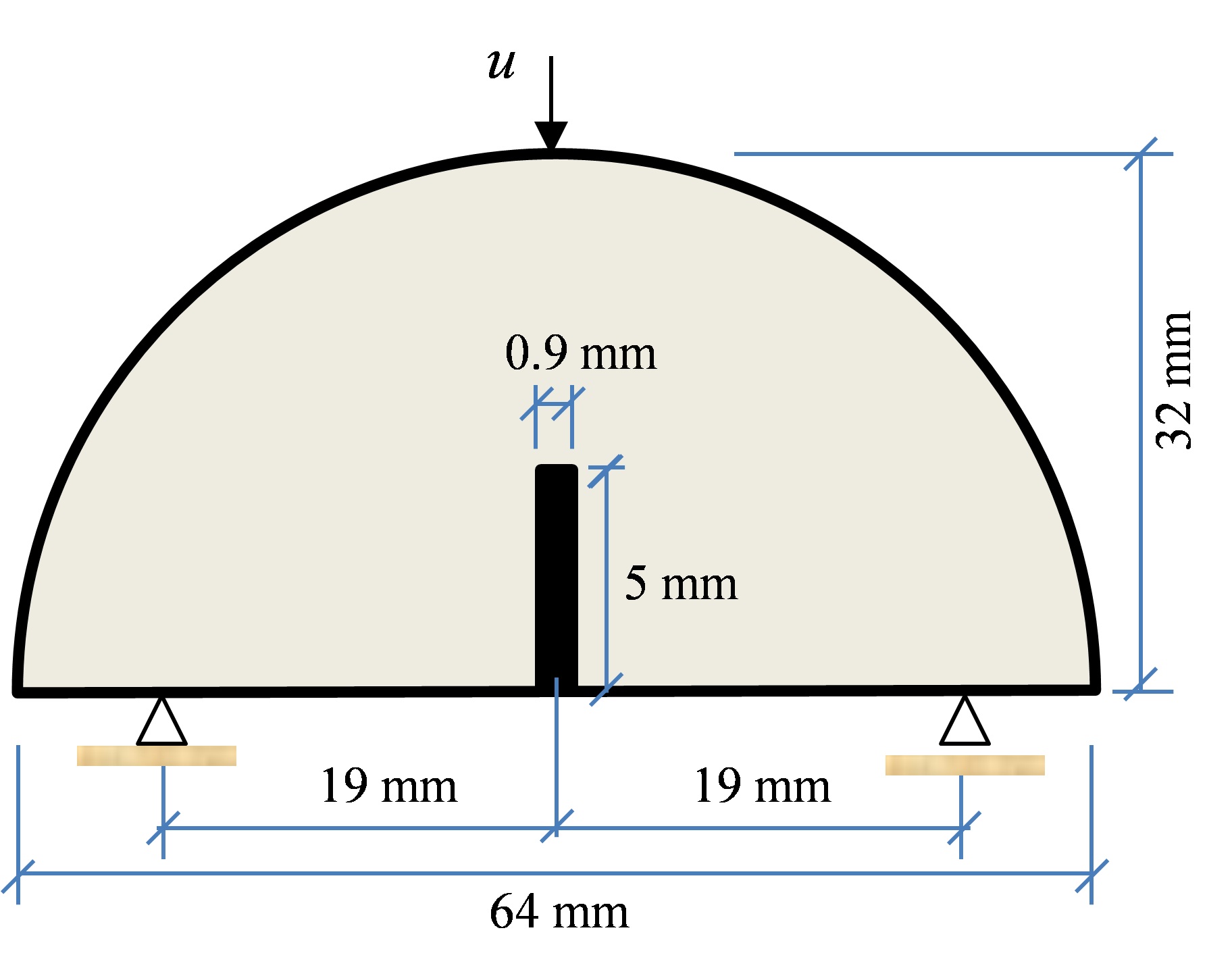}}
	\subfigure[Fracture pattern]{\includegraphics[height = 4cm]{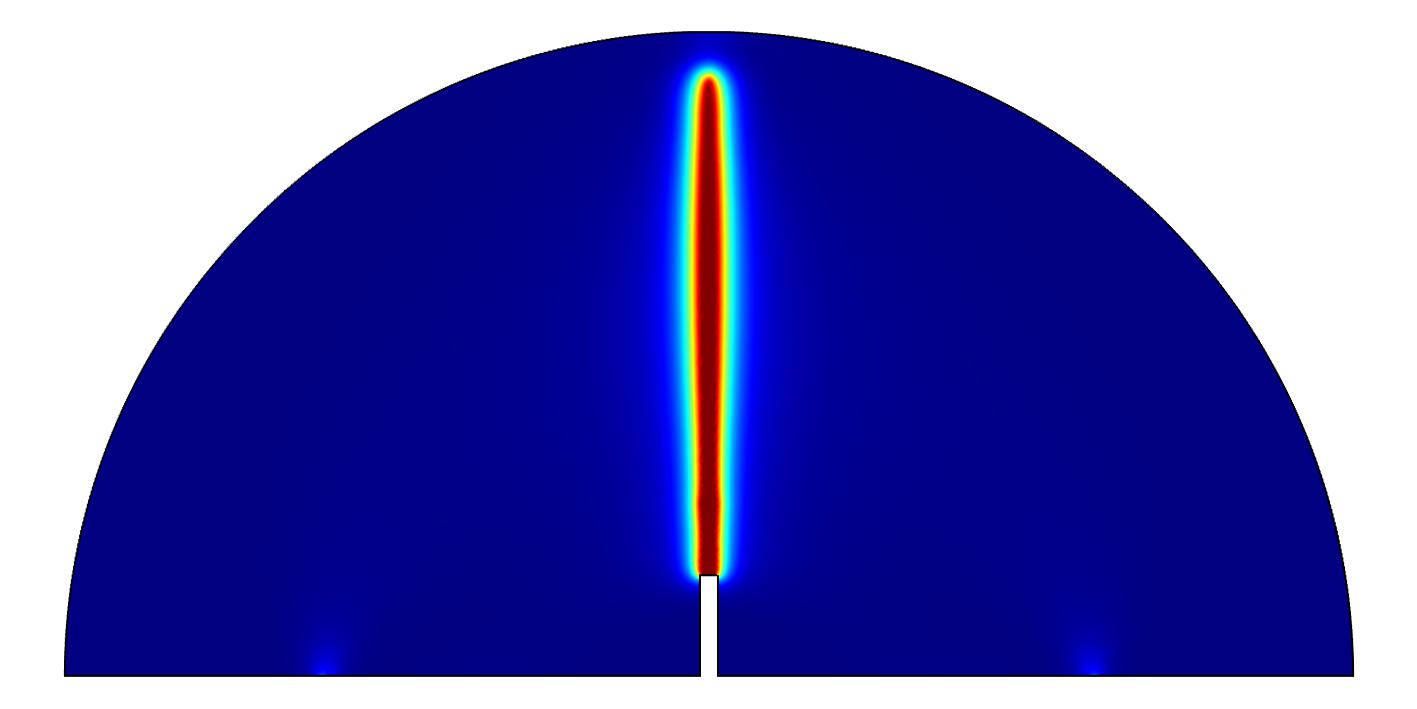}}\\
	\caption{Fracture patterns of a 2D notched semi-circular bend (NSCB) test}\citep{zhou2018phase3}
	\label{Fracture patterns of a 2D notched semi-circular bend (NSCB) test}
	\end{figure}

Fracture patterns in a 2D notched semi-circular bend (NSCB) test are simulated by \citet{zhou2018phase3} and the predicted results are presented in Fig. \ref{Fracture patterns of a 2D notched semi-circular bend (NSCB) test}. The geometry and loading condition are shown in Fig. \ref{Fracture patterns of a 2D notched semi-circular bend (NSCB) test}a. As shown in Fig. \ref{Fracture patterns of a 2D notched semi-circular bend (NSCB) test}b, the fracture initiates from the upper tip of the pre-existing notch and propagates along the vertical direction. This observation is in good agreement with those in experimental tests \citep{gao2015application}.

\paragraph {5. Propagation of multiple echelon flaws}

\begin{figure}[htbp]
	\centering
	\subfigure[Geometry and boundary conditions]{\includegraphics[height = 5cm]{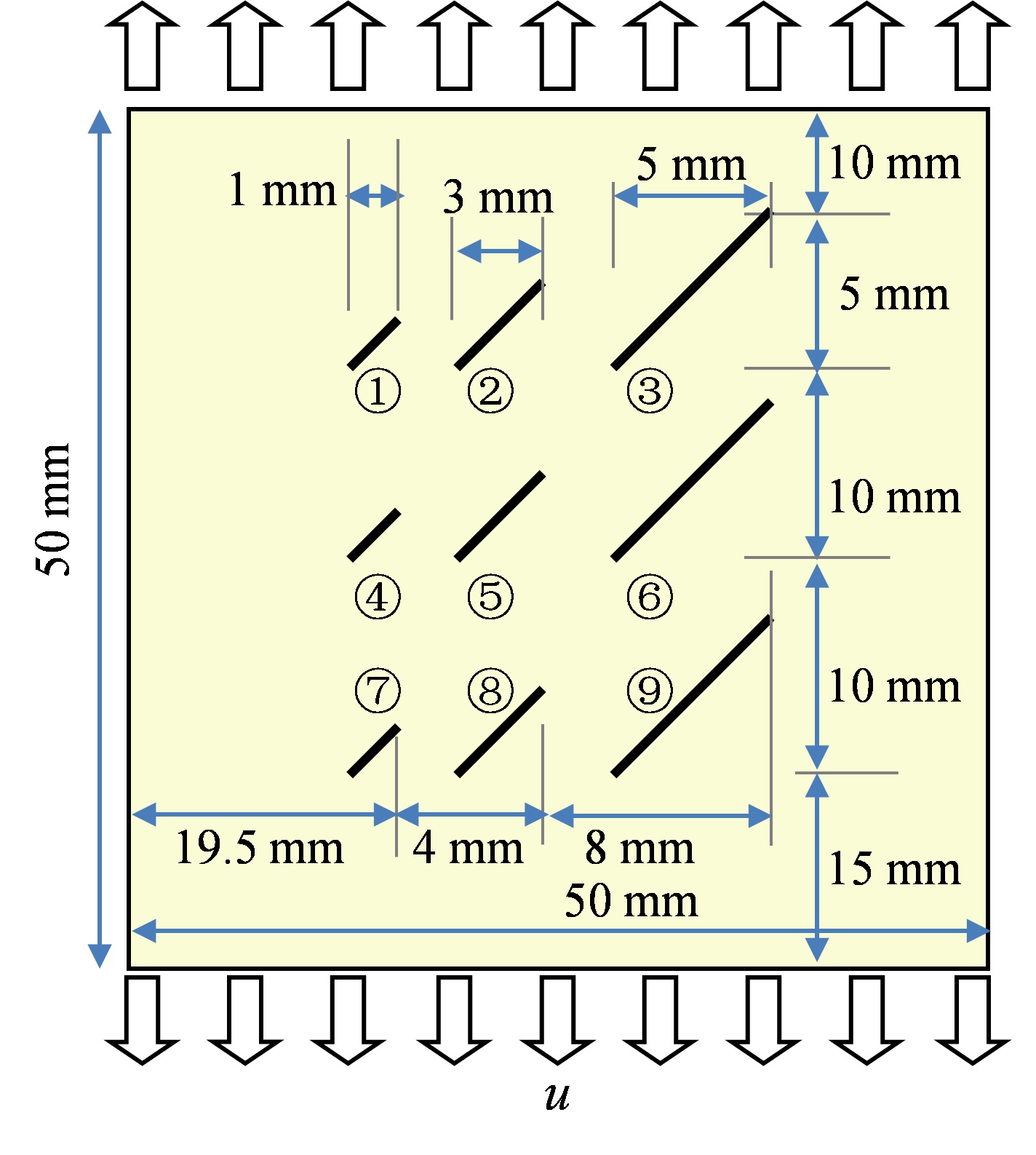}}
	\subfigure[Fracture pattern]{\includegraphics[width = 5cm]{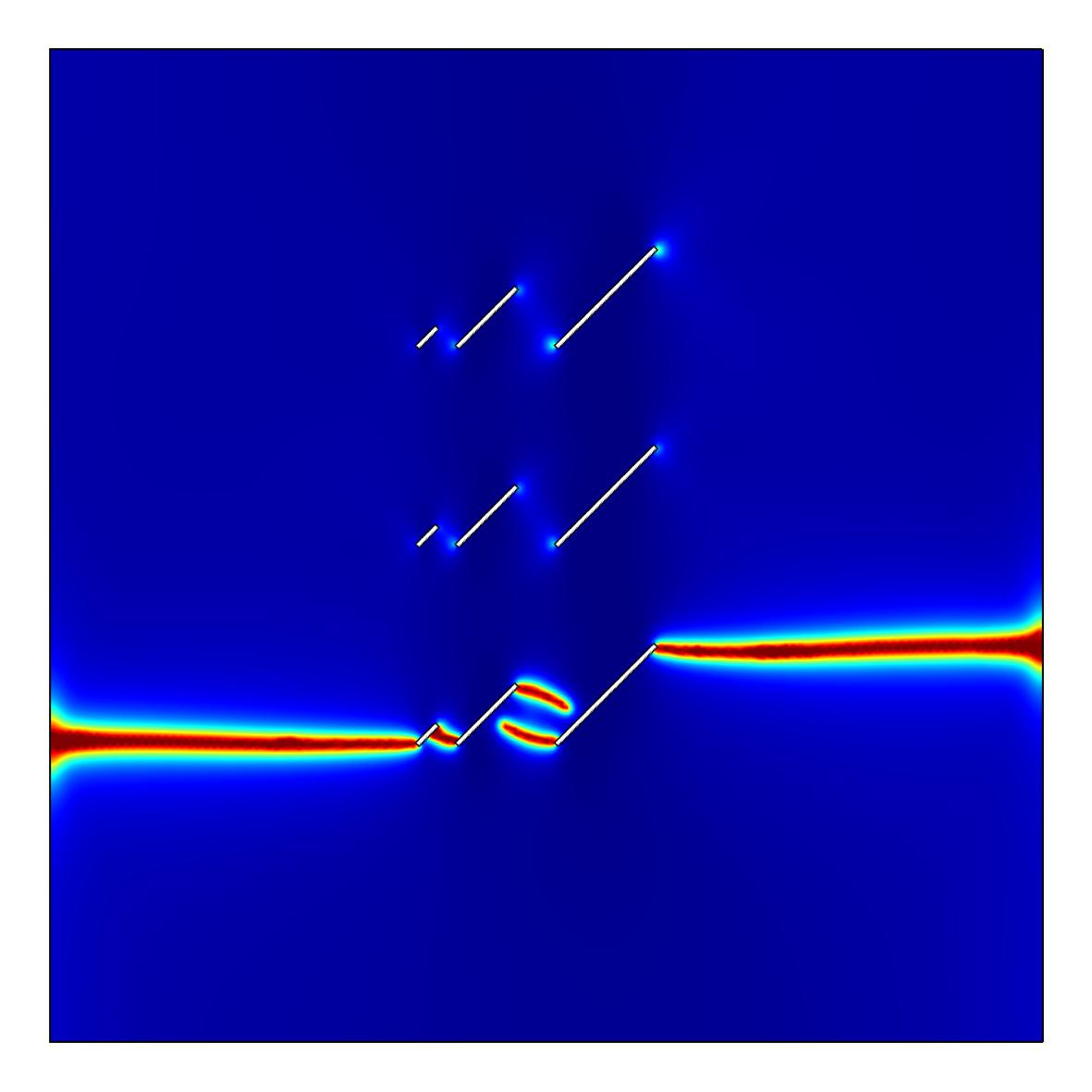}}\\
	\caption{Fracture patterns of a square plate with nine echelon flaws subjected to tension}\citep{zhou2018phase3}
	\label{Fracture patterns of a square plate with nine echelon flaws subjected to tension}
	\end{figure}

This example shows a square plate with nine echelon flaws subjected to tension. The geometry and boundary conditions are shown in Fig. \ref{Fracture patterns of a square plate with nine echelon flaws subjected to tension}a. As observed, these flaws have the same inclination angle of $45 ^\circ$ and a varying length and spacing. The final fracture pattern is shown in Fig. \ref{Fracture patterns of a square plate with nine echelon flaws subjected to tension}b. Fractures from the bottom flaws are dominated due to the stress shielding and amplification effects from flaw interaction. 

\paragraph {6. Propagation and coalescence of twenty parallel flaws}

\begin{figure}[htbp]
	\centering
	\subfigure[Geometry and boundary conditions]{\includegraphics[height = 5cm]{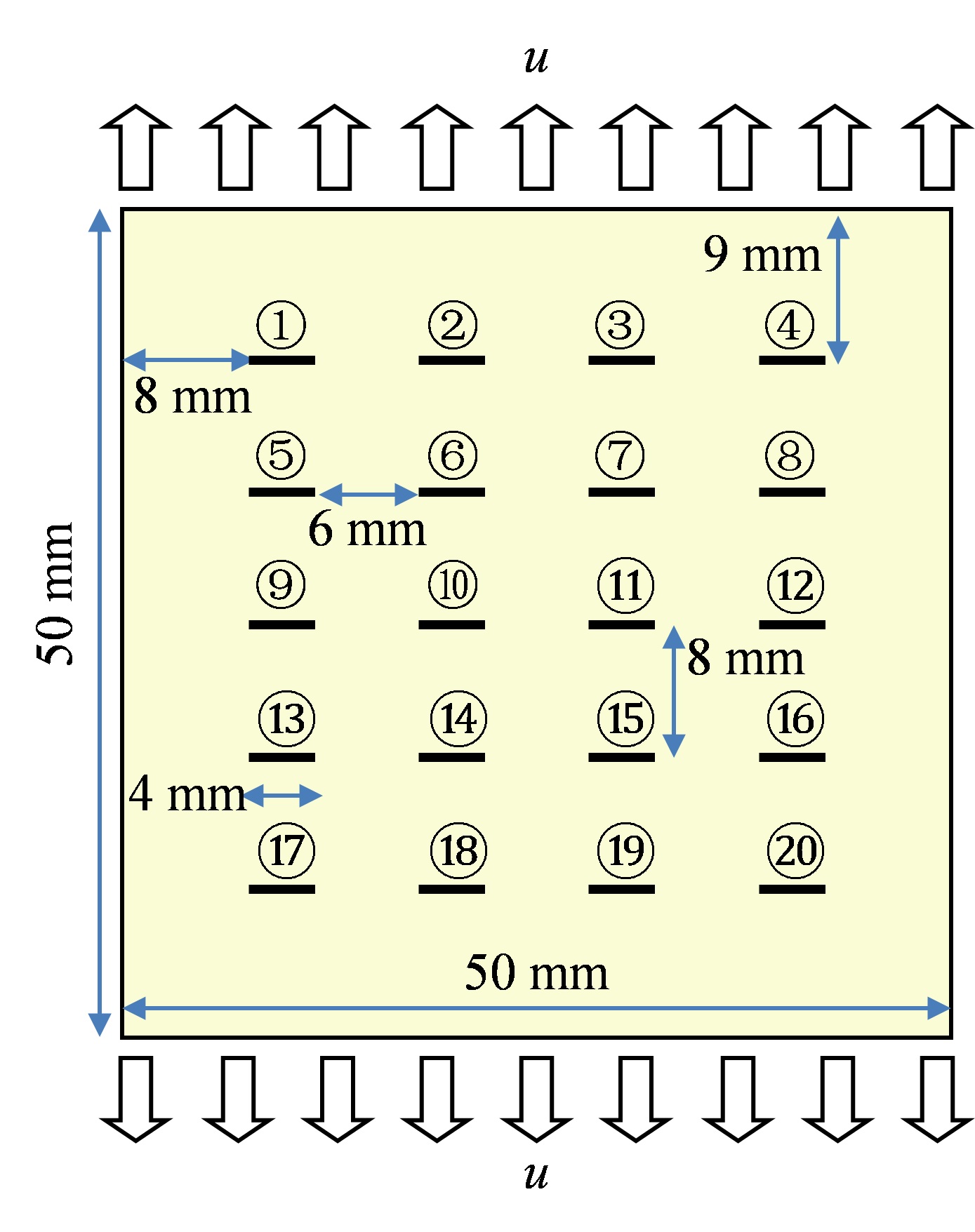}}
	\subfigure[Fracture pattern]{\includegraphics[width = 5cm]{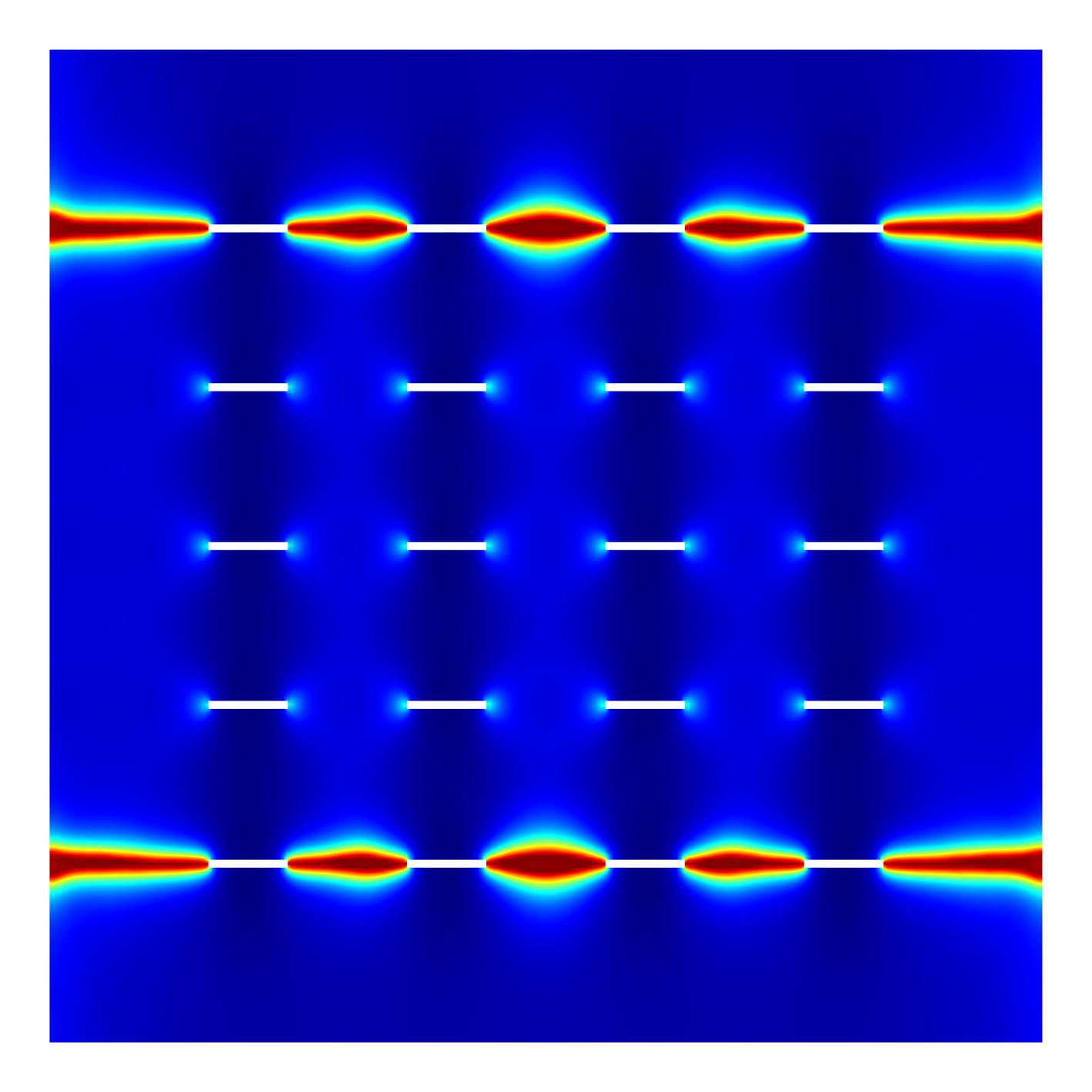}}\\
	\caption{Fracture patterns of a square plate with twenty parallel flaws subjected to tension}\citep{zhou2018phase3}
	\label{Fracture patterns of a square plate with twenty parallel flaws subjected to tension}
	\end{figure}

This example shows a square plate with twenty parallel flaws subjected to tension and the fracture patterns are simulated by \citet{zhou2018phase3} by using the anisotropic phase field model. With the geometry and boundary conditions being shown in Fig. \ref{Fracture patterns of a square plate with twenty parallel flaws subjected to tension}a, symmetric fractures are depicted in Fig. \ref{Fracture patterns of a square plate with twenty parallel flaws subjected to tension}b. In addition, fractures only initiate from the upper and bottom flaws and there are no interior fractures.

\paragraph {7. Propagation and coalescence of three parallel flaws in Brazilian discs}

\begin{figure}[htbp]
	\centering
	\subfigure[$S = 1$ cm]{\includegraphics[width = 5cm]{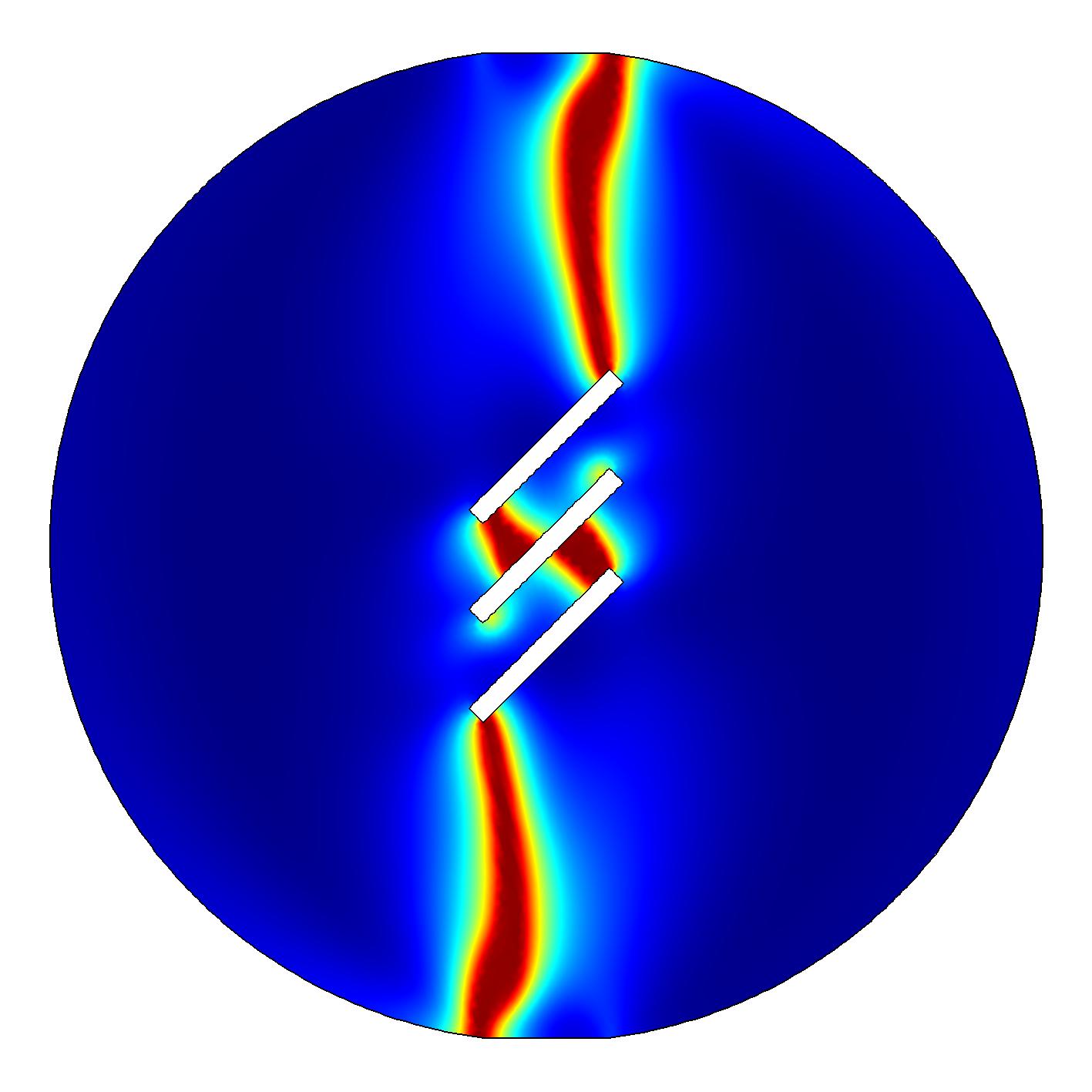}}
	\subfigure[$S = 2$ cm]{\includegraphics[width = 5cm]{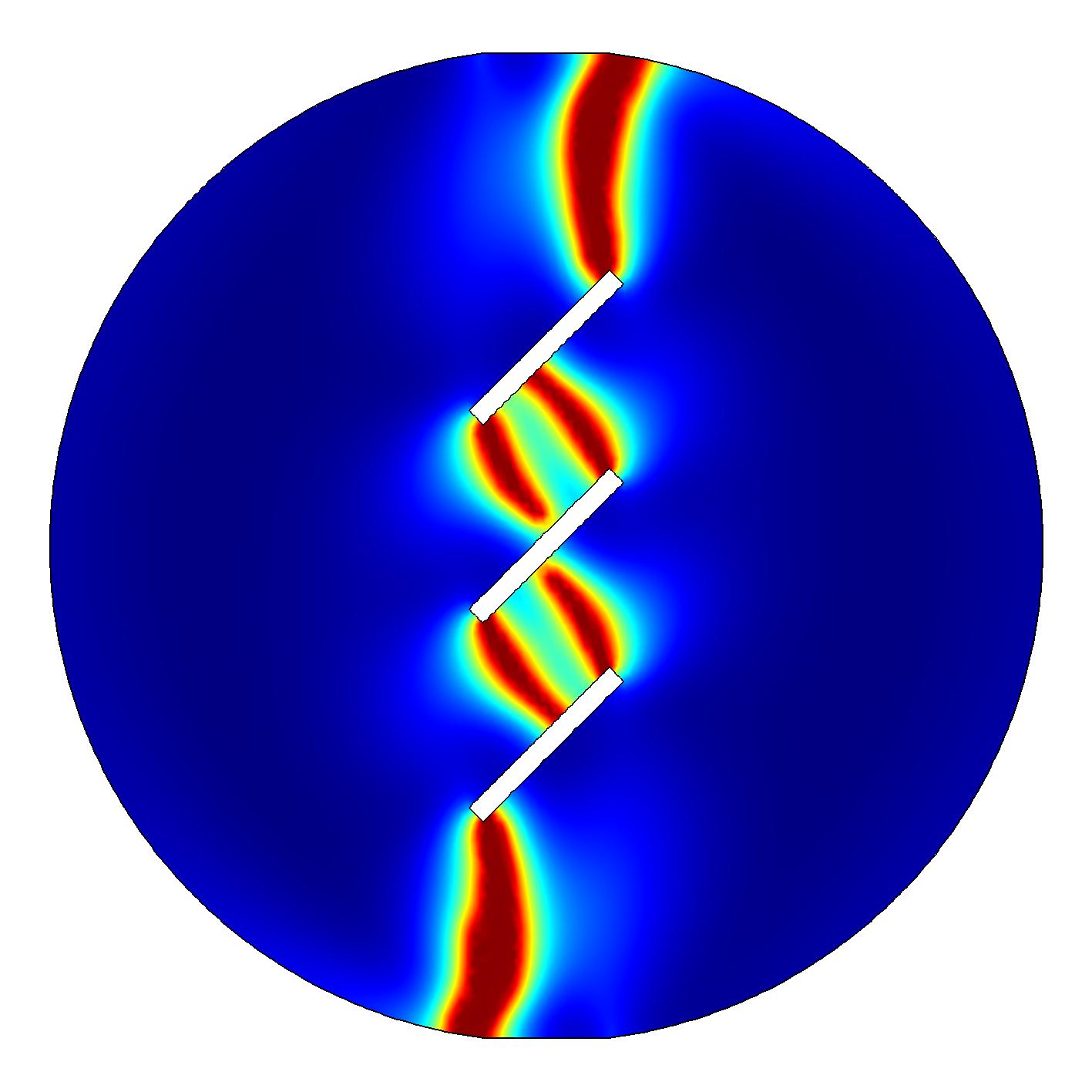}}
	\subfigure[$S = 3$ cm]{\includegraphics[width = 5cm]{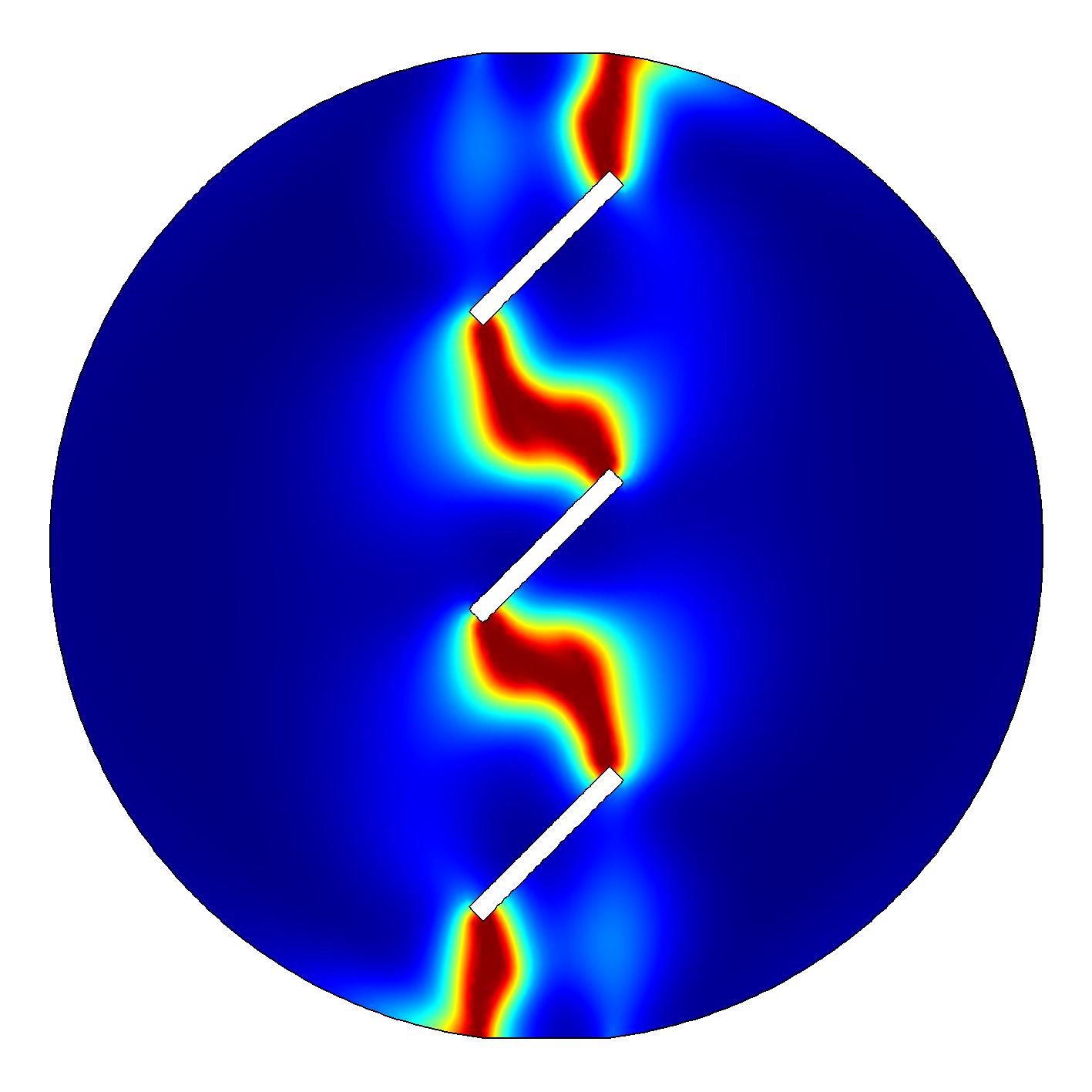}}\\
	\caption{Fracture patterns of a Brazilian disc with three vertical parallel flaws}\citep{zhou2018fracture}
	\label{Fracture patterns of a Brazilian disc with three vertical parallel flaws}
	\end{figure}

In a Brazilian splitting test, a plane cylinder specimen is subjected to diametral compression \citep{zhou2018fracture}. \citet{zhou2018fracture} used the phase field method to investigate the fracture propagation in Brazilian discs with multiple pre-existing notches. The specimen and flaw dimensions can be found in \citet{zhou2018fracture}. Figure \ref{Fracture patterns of a Brazilian disc with three vertical parallel flaws} shows the final crack patterns of the Brazilian discs with three vertically arranged notches under different notch spacing. Outer cracks initiate from the notch tips and propagate towards the disc ends. The crack propagation intersects with the vertical direction at a small angle.  When the spacing $S = 1$ cm, only one inner crack occurs between two adjoining notches. However, for $S = 3$ cm, the two inner cracks from the inner tips of the notches coalesce.

\begin{figure}[htbp]
	\centering
	\subfigure[$S = 1$ cm]{\includegraphics[width = 5cm]{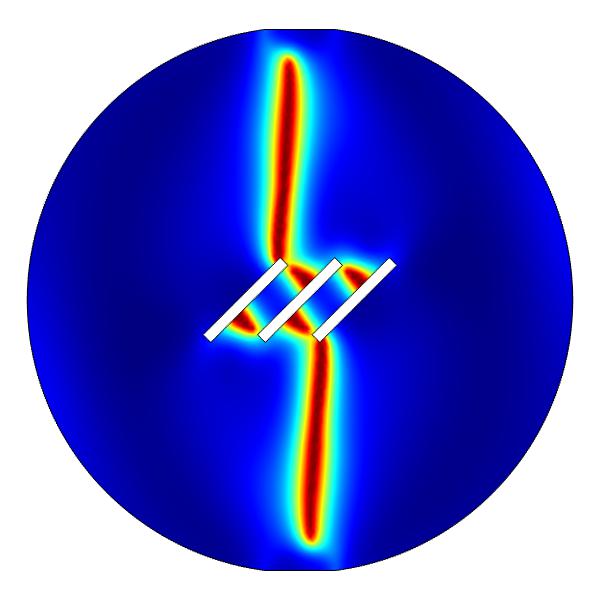}}
	\subfigure[$S = 2$ cm]{\includegraphics[width = 5cm]{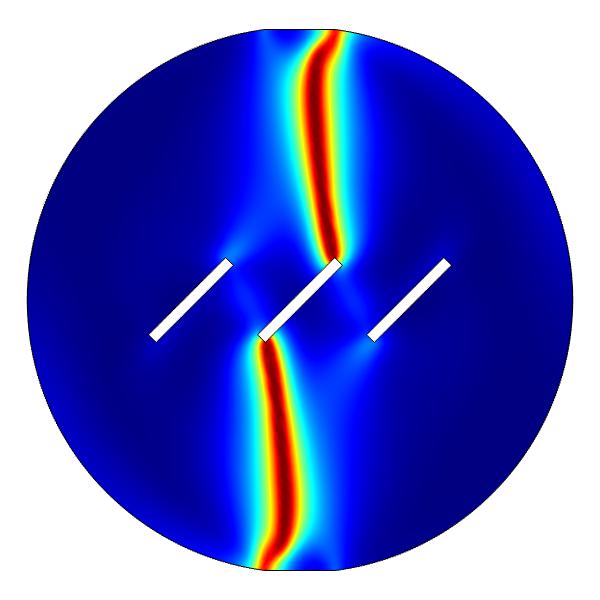}}
	\subfigure[$S = 3$ cm]{\includegraphics[width = 5cm]{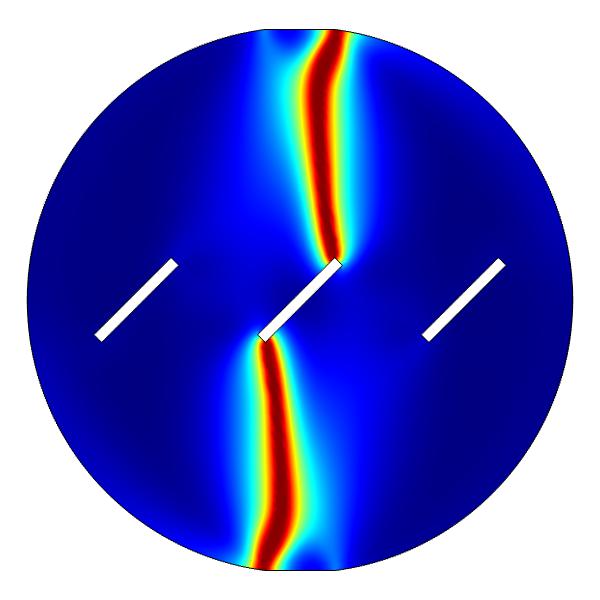}}\\
	\caption{Fracture patterns of a Brazilian disc with three horizontal parallel flaws}\citep{zhou2018fracture}
	\label{Fracture patterns of a Brazilian disc with three horizontal parallel flaws}
	\end{figure}

Figure \ref{Fracture patterns of a Brazilian disc with three horizontal parallel flaws} shows the final crack patterns of the Brazilian discs with three horizontal notches and the influence of different notch spacing. For $S = 2$ cm and $S = 3$ cm, the cracks are similar. The phase field modelling only shows two outer cracks that propagate towards the two ends of the specimen. However, for $S=1$ cm, additional inner cracks initiate from the notch tips and evolves between two adjoining notches.

\paragraph {8. Compressive-shear fracture}

\begin{figure}[htbp]
	\centering
	\includegraphics[width = 12cm]{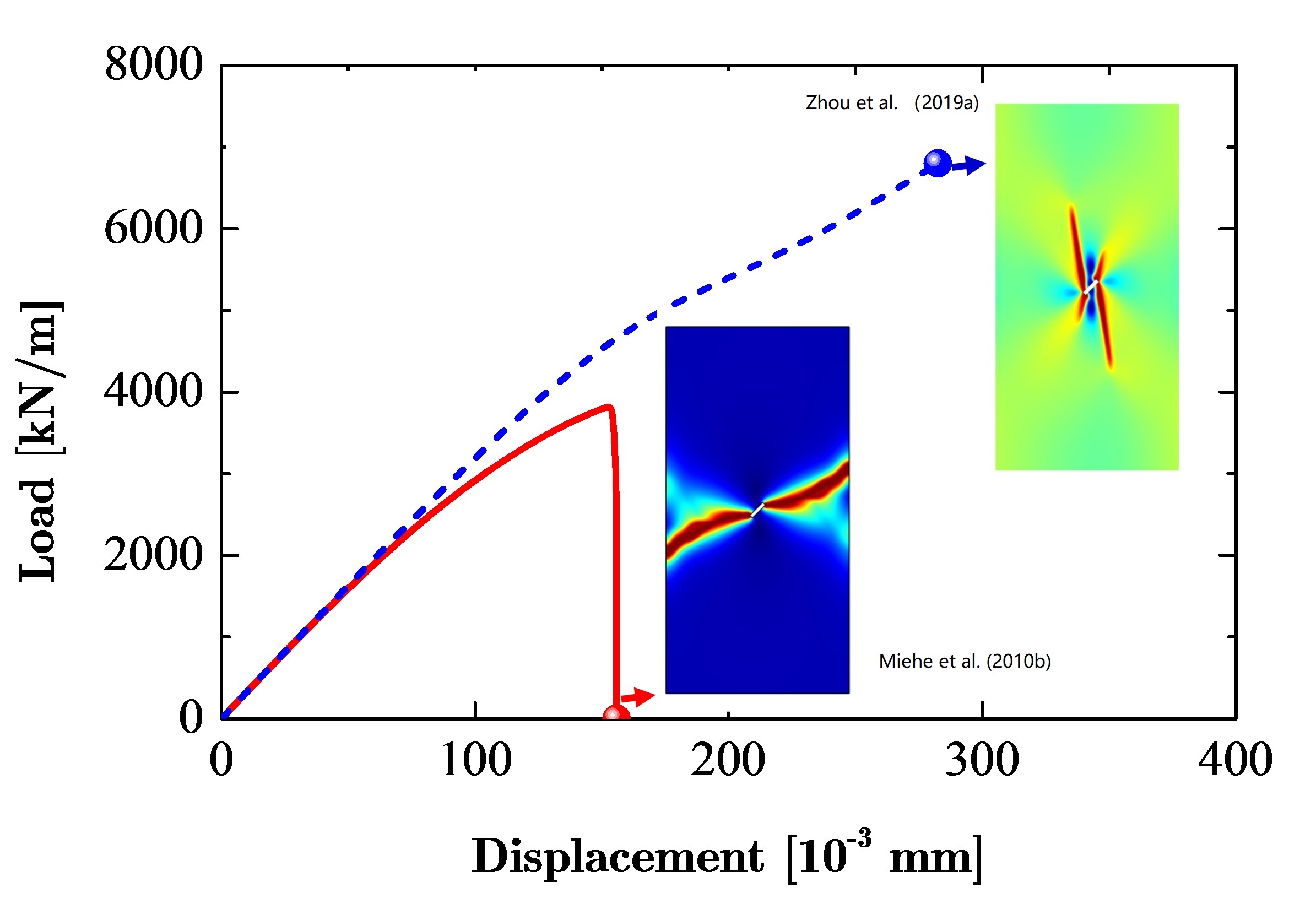}
	\caption{Comparison of fracture pattern and load-displacement curve}
	\label{Comparison of fracture pattern and load-displacement curve obtained by the proposed PFM and the anisotropic model of}
\end{figure}

Rock is a typical geological material and compressive-shear fractures can be formed in rocks during loading. \citet{zhou2019phase2} proposed a phase field model for simulating compressive-shear fracture. Here, we compare the model of \citet{zhou2019phase2} and the anisotropic model of \citet{miehe2010thermodynamically}. The fracture pattern and load-displacement curves are shown in \ref{Comparison of fracture pattern and load-displacement curve obtained by the proposed PFM and the anisotropic model of} under the same elastic and fracture parameters. In Fig. \ref{Comparison of fracture pattern and load-displacement curve obtained by the proposed PFM and the anisotropic model of}, the anisotropic model of \citep{miehe2010thermodynamically} does not have a drop stage in the load-displacement curve while the model of \citet{zhou2019phase2} has an obvious drop stage. Moreover, only wing and anti-wing tensile cracks are simulated in the model of \citep{miehe2010thermodynamically} while the model of \citet{zhou2019phase2} predicts well the compressive-shear fracture in rocks.

	\subsubsection{Dynamic fracture}
\paragraph {1. 2D dynamic shear loading of Kalthoff experiment}

\begin{figure}[htbp]
	\centering
	\subfigure[Geometry and boundary conditions]{\includegraphics[height = 4cm]{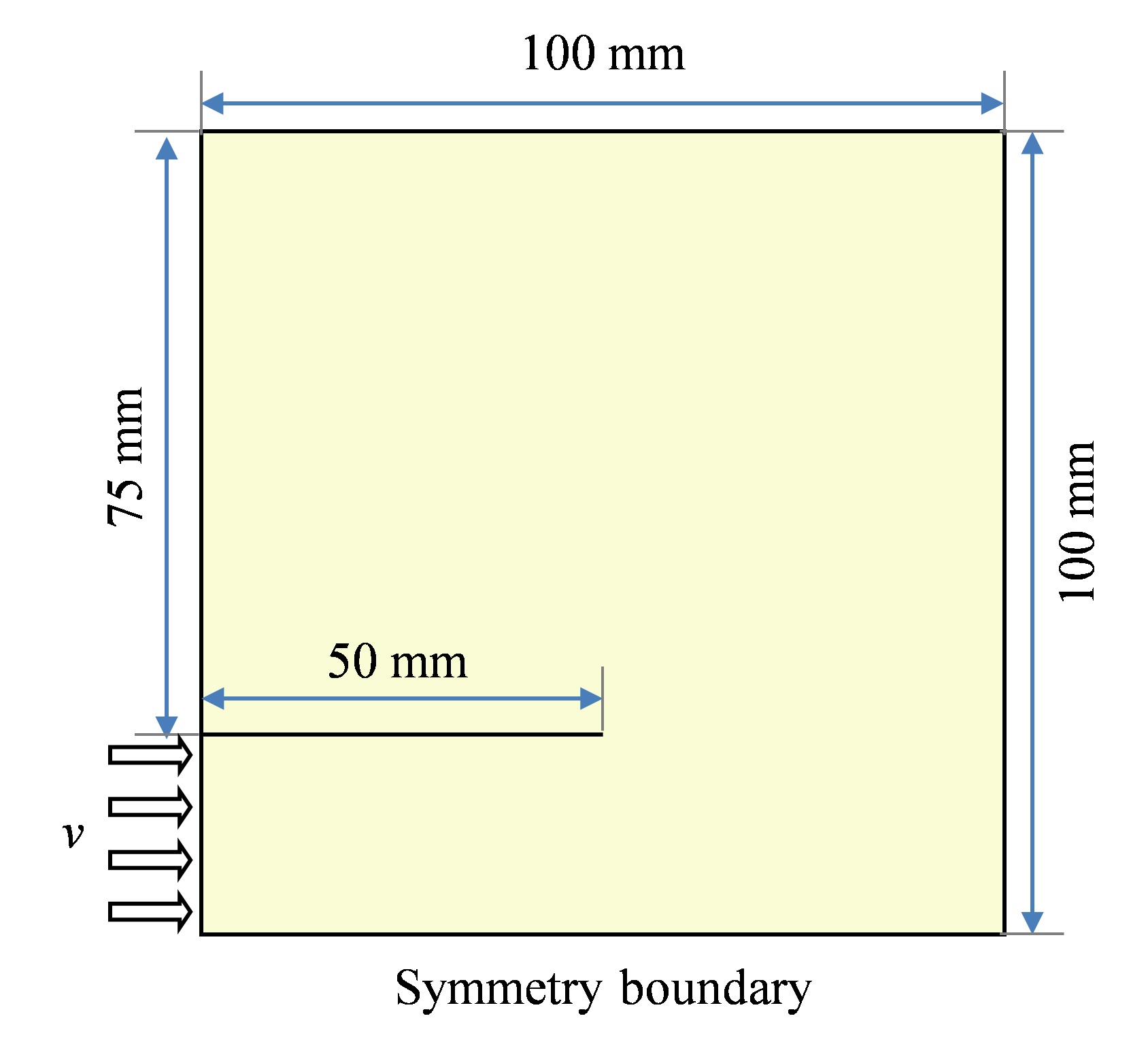}}
	\subfigure[$G_c=5\times10^3$ J/m$^2$]{\includegraphics[width = 4cm]{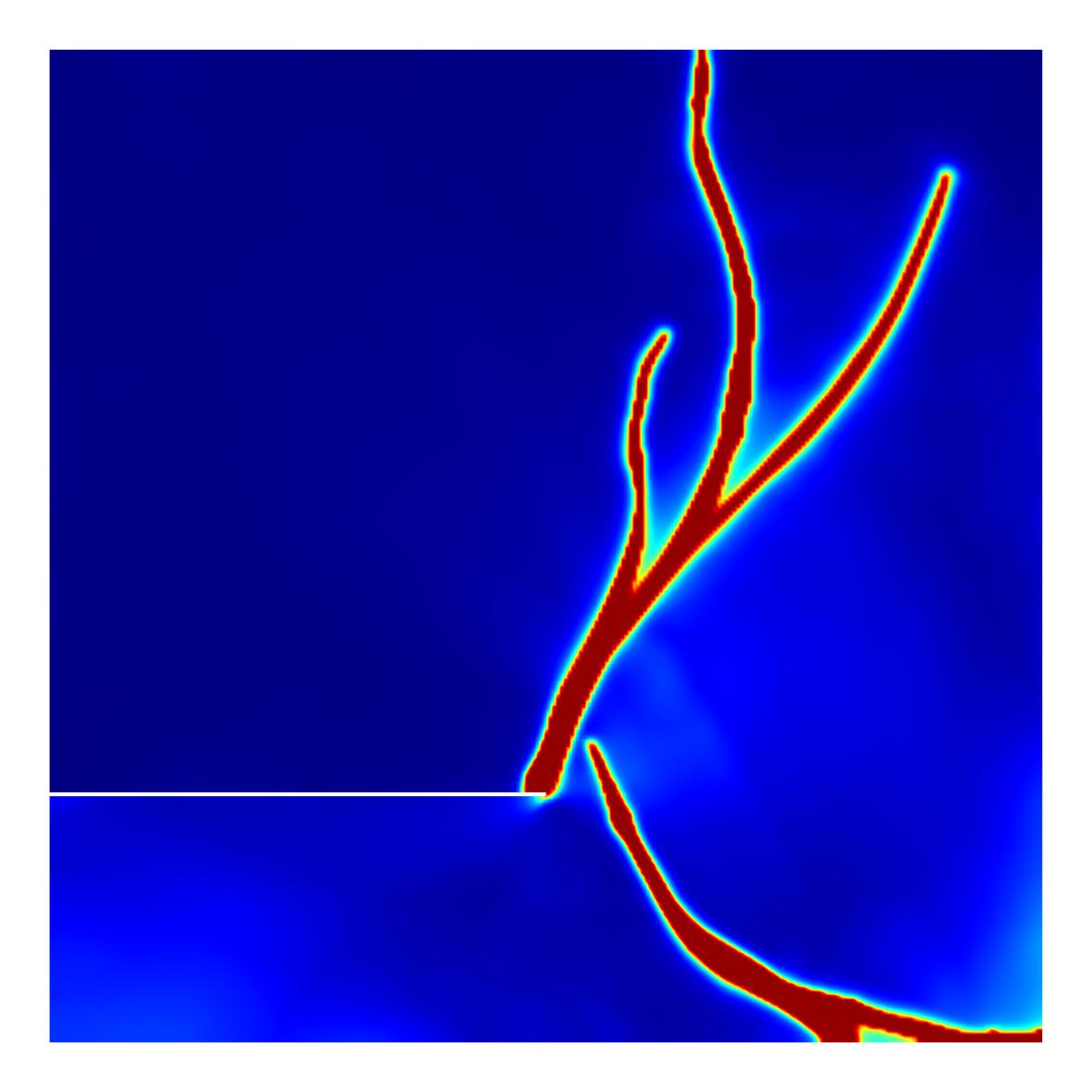}}
	\subfigure[$G_c=1\times10^4$ J/m$^2$]{\includegraphics[width = 4cm]{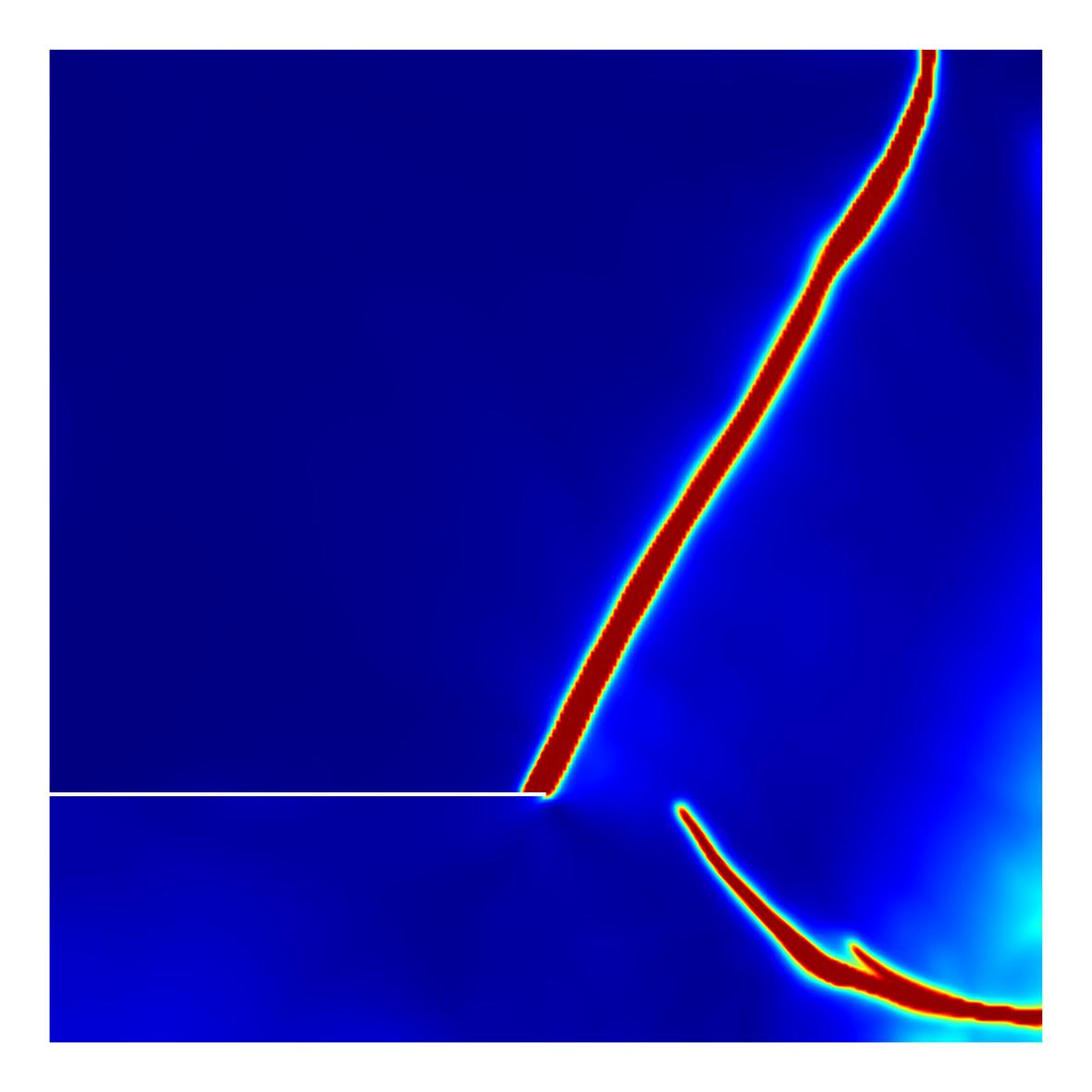}}\\
	\subfigure[$G_c=2.213\times10^4$ J/m$^2$]{\includegraphics[width = 4cm]{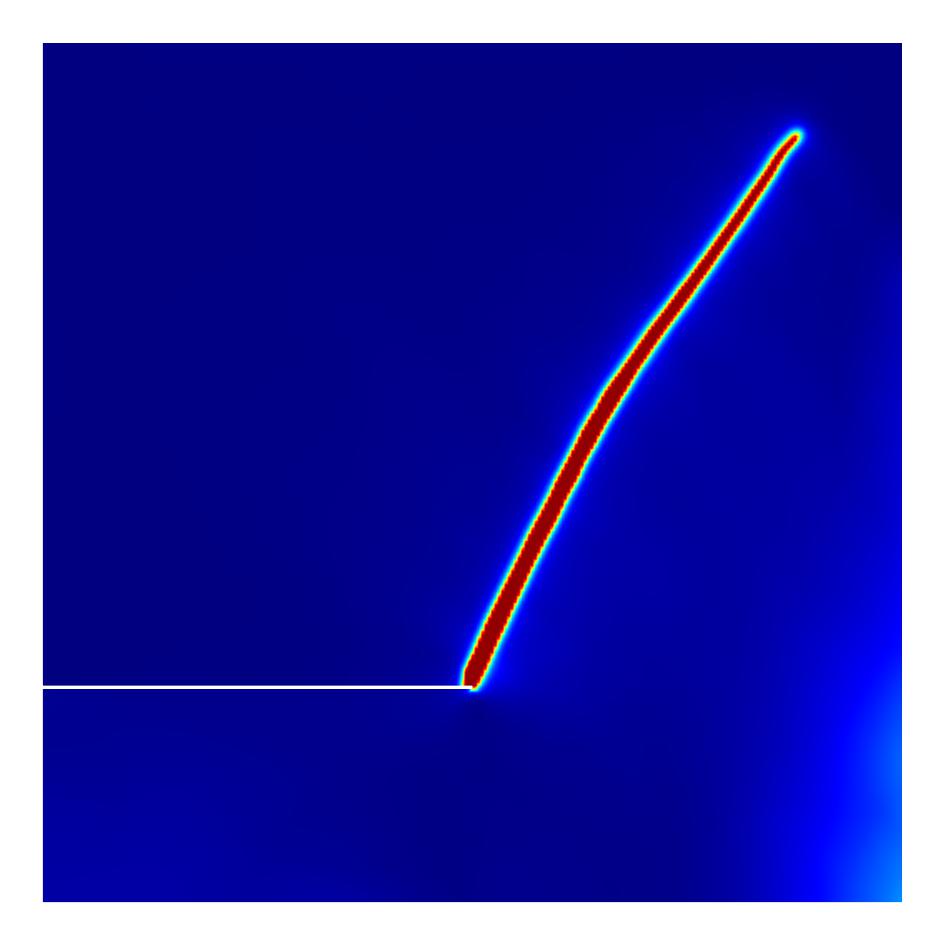}}
	\subfigure[$G_c=3\times10^4$ J/m$^2$]{\includegraphics[width = 4cm]{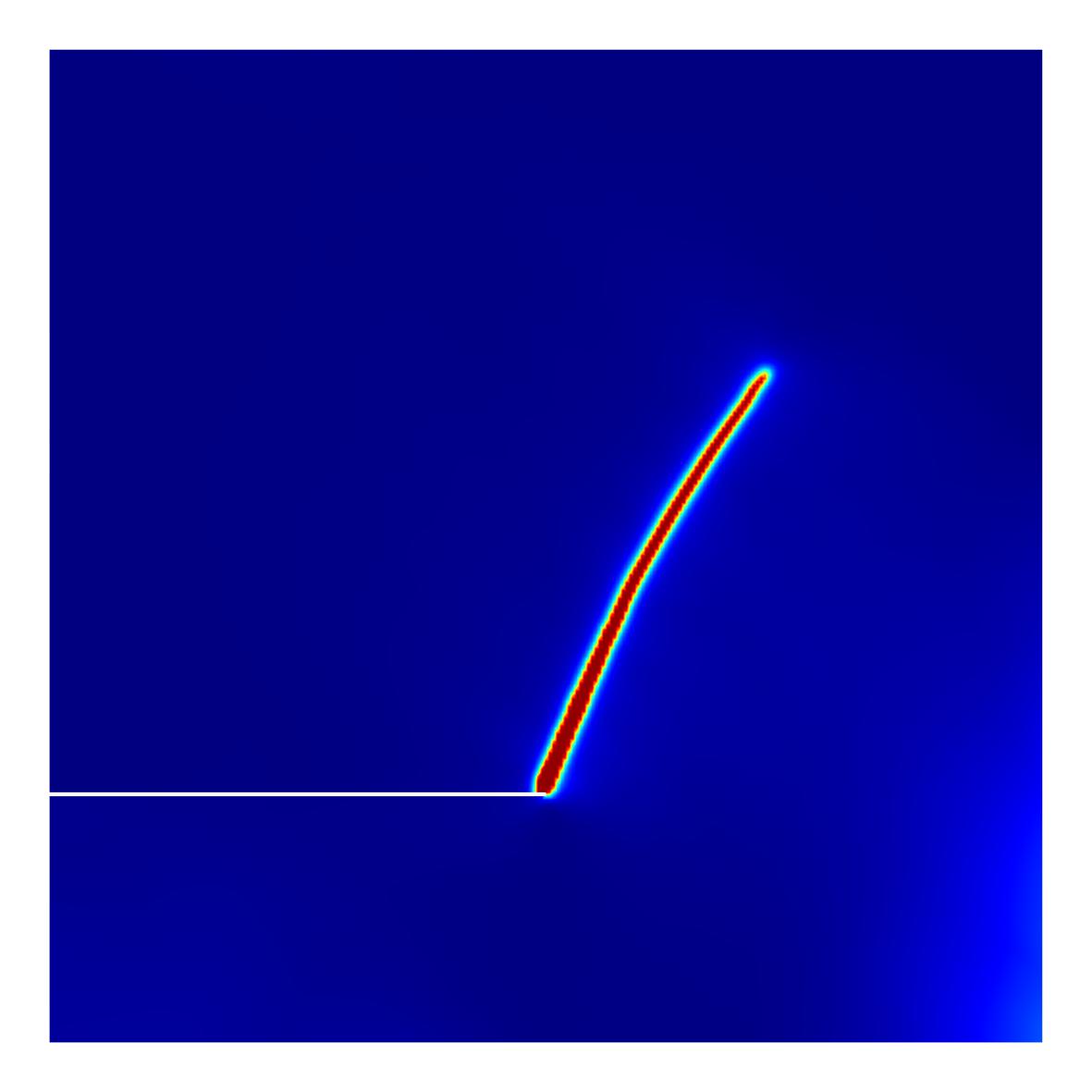}}
	\caption{Phase field of dynamic shear loading tests at 90 $\mu$s for different $G_c$}\citep{zhou2018phase}
	\label{Phase field of dynamic shear loading tests at 90 for different G_c}
	\end{figure}

\begin{figure}[htbp]
	\centering
	\includegraphics[width = 8cm]{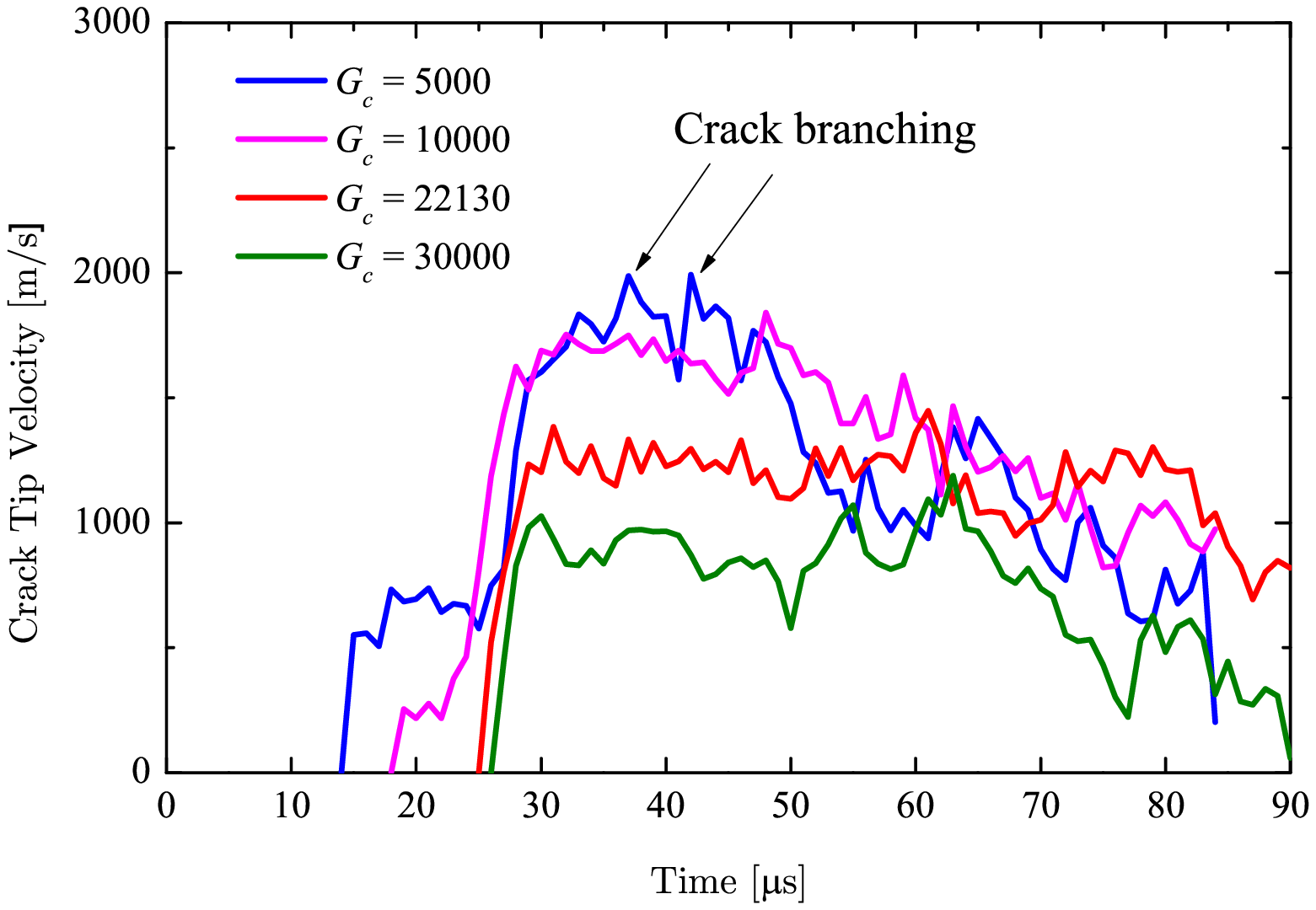}
	\caption{Crack-tip velocity of the dynamic shear loading example for different $G_c$}\citep{zhou2018phase}
	\label{Crack-tip velocity of the dynamic shear loading example for different G_c}
	\end{figure}

The phase field method has been used in Kalthoff experiment \citep{borden2012phase, liu2016abaqus, zhou2018phase} and its geometry and boundary condition are shown in Fig. \ref{Phase field of dynamic shear loading tests at 90 for different G_c}a. The influence of the critical energy release rate $G_c$ on the phase field is shown in Fig. \ref{Phase field of dynamic shear loading tests at 90 for different G_c}b to \ref{Phase field of dynamic shear loading tests at 90 for different G_c}e according to the results of \citet{zhou2018phase}. As observed, a smaller $G_c$ produces more complex crack patterns. Crack branching occurs for $G_c=5\times10^3$ and $1\times10^4$ J/m$^2$. Whereas, for a larger $G_c$, the phase field simulation only show a single crack. The distance of the crack tip from the upper boundary of the plate increases as $G_c$ increases. Figure \ref{Crack-tip velocity of the dynamic shear loading example for different G_c} shows the crack-tip velocity under different $G_c$. As observed, the maximum crack-tip velocity decreases with an increasing $G_c$ but the time of crack initiation increases \citep{zhou2018phase}.

\paragraph{2. 2D dynamic crack branching under tension}

\begin{figure}[htbp]
	\centering
	\includegraphics[width = 8cm]{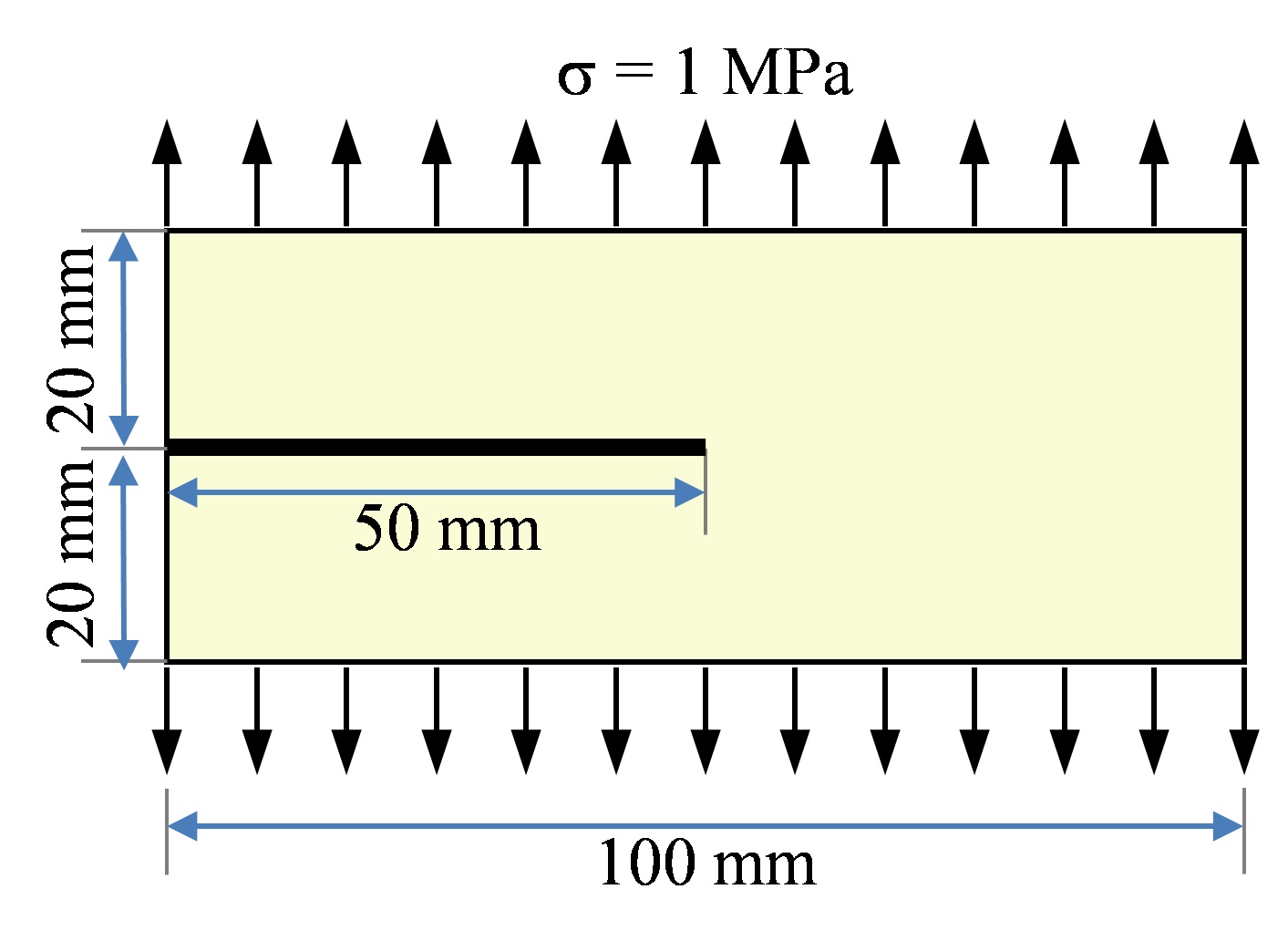}
	\caption{Geometry and boundary conditions for the case of dynamic crack branching}\citep{zhou2018phase}
	\label{Geometry and boundary conditions for the case of dynamic crack branching}
	\end{figure}

\begin{figure}[htbp]
	\centering
	\subfigure[$G_c=0.5$ J/m$^2$, $t=56$ $\mu$s]{\includegraphics[width = 8cm]{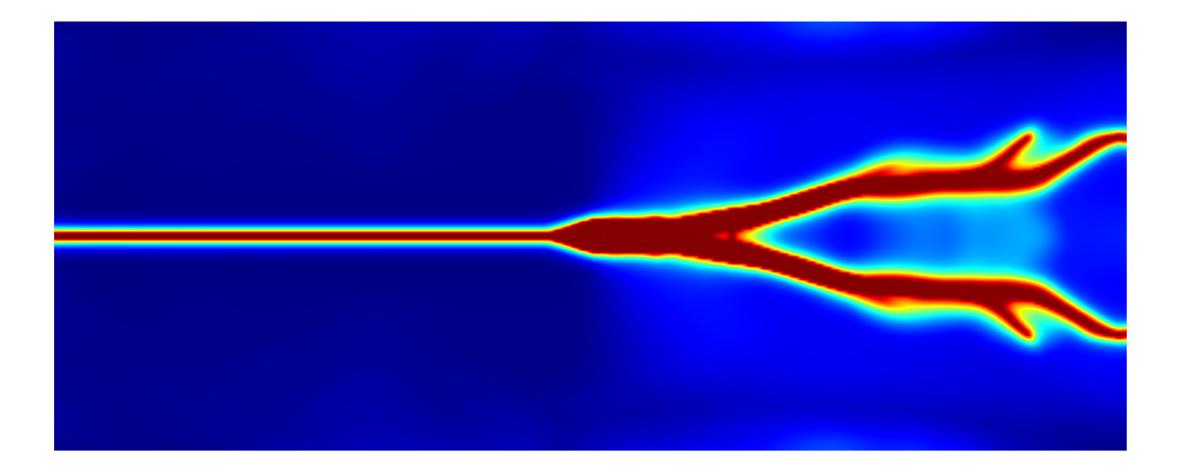}}
	\subfigure[$G_c=1$ J/m$^2$, $t=66$ $\mu$s]{\includegraphics[width = 8cm]{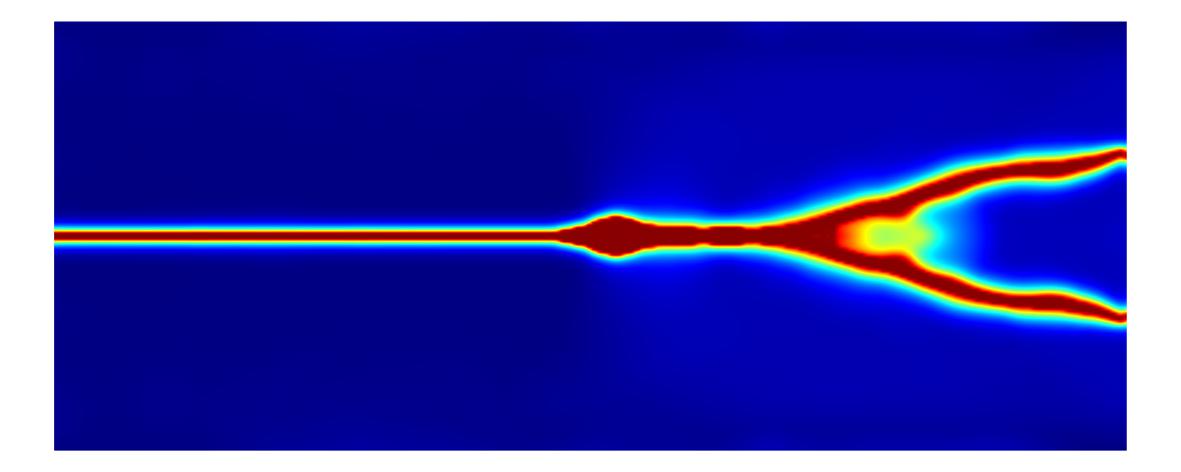}}
	\subfigure[$G_c=5$ J/m$^2$, $t=113$ $\mu$s]{\includegraphics[width = 8cm]{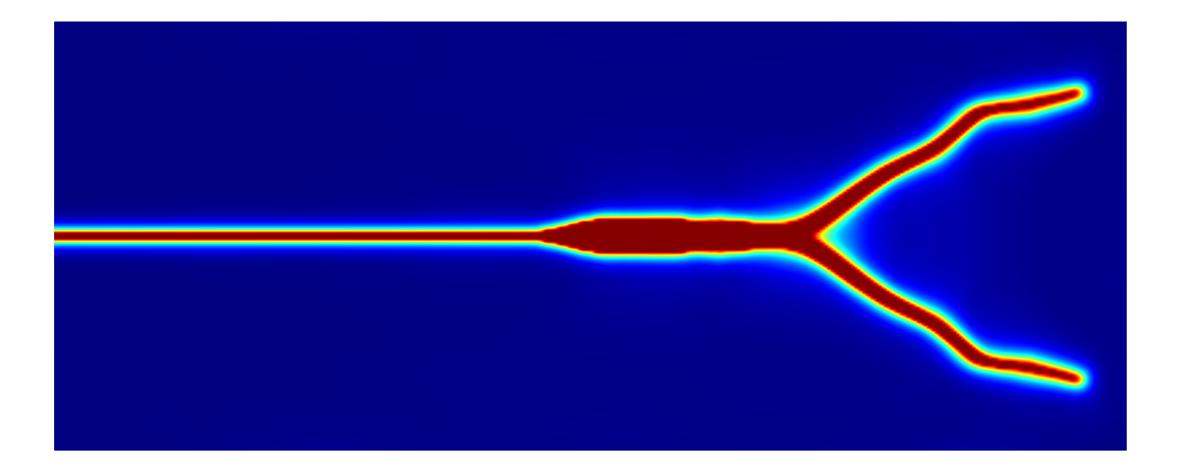}}
	\subfigure[$G_c=10$ J/m$^2$, $t=142$ $\mu$s]{\includegraphics[width = 8cm]{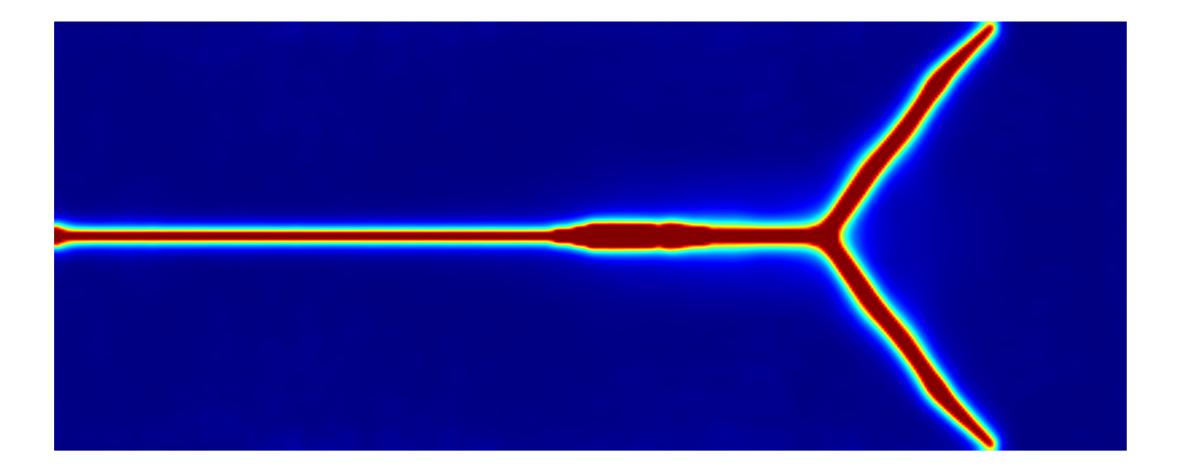}}
	\caption{Crack patterns of the 2D crack branching example for different $G_c$}\citep{zhou2018phase}
	\label{Crack patterns of the 2D crack branching example for different G_c}
	\end{figure}

\begin{figure}[htbp]
	\centering
	\includegraphics[width = 8cm]{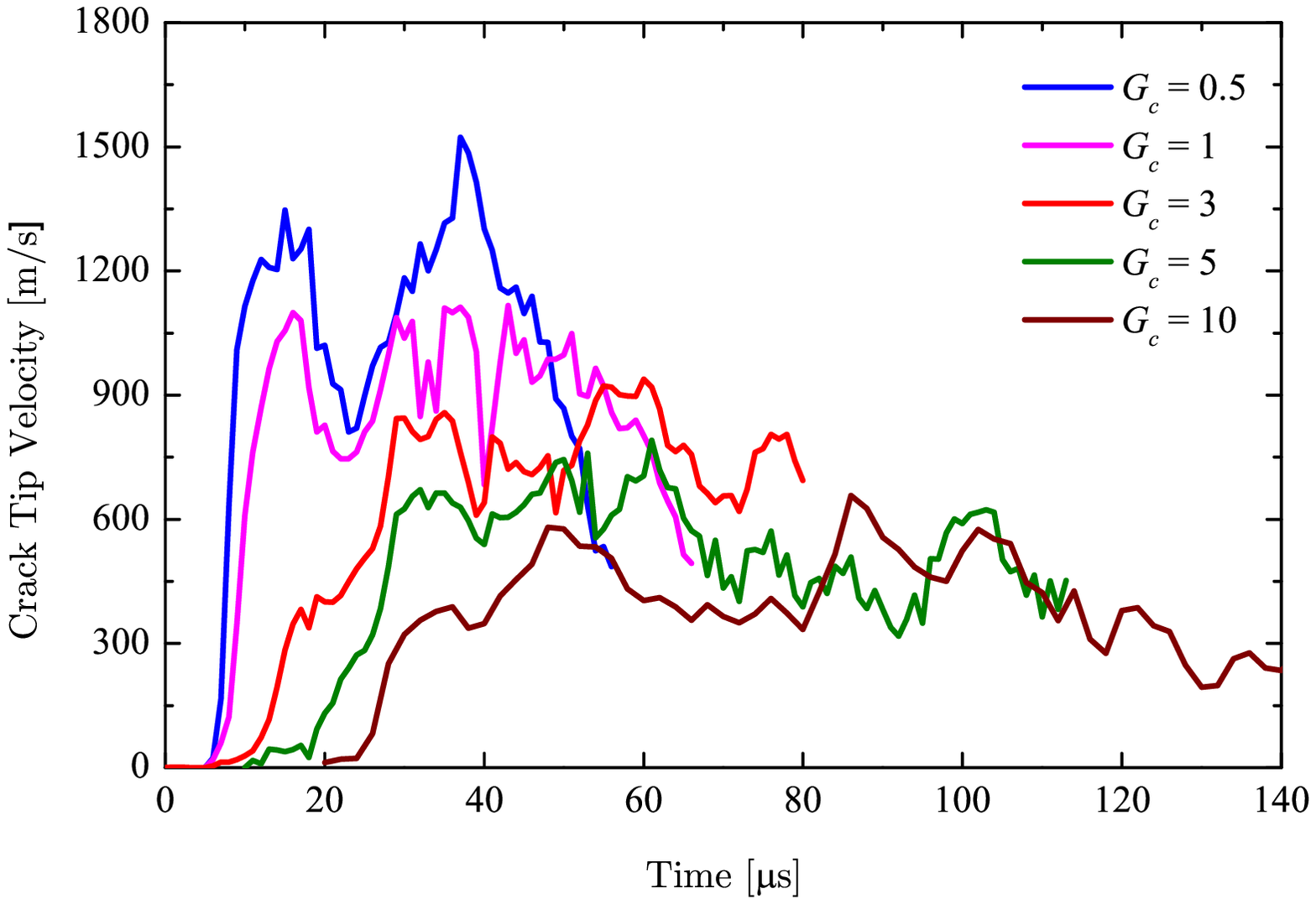}
	\caption{Crack-tip velocity of the 2D crack branching example for different $G_c$}\citep{zhou2018phase}
	\label{Crack-tip velocity of the 2D crack branching example for different G_c}
	\end{figure}

This example is a pre-notched rectangular plate subjected to uniaxial traction and has been investigated by \citet{borden2012phase, liu2016abaqus, zhou2018phase}. Geometry and boundary condition of the pre-notched plate are given in Fig. \ref{Geometry and boundary conditions for the case of dynamic crack branching}. The influences of the critical energy release rate $G_c$ on the crack pattern and the crack-tip velocity are shown in Figs. \ref{Crack patterns of the 2D crack branching example for different G_c} and \ref{Crack-tip velocity of the 2D crack branching example for different G_c}. As observed, multiple crack branching occurs for different $G_c$. In addition, the crack propagates at a larger angle with the horizontal after the first crack branching under a larger $G_c$. The maximum crack-tip velocity in Fig. \ref{Crack-tip velocity of the 2D crack branching example for different G_c} decreases with the increase in $G_c$.

% \subsection{Thermal fracture}
\subsection{Hydraulic fracture}
\subsubsection{Propagation of a single crack by internal fluid injection}

This example shows propagation of one single pre-existing crack when fluid is injected. The crack is placed horizontally at the center of a square specimen of 4 m $\times$ 4 m. The initial length of the crack is 0.4 m and the displacements on the outer boundaries of the square are fixed with the fluid pressure $p=0$. Q4 elements are used to discretize all the fields with element size $h=$ $2\times 10^{-2}$ m. Here, we used the same method and parameters in \citet{zhou2018phase2} to simulate the crack pattern and pressure field.

\begin{figure}[htbp]
	\centering
	\subfigure[Crack pattern]{\includegraphics[height = 6cm]{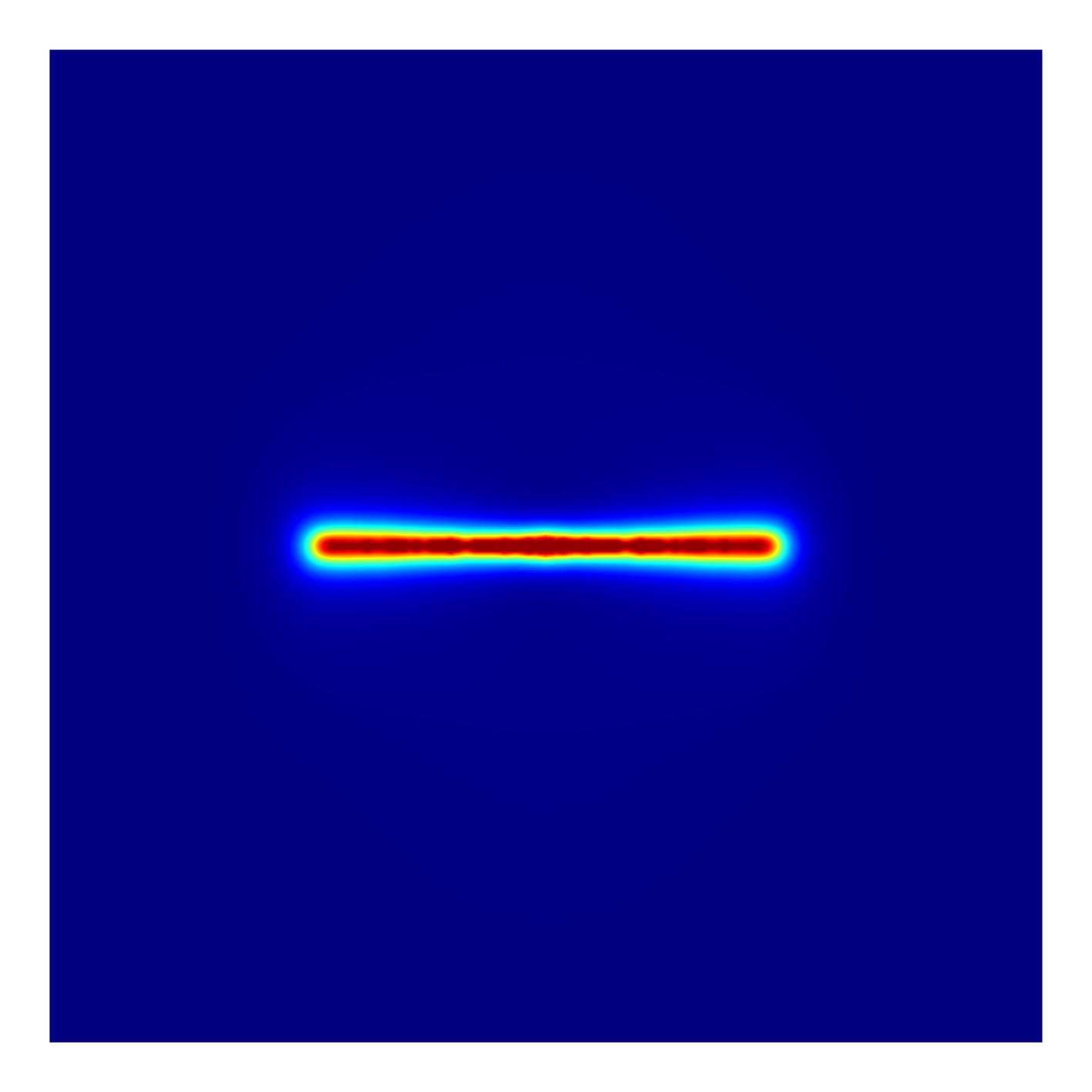}}
	\subfigure[Fluid pressure]{\includegraphics[height = 6cm]{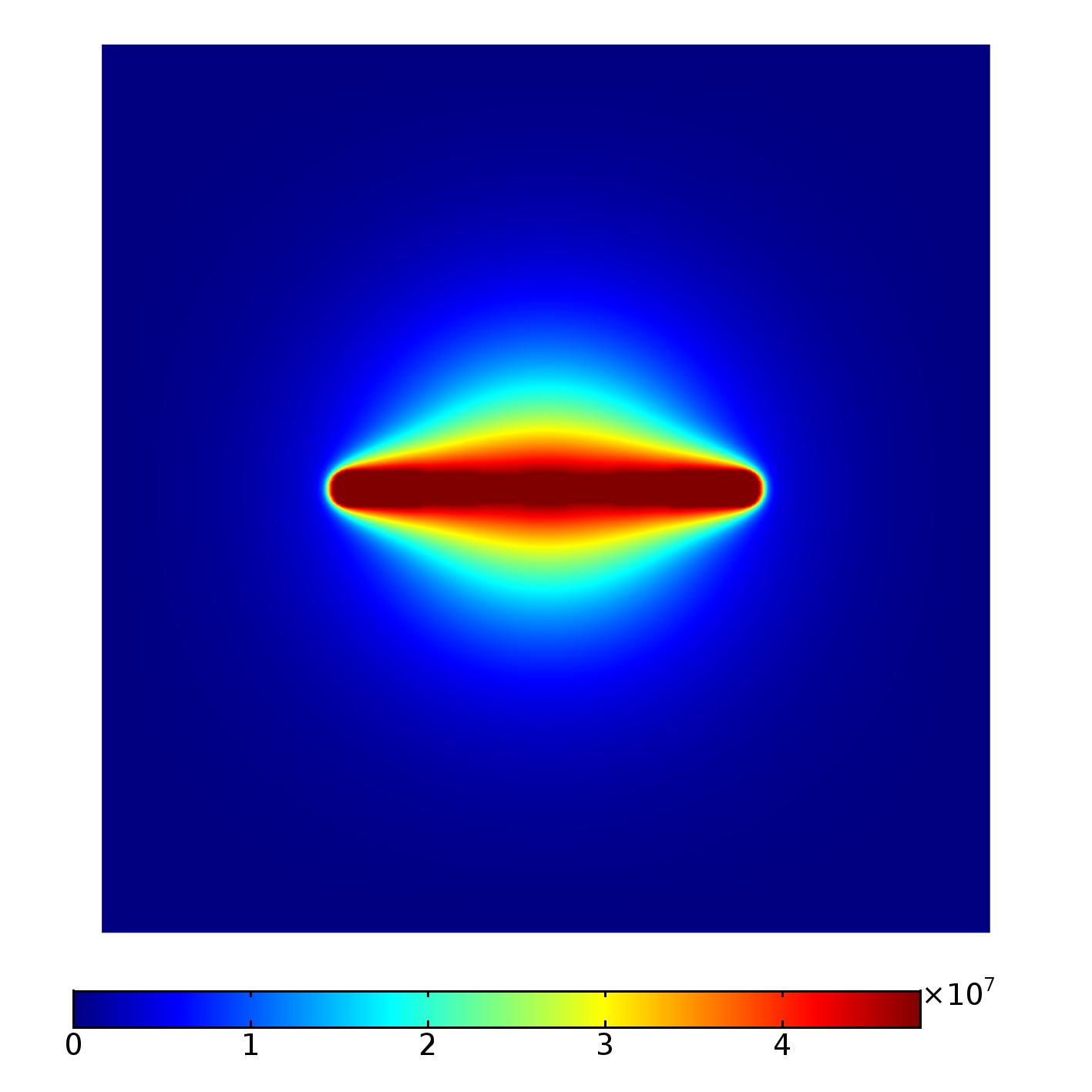}}\\

	\caption{Crack pattern and fluid pressure for the example of one single pre-existing crack subjected to internal fluid injection at $t=8.8$ s}
	\label{Crack pattern and fluid pressure for the example of one single pre-existing crack subjected to internal fluid injection}
	\end{figure}

The fracture pattern at $t=8.8$ s is shown in Fig. \ref{Crack pattern and fluid pressure for the example of one single pre-existing crack subjected to internal fluid injection}a. As expected, the single pre-existing crack propagates along the horizontal direction, in good agreement with the results of \citet{mikelic2015phase} and \citet{mikelic2015phase2}. The pressure distribution at $t=8.8$ s is presented in Fig. \ref{Crack pattern and fluid pressure for the example of one single pre-existing crack subjected to internal fluid injection}b. The pressure field is similar to the phase field in Fig. \ref{Crack pattern and fluid pressure for the example of one single pre-existing crack subjected to internal fluid injection}a. However, the region where the pressure concentrates is larger than the cracked region because of the radial penetration of the fluid is used in \citet{zhou2018phase2}. 

\subsubsection{Two parallel propagating cracks subjected to internal fluid injection}
 
\begin{figure}[htbp]
	\centering
	\subfigure[Geometry and boundary condition]{\includegraphics[height = 6cm]{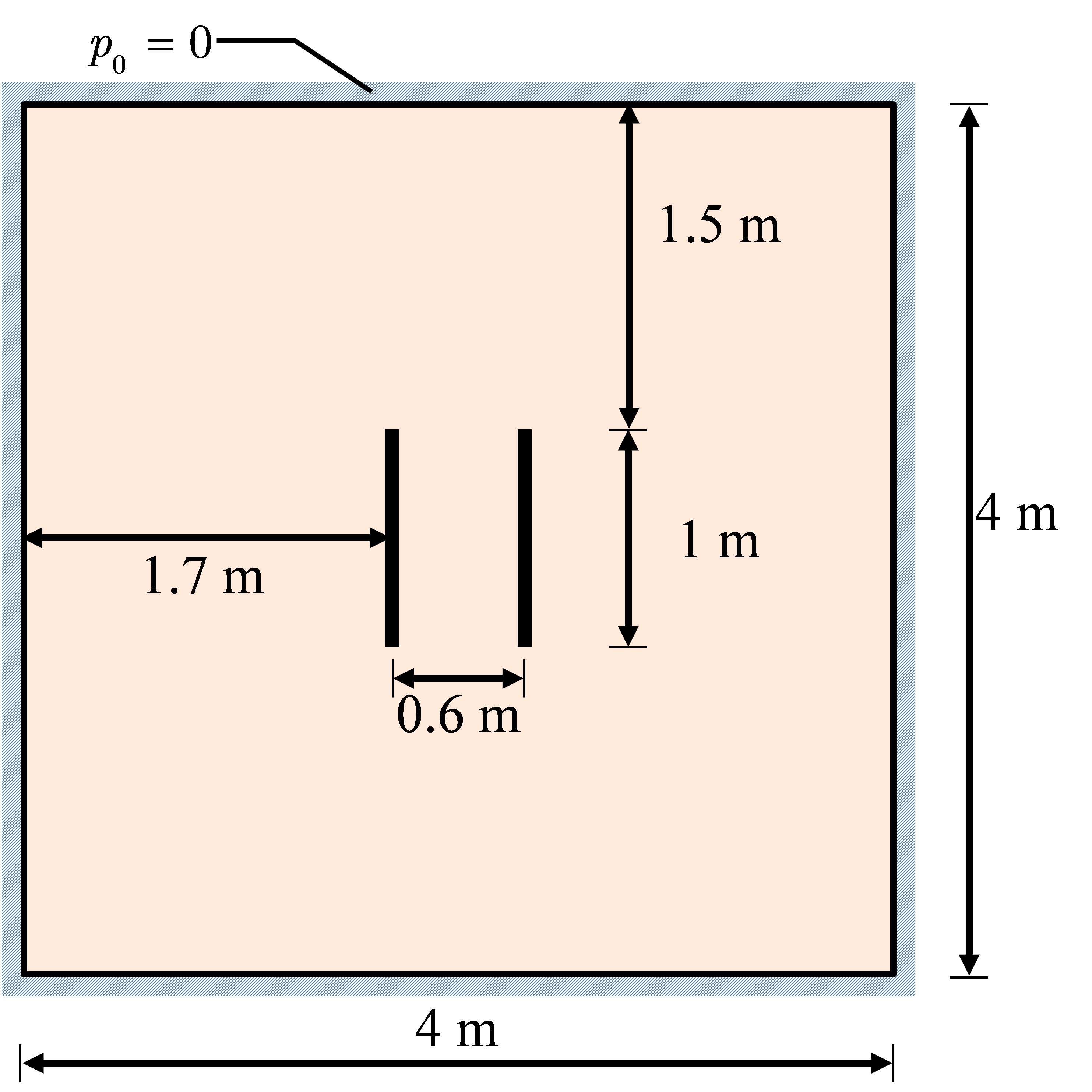}}
	\subfigure[Crack pattern]{\includegraphics[height = 6cm]{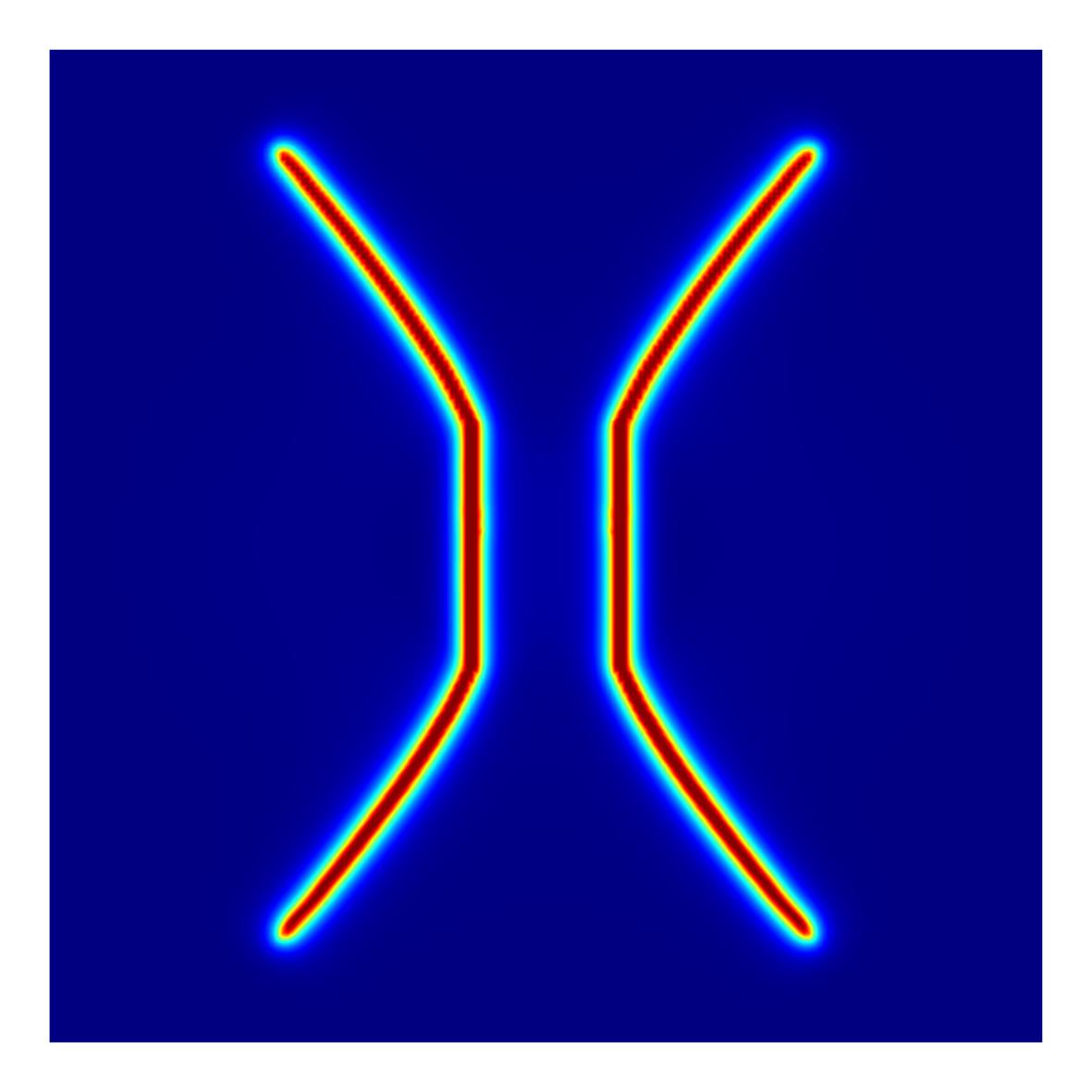}}
	\subfigure[Pressure field]{\includegraphics[height = 6cm]{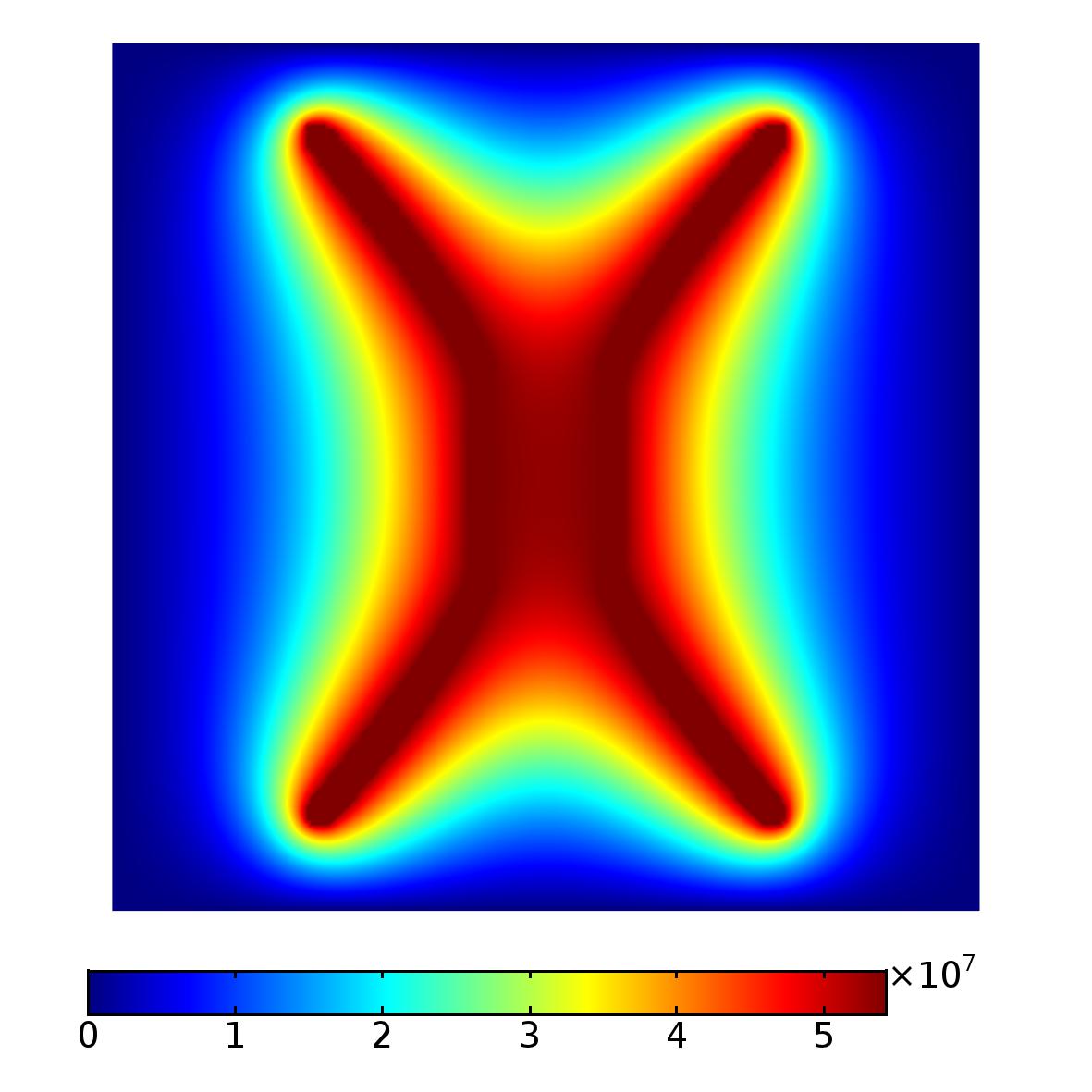}}\\
	\caption{The example of two parallel pre-existing cracks}
	\label{The example of two parallel pre-existing cracks}
	\end{figure}

This example shows propagation of two parallel pre-existing cracks when fluid is injected. Geometry and boundary conditions of this example are shown in Fig. \ref{The example of two parallel pre-existing cracks}a. The two pre-existing cracks are 1 m in length and have a spacing of 0.6 m. Q4 elements are used to discretize all the fields with the element size $h=$ $2\times 10^{-2}$ m. The evolution of the phase field is presented in Fig. \ref{The example of two parallel pre-existing cracks}b.  The spacing of the propagating cracks increases with increasing time. The pressure distribution is shown in Fig. \ref{The example of two parallel pre-existing cracks}c. The pressure is similar to the phase field but has a larger transition band. 

\subsubsection{Three parallel propagating cracks in 2D}

\begin{figure}[htbp]
	\centering
	\subfigure[Geometry and boundary condition]{\includegraphics[height = 6cm]{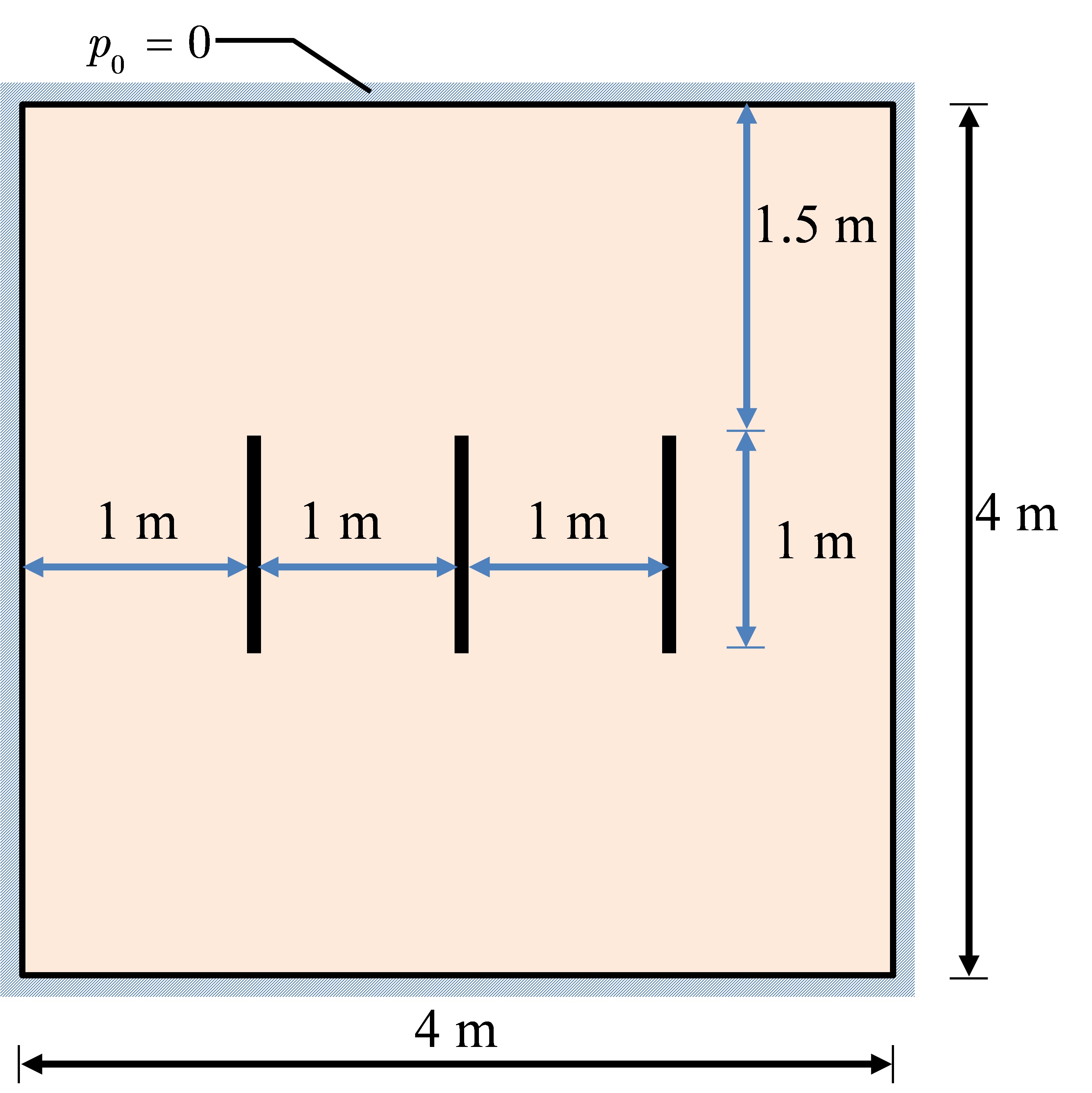}}
	\subfigure[Crack pattern]{\includegraphics[height = 6cm]{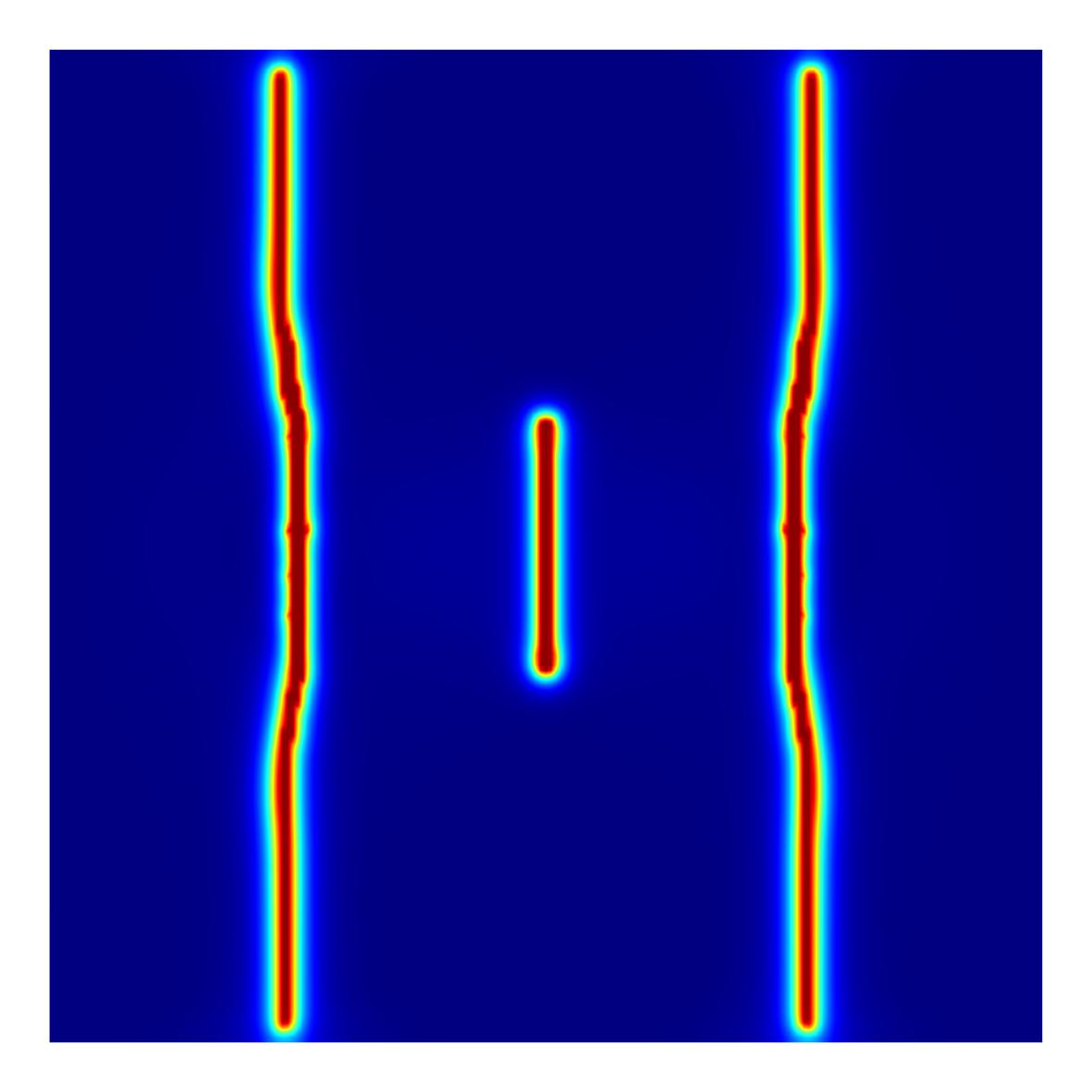}}
	\subfigure[Pressure field]{\includegraphics[height = 6cm]{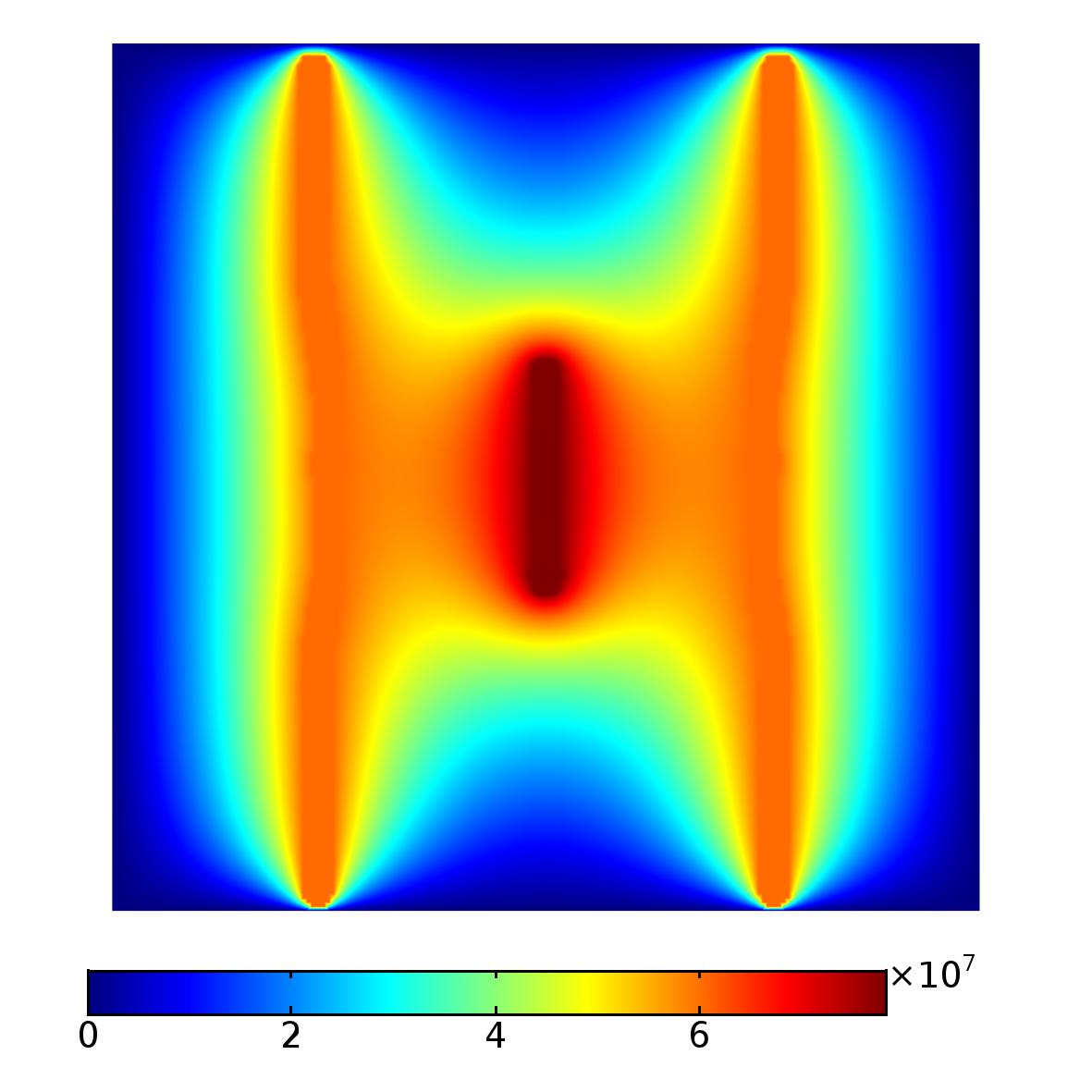}}\\
	\caption{The example of three parallel pre-existing cracks in 2D}
	\label{The example of three parallel pre-existing cracks in 2D}
	\end{figure}

Crack propagation of three propagating cracks in 2D is simulated by the phase field model. The geometry and boundary conditions are given in Fig. \ref{The example of three parallel pre-existing cracks in 2D}a. The three cracks have the same length of 1 m and a spacing of 1 m. Uniformly spaced Q4 elements with size $h=2\times 10^{-2}$ m are employed. The crack pattern is shown in Fig. \ref{The example of three parallel pre-existing cracks in 2D}b. No propagation is found for the middle crack and only the left and right cracks propagate. The pressure distribution  is shown in Fig. \ref{The example of three parallel pre-existing cracks in 2D}c. The pressure in the middle crack is much larger than that in the left and right cracks because the middle crack barely propagates. 

\subsubsection{Propagation of two parallel penny-shaped cracks in 3D}

\begin{figure}[htbp]
	\centering
	\subfigure[$t=13.2$ s]{\includegraphics[height = 6cm]{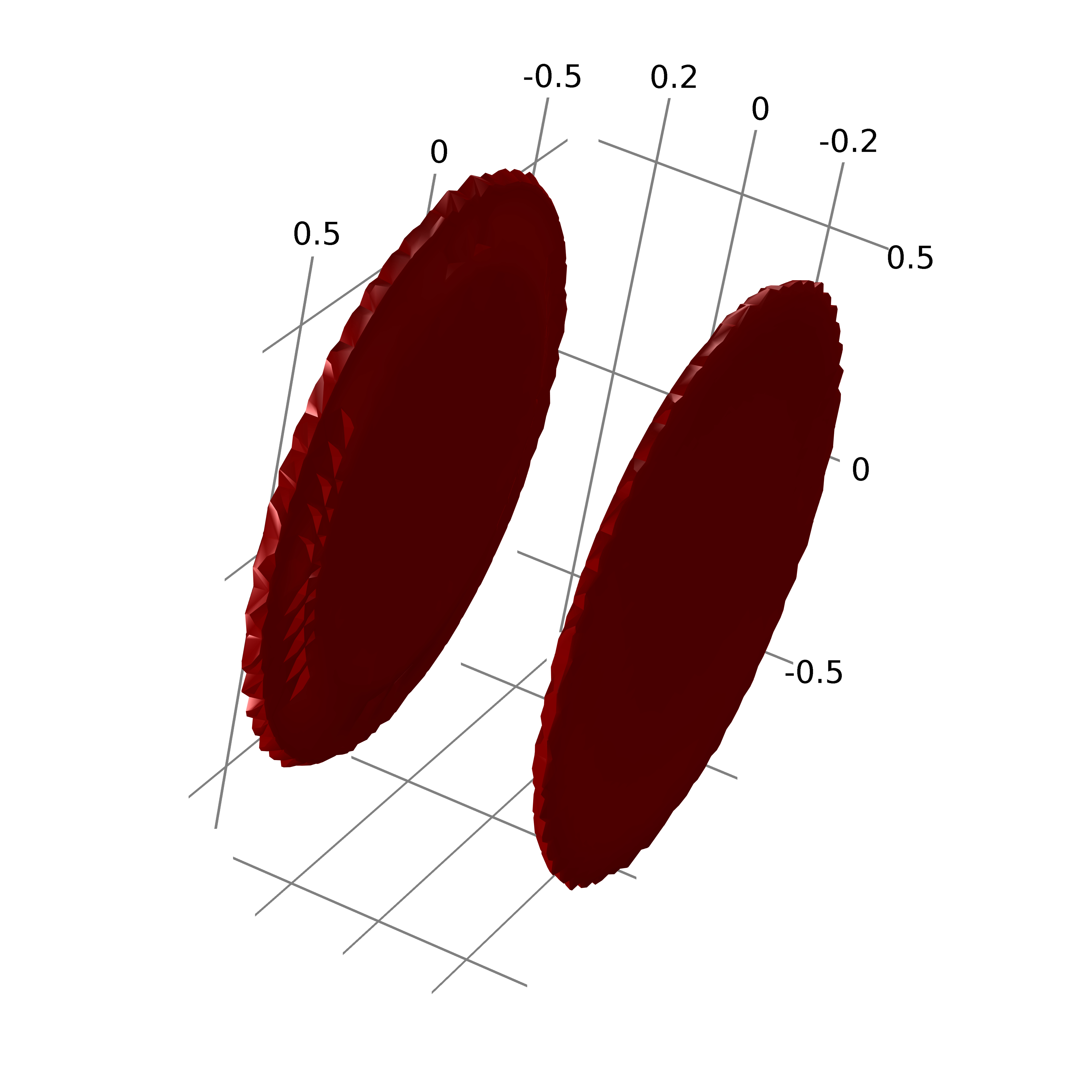}}
	\subfigure[$t=13.3$ s]{\includegraphics[height = 6cm]{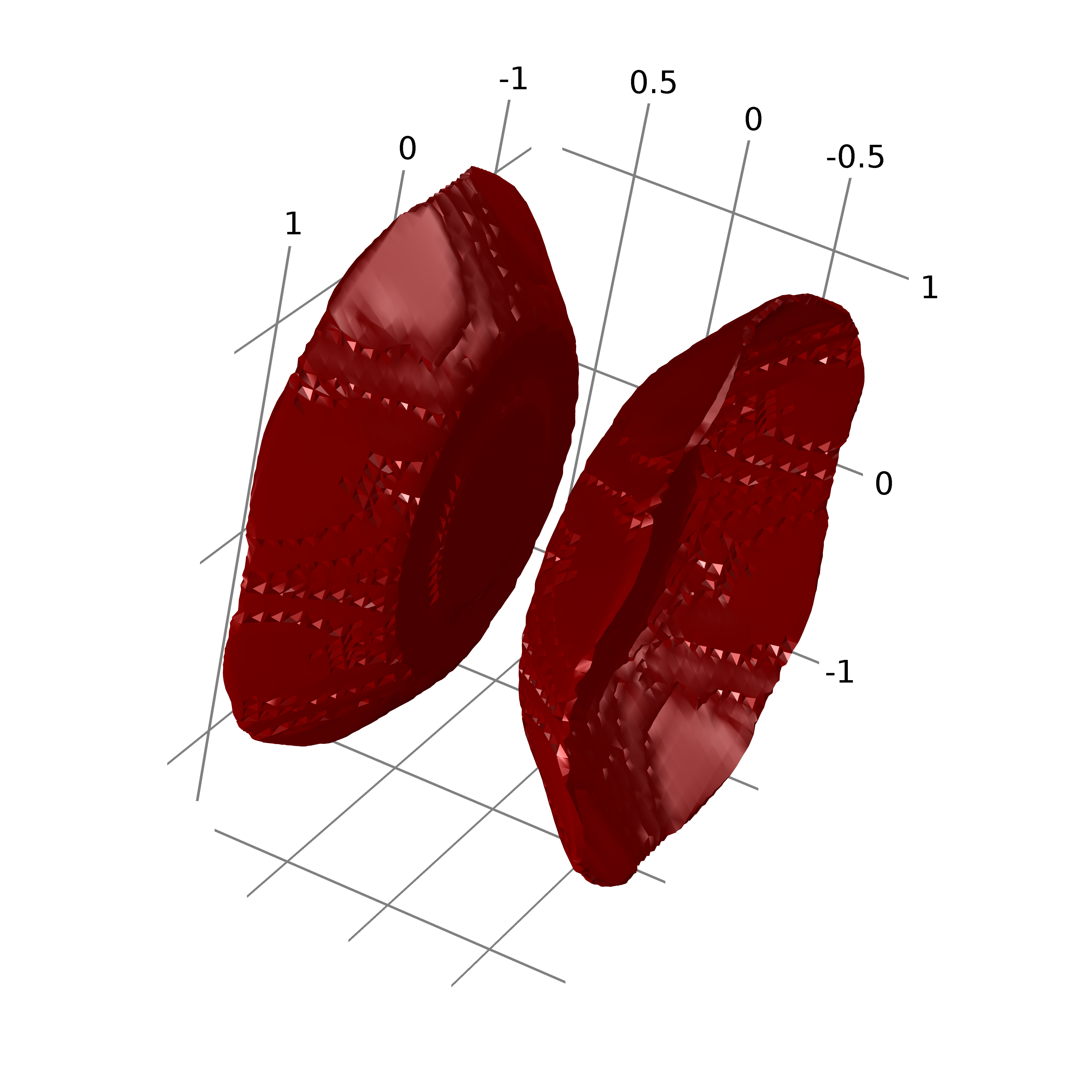}}\\
	\caption{Crack patterns for the example of two parallel penny-shaped cracks in 3D}
	\label{Crack patterns for the example of two parallel  penny-shaped cracks in 3D}
	\end{figure}

This example verifies the capability of the phase field modeling for hydraulic fractures in 3D. Two parallel penny-shaped cracks are set in the domain of (-2 m, 2 m)$^3$. The initial crack radius is 0.5 m. More details can be referred to \citet{zhou2018phase2}. Crack propagation patterns of the two parallel penny-shape cracks in 3D are shown in Fig. \ref{Crack patterns for the example of two parallel  penny-shaped cracks in 3D}. The radii of the two cracks exceed 0.5 m when $t=13.2$ s and bowl-shaped cracks are observed when $t=13.3$ s.

\subsection{Fractures in plates and shells}

To illustrate the capability of phase field modeling on capturing cracking in thin structures, three representative numerical examples are given with a focus on qualitatively crack paths formed from these examples in this subsection. The employed approach is based on \citet{Kiendl2016}, where Kirchhoff-Love thin shell and brittle fracture are considered in linear regime. To accommodate  C0-continuity requirement from the shell theory, quadratic NURBS basis functions are used to approximate solution fields. 

\subsubsection{Single edge notched tension test}

\begin{figure}[!h]
\centering
\includegraphics[width=0.3\textwidth]{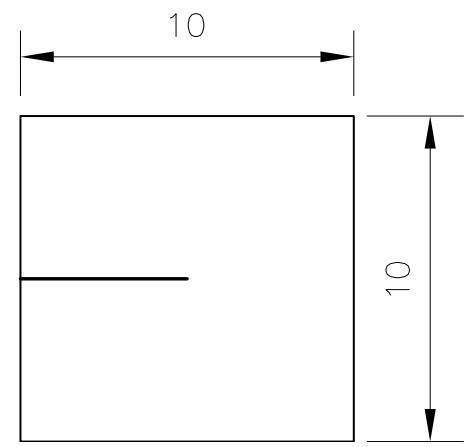}
  \caption{Tension test. Geometry}.
 \label{TensionTest_geometry}
\end{figure}

\begin{figure}[h]
\centering
  \begin{tabular}{@{}cccc@{}}
    \includegraphics[width=0.2\textwidth]{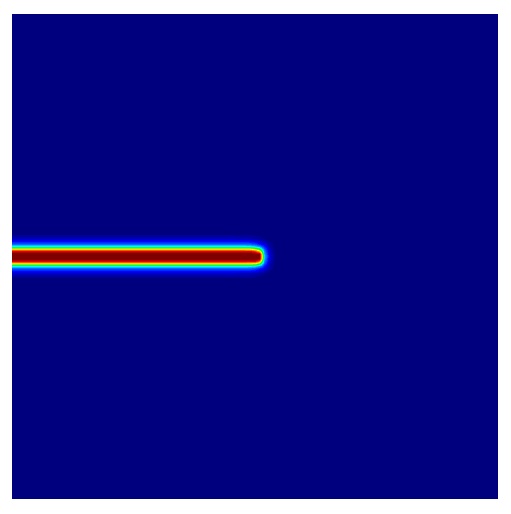}  
    \includegraphics[width=0.2\textwidth]{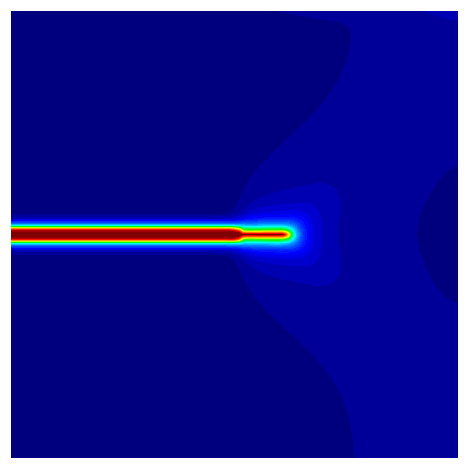} &  \\[0.1cm]
    \includegraphics[width=0.2\textwidth]{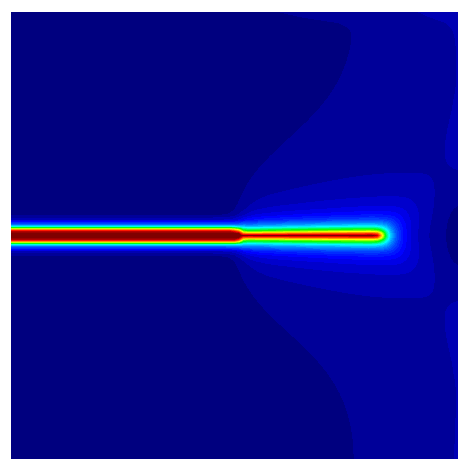}  
    \includegraphics[width=0.2\textwidth]{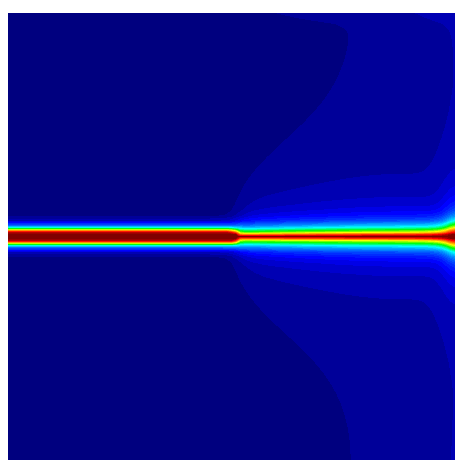} &  \\[0.1cm]
  \end{tabular}
  \caption{Tension test. Crack patterns at various loading stages}.
 \label{TensionTest_crack_pattern}
\end{figure}

\begin{figure}[!h]
\centering
\includegraphics[width=0.6\textwidth]{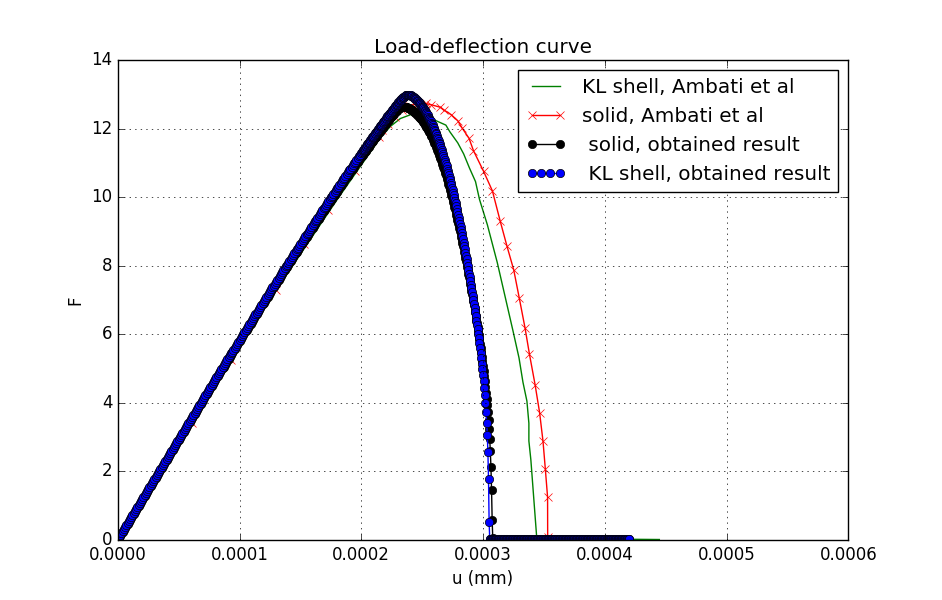}
  \caption{Tension test. Load-deflection curve}.
 \label{TensionTest_l_d_curve}
\end{figure}

First, a square plate consisting of a horizontal notch is considered. The top edge of the specimen is applied vertical displacement. The geometric properties of the plate are depicted in Fig. \ref{TensionTest_geometry} . The material parameters are chosen such that $E = 1$ GPa, $ \rho = 0.3$, $G_c = 2$ N/mm and $l_0 = 0.02$ mm. 23,000 elements are used to discretize the analysis domain and refine a priori in expected crack zones. Displacement control method is applied to conduct the simulation with displacement increment $\Delta u= 10^{-6}$ mm for the first 200 loading steps and $\Delta u=0.5\times10^{-6}$ mm for the remaining steps until the specimen are broken completely. Fig. \ref{TensionTest_crack_pattern} shows the crack patterns at different loading stages. The resulting crack paths are coincident with those obtained from previous works of \citet{ambati2016phase} and \citet{Kiendl2016}. The load-deflection curves for solid and Kirchhoff-Love shell elements in Fig. \ref{TensionTest_l_d_curve} are comparable to each other and in good agreement with those from \citet{ambati2016phase}.

\subsubsection{Simply supported Plate}

\begin{figure}[!t]
	\centering
	\includegraphics[width=0.25\textwidth]{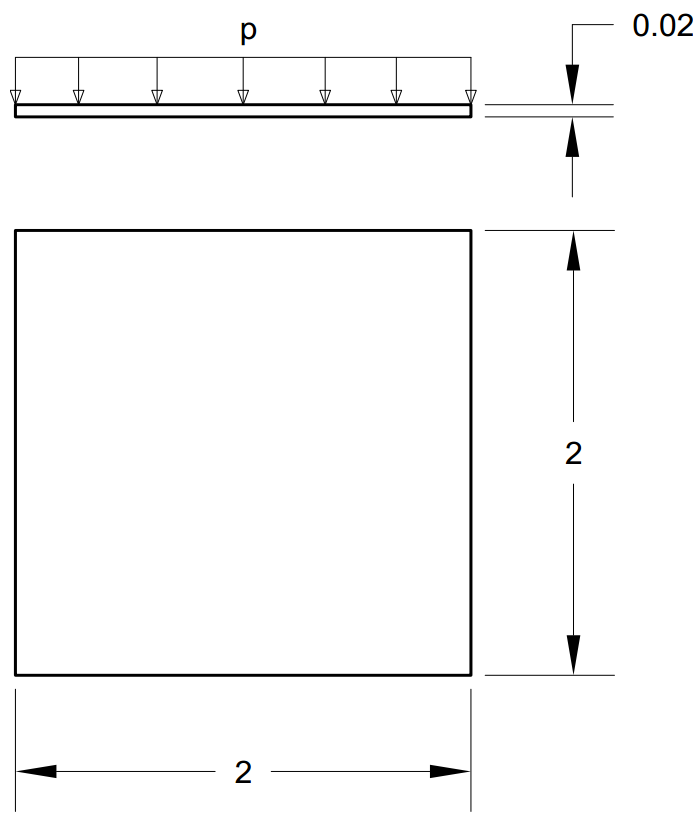}
	  \caption{Square plate. Geometry and boundary condition setups}.
	 \label{SquarePlate_geometry}
	\end{figure}

\begin{figure}[h]
	\centering
	  \begin{tabular}{@{}cccc@{}}
	    \includegraphics[width=0.2\textwidth]{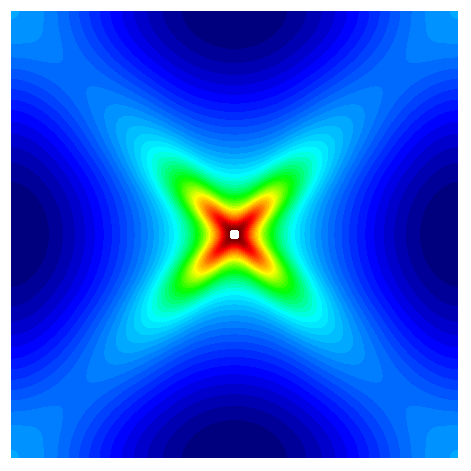}  
	    \includegraphics[width=0.2\textwidth]{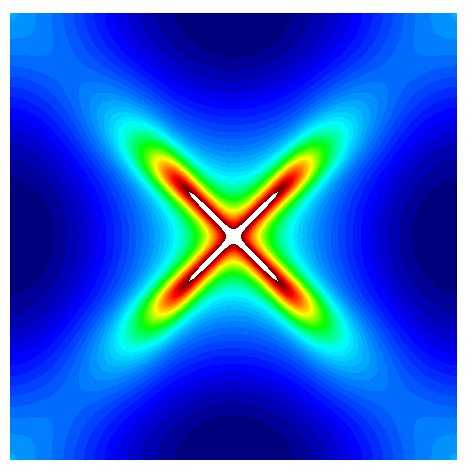} &  \\[0.1cm]
	    \includegraphics[width=0.2\textwidth]{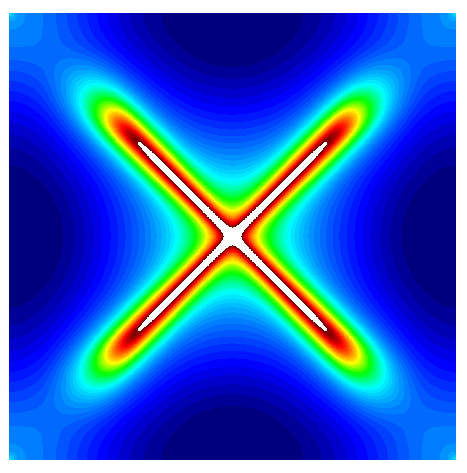}  
	    \includegraphics[width=0.2\textwidth]{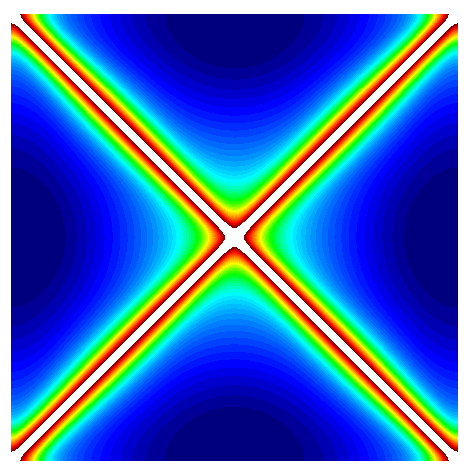} &  \\[0.1cm]
	  \end{tabular}
	  \caption{Simply supported plate. Crack patterns at various loading stages}.
	 \label{SquarePlate_crack_pattern}
	\end{figure}

\begin{figure}[!h]
	\centering
	\includegraphics[width=0.5\textwidth]{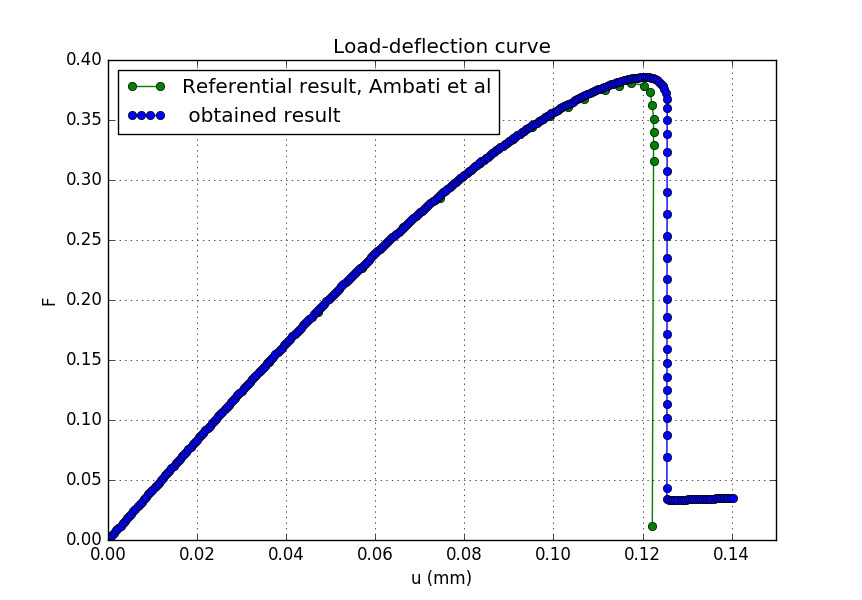}
	  \caption{Square plate. Load-deflection curve}.
	 \label{SquarePlate_l_d_curve}
	\end{figure}

Next, a square plate without initial cracks subjected to uniform pressure is simply supported on its four edges. The geometry and boundary conditions of the problem are depicted in Fig. \ref{SquarePlate_geometry} and  material parameters are chosen as $E = 190$ GPa, $\rho = 0.29$, $G_c = 0.295$ N/mm and $l_0 = 0.02$ mm. The model is discretized with 28,000 elements and $h/l_0 = 1.7$. With such applied load, arc-length control is utilized as a solution strategy to track mechanical response of the structure when crack starts to propagate. The resulting crack patterns are depicted at various stages of deformation as in Fig. \ref{SquarePlate_crack_pattern}. As can be seen, crack initiation arises at the center and then crack branching occurs, leading to the evolution of cracks towards the corners, which is coincident with observations from previous investigations \citep{ambati2016phase,Kiendl2016}.  Figure \ref{SquarePlate_l_d_curve} shows the corresponding load-deflection curve, which is in agreement with one obtained by \citet{ambati2016phase}.

\subsubsection{Notched cylinder with internal pressure}

\begin{figure}[!h]
	\centering
	\includegraphics[width=0.7\textwidth]{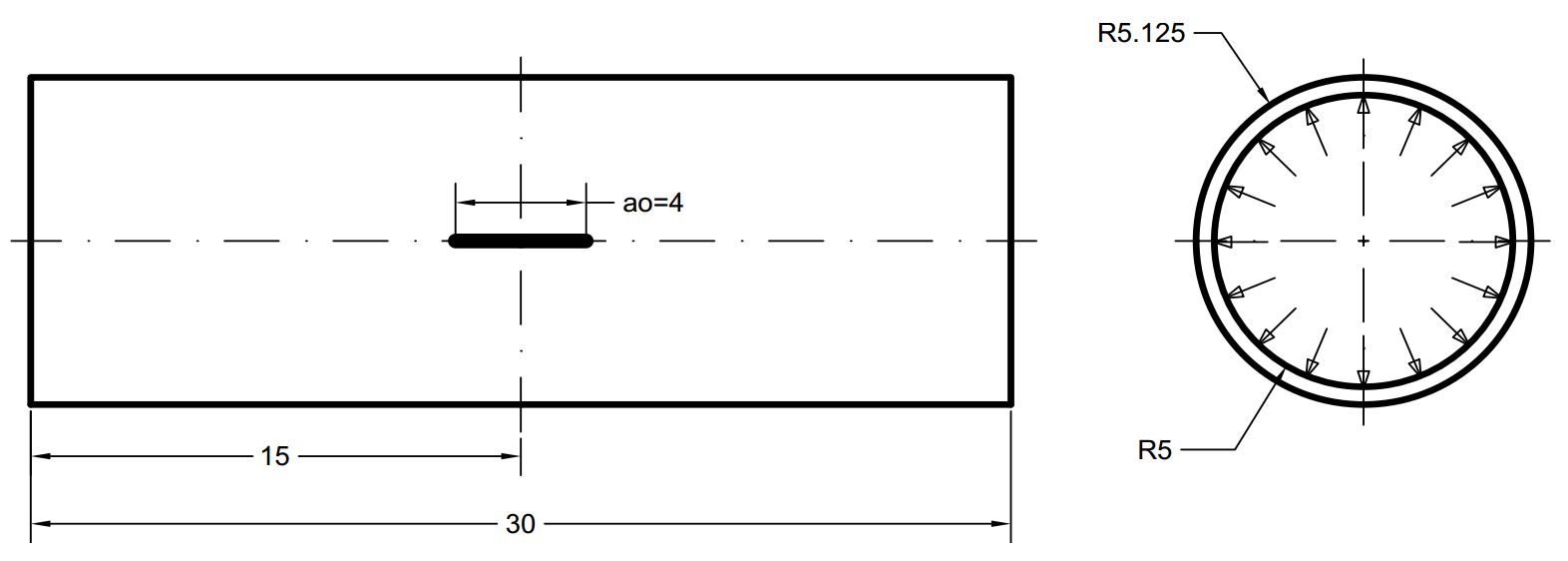}
	  \caption{Notched cylinder with internal pressure. Geometry and boundary condition setups}.
	 \label{NotchedCylinder_geometry}
	\end{figure}

\begin{figure}[!h]
	\centering
	  \begin{tabular}{@{}cccc@{}}
	    \includegraphics[width=0.4\textwidth]{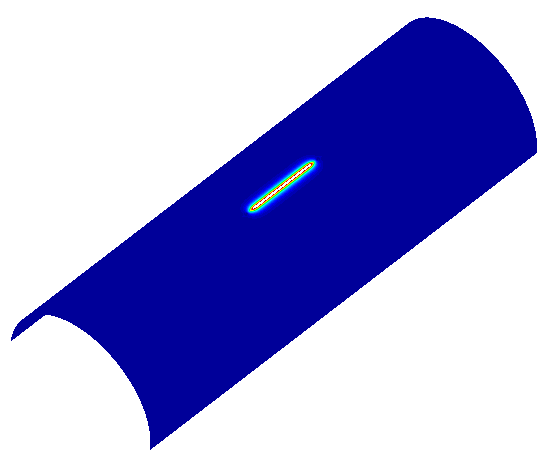}  
	    \includegraphics[width=0.4\textwidth]{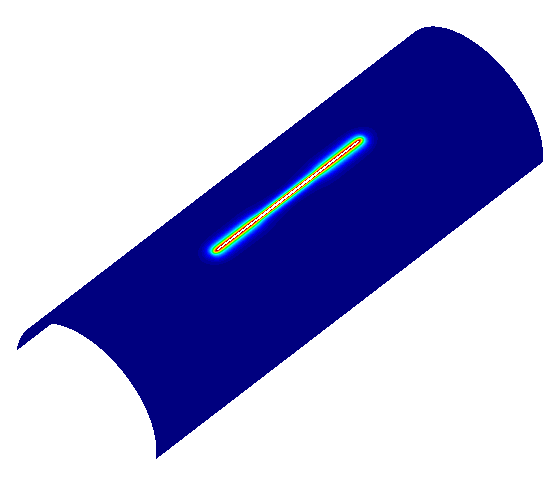} &  \\[0.1cm]
	    \includegraphics[width=0.4\textwidth]{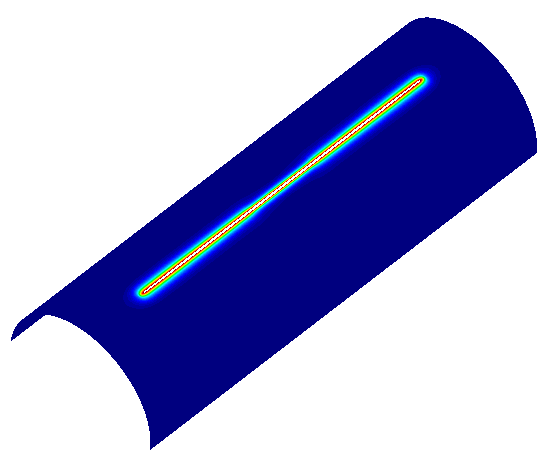}  
	    \includegraphics[width=0.4\textwidth]{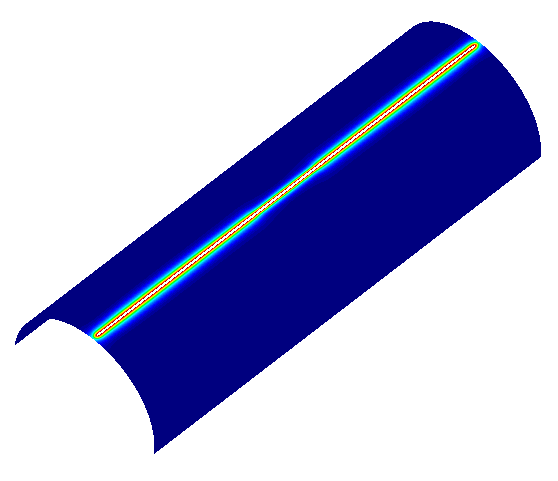} &  \\[0.1cm]
	  \end{tabular}
	  \caption{Notched plate with hole. Crack patterns at different loading stages}.
	 \label{NotchedPlateCrackPattern}
	\end{figure}

Finally, a curved shell structure is considered. It consists of two notches in the axial direction, which are located on opposite sides and the shell is subjected to internal pressure. Geometry and boundary condition setups are given in Fig. \ref{NotchedCylinder_geometry}. The material parameters are chosen as $E = 70$ GPa , $\rho = 0.3$, $G_c = 1.5$ N/mm and $l_0 = 0.05$ mm. A mesh with 32,000 elements, which is refined a priori in the regions where cracks are expected to propagate,  is used to discretize the shell structure . Arc-length control is adopted for the simulation. Fig. \ref{NotchedPlateCrackPattern} shows crack phase field at different stages of deformation. As expected, a straight crack propagating axially  is observed, which is coincident with those obtained from previous investigations \citep{ambati2016phase,Kiendl2016}.

\section {Conclusions}\label{Conclusions}

This work presents an overall picture and some recent progress of the phase field method in modeling fracture initiation and propagation. This work sums up the significant stages during the development of the phase field models of fracture in both physics and mechanics communities. The theory and the use of the phase field models are also shown to demonstrate their advantages of handling complex fracture problems. The phase field models use an additional scalar field (phase field) to represent the discrete cracks rather than introduce directly physical discontinuity. The fracture shape and propagation are only determined from the evolution equations of the phase-field. Compared with other fracture methods, the phase field model does not require additional work to track the fracture surfaces algorithmically and thereby less computational effort is needed for numerical implementation of fracture. Crack branching and merging in materials with arbitrary 2D and 3D geometries can be easily simulated by the phase field models.

Although the phase field model can be coupled with different discretization methods such as isogeometric analysis and mesh-free methods, most phase field models use the finite element discretization. This work also presents details about computer implementation of the phase field models coupled with finite element methods. Because of the great benefits from the phase field modeling, the phase field models for brittle fractures are extended for solving more complicated problems such as ductile fractures and multi-field problems. In addition, some representative 2D and 3D examples for quasi-static and dynamic fractures are presented to show practicability and capability of the phase field modeling. For future work, the phase field models can be applied in fracture modeling in caverns used for compressed air energy storage (CAES) \citep{zhou2015analytical, zhou2017numerical}, which shows strong effect of cyclic stress and temperature \citep{xia2015strength, zhou2015damage, zhou2017statistical, zhou2018theoretical}.

\section*{Acknowledgement}
The authors gratefully acknowledge financial support provided by the Natural Science Foundation of China (51474157), and RISE-project BESTOFRAC (734370).

\bibliography{references}

%\onecolumn
%\tableofcontents
\end{document}